\documentclass[twocolumn, tighten]{aastex63}

\usepackage{hyperref}
\usepackage{textcomp}
\usepackage{bm,color}
\usepackage{amsmath}
\usepackage{ulem}
\usepackage{csquotes}
\usepackage[T1]{fontenc}

\newcommand{\pb}{\textsc{Polarbear}}

\newcommand{\sptpol}{{\sc SPTpol}}
\newcommand{\act}{{\sc ACT}}
\newcommand{\actpol}{{\sc ACTPol}}

\newcommand{\planckeight}{Planck~2018}
\newcommand{\planck}{Planck}
\newcommand{\apex}{{\sc APEX}}
\newcommand{\abs}{{\sc ABS}}
\newcommand{\quiet}{{\sc QUIET}}
\newcommand{\bicep}{{BICEP}}
\newcommand{\wmap}{{\sc WMAP}}
\newcommand{\taua}{Tau~A}
\newcommand{\lcdm}{$\Lambda$CDM}
\newcommand{\keckarray}{\textit{Keck Array}}

\newcommand{\nbolo}{1274}
\newcommand{\rotvalanderr}{$-0 \fdg 61 \pm 0 \fdg 22$}
\newcommand{\gainvalanderr}{$1.08 \pm 0.04$}
\newcommand{\rupperlimit}{$r < 0.90$}
\newcommand{\mapdepth}{$32\,\mu\mathrm{K}$-$\mathrm{arcmin}$}
\newcommand{\ellknee}{$\ell = 90$}
\newcommand{\arraynet}{$\mathrm{NET}_\mathrm{array}=23\, \mu \mathrm{K} \sqrt{\mathrm{s}}$}
\newcommand{\patchloc}{(RA, Dec)=($+0^\mathrm{h}12^\mathrm{m}0^\mathrm{s},-59^\circ18^\prime$)}

\begin{document}

\title{A Measurement of the Degree Scale CMB $B$-mode Angular Power Spectrum with \pb} 

\correspondingauthor{Neil Goeckner-Wald}
\email{ngoecknerwald@stanford.edu}

\author[0000-0002-0400-7555]{S. Adachi}
\affiliation{Department of Physics, Kyoto University, Kyoto 606-8502, Japan}

\author[0000-0002-1571-663X]{M. A. O. Aguilar Fa\'undez}
\affiliation{Department of Physics and Astronomy, Johns Hopkins University, Baltimore, MD 21218, USA}
\affiliation{Departamento de F\'isica, FCFM, Universidad de Chile, Blanco Encalada 2008, Santiago, Chile}

\author{K. Arnold}
\affiliation{Department of Physics, University of California, San Diego, CA 92093-0424, USA}

\author[0000-0002-8211-1630]{C. Baccigalupi}
\affiliation{International School for Advanced Studies (SISSA), Via Bonomea 265, 34136, Trieste, Italy}
\affiliation{Institute for Fundamental Physics of the Universe (IFPU), Via Beirut 2, 34014 Trieste, Italy}
\affiliation{National Institute for Nuclear Physics (INFN) Sezione di Trieste, Padriciano, 99, 34149 Trieste, Italy}

\author[0000-0002-1623-5651]{D. Barron}
\affiliation{Department of Physics and Astronomy, University of New Mexico, Albuquerque, NM 87131, USA}

\author[0000-0003-0848-2756]{D. Beck}
\affiliation{AstroParticule et Cosmologie (APC), Univ Paris Diderot, CNRS/IN2P3, CEA/Irfu, Obs de Paris, Sorbonne Paris Cit\'e, France}

\author{S. Beckman}
\affiliation{Department of Physics, University of California, Berkeley, CA 94720, USA}

\author[0000-0003-4847-3483]{F. Bianchini}
\affiliation{School of Physics, University of Melbourne, Parkville, VIC 3010, Australia}

\author{D. Boettger}
\affiliation{Instituto de Astrof\'isica and Centro de Astro-Ingenier\'ia, Facultad de F\'isica, Pontificia Universidad Cat\'olica de Chile, Av. Vicuna Mackenna 4860, 7820436 Macul, Santiago, Chile}

\author{J. Borrill}
\affiliation{Computational Cosmology Center, Lawrence Berkeley National Laboratory, Berkeley, CA 94720, USA}
\affiliation{Space Sciences Laboratory, University of California, Berkeley, CA 94720, USA}

\author[0000-0002-5751-1392]{J. Carron}
\affiliation{Department of Physics \& Astronomy, University of Sussex, Brighton BN1 9QH, UK}

\author{S. Chapman}
\affiliation{Department of Physics and Atmospheric Science, Dalhousie University, Halifax, NS, B3H 4R2, Canada}

\author[0000-0002-7764-378X]{K. Cheung}
\affiliation{Department of Physics, University of California, Berkeley, CA 94720, USA}

\author[0000-0002-3266-857X]{Y. Chinone}
\affiliation{Department of Physics, University of California, Berkeley, CA 94720, USA}
\affiliation{Kavli Institute for the Physics and Mathematics of the Universe (Kavli IPMU, WPI), UTIAS, The University of Tokyo, Kashiwa, Chiba 277-8583, Japan}
\affiliation{Kavli Institute for the Physics and Mathematics of the Universe (Kavli IPMU, WPI), Berkeley Satellite, the University of California, Berkeley 94720, USA}

\author[0000-0001-5068-1295]{K. Crowley}
\affiliation{Department of Physics, University of California, Berkeley, CA 94720, USA}

\author[0000-0002-7471-719X]{A. Cukierman}
\affiliation{Kavli Institute for Particle Astrophysics and Cosmology,SLAC National Accelerator Laboratory,2575 Sand Hill Rd, Menlo Park , CA 94025}
\affiliation{Department of Physics, Stanford University, Stanford, CA, 94305}

\author{M. Dobbs}
\affiliation{Physics Department, McGill University, Montreal, QC H3A 0G4, Canada}
\affiliation{Canadian Institute for Advance Research (CIfAR), Toronto, Canada, M5G 1M1}

\author[0000-0001-5471-3434]{H. El Bouhargani}
\affiliation{AstroParticule et Cosmologie (APC), Univ Paris Diderot, CNRS/IN2P3, CEA/Irfu, Obs de Paris, Sorbonne Paris Cit\'e, France}

\author[0000-0002-5166-5614]{T. Elleflot}
\affiliation{Department of Physics, University of California, San Diego, CA 92093-0424, USA}

\author[0000-0002-1419-0031]{J. Errard}
\affiliation{AstroParticule et Cosmologie (APC), Univ Paris Diderot, CNRS/IN2P3, CEA/Irfu, Obs de Paris, Sorbonne Paris Cit\'e, France}

\author[0000-0002-3255-4695]{G. Fabbian}
\affiliation{Department of Physics \& Astronomy, University of Sussex, Brighton BN1 9QH, UK}

\author{C. Feng}
\affiliation{Department of Physics, University of Illinois at Urbana-Champaign, 1110 W Green St, Urbana, IL, 61801, USA}

\author{T. Fujino}
\affiliation{Yokohama National University, Yokohama, Kanagawa 240-8501, Japan}

\author{N. Galitzki}
\affiliation{Department of Physics, University of California, San Diego, CA 92093-0424, USA}

\author{N. Goeckner-Wald}
\affiliation{Department of Physics, University of California, Berkeley, CA 94720, USA}
\affiliation{Department of Physics, Stanford Unviersity, Stanford, CA 94305, USA}

\author{J. Groh}
\affiliation{Department of Physics, University of California, Berkeley, CA 94720, USA}

\author{G. Hall}
\affiliation{Minnesota Institute for Astrophysics, University of Minnesota, Minneapolis, MN 55455, USA}

\author{N. Halverson}
\affiliation{Center for Astrophysics and Space Astronomy, University of Colorado, Boulder, CO 80309, USA}
\affiliation{Department of Astrophysical and Planetary Sciences, University of Colorado, Boulder, CO 80309, USA}
\affiliation{Department of Physics, University of Colorado, Boulder, CO 80309, USA}

\author{T. Hamada}
\affiliation{Astronomical Institute, Tohoku University, Sendai, Miyagi 980-0845, Japan}

\author[0000-0003-1443-1082]{M. Hasegawa}
\affiliation{High Energy Accelerator Research Organization (KEK), Tsukuba, Ibaraki 305-0801, Japan}

\author{M. Hazumi}
\affiliation{High Energy Accelerator Research Organization (KEK), Tsukuba, Ibaraki 305-0801, Japan}
\affiliation{Kavli Institute for the Physics and Mathematics of the Universe (Kavli IPMU, WPI), UTIAS, The University of Tokyo, Kashiwa, Chiba 277-8583, Japan}
\affiliation{Institute of Space and Astronautical Science (ISAS), Japan Aerospace Exploration Agency (JAXA), Sagamihara, Kanagawa 252-0222, Japan}
\affiliation{SOKENDAI (The Graduate University for Advanced Studies), Shonan Village, Hayama, Kanagawa 240-0193, Japan}

\author{C. A. Hill}
\affiliation{Department of Physics, University of California, Berkeley, CA 94720, USA}
\affiliation{Physics Division, Lawrence Berkeley National Laboratory, Berkeley, CA 94720, USA}

\author{L. Howe}
\affiliation{Department of Physics, University of California, San Diego, CA 92093-0424, USA}

\author{Y. Inoue}
\affiliation{Department of Physics, National Central University, Taoyuan 32002, Taiwan}
\affiliation{Center for High Energy and High Field Physics, National Central University, Taoyuan 32002, Taiwan}
\affiliation{High Energy Accelerator Research Organization (KEK), Tsukuba, Ibaraki 305-0801, Japan}

\author[0000-0001-8697-0064]{G. Jaehnig}
\affiliation{Center for Astrophysics and Space Astronomy, University of Colorado, Boulder, CO 80309, USA}
\affiliation{Department of Astrophysical and Planetary Sciences, University of Colorado, Boulder, CO 80309, USA}

\author[0000-0001-5893-7697]{O. Jeong}
\affiliation{Department of Physics, University of California, Berkeley, CA 94720, USA}

\author{D. Kaneko}
\affiliation{Kavli Institute for the Physics and Mathematics of the Universe (Kavli IPMU, WPI), UTIAS, The University of Tokyo, Kashiwa, Chiba 277-8583, Japan}

\author{N. Katayama}
\affiliation{Kavli Institute for the Physics and Mathematics of the Universe (Kavli IPMU, WPI), UTIAS, The University of Tokyo, Kashiwa, Chiba 277-8583, Japan}

\author[0000-0003-3118-5514]{B. Keating}
\affiliation{Department of Physics, University of California, San Diego, CA 92093-0424, USA}

\author{R. Keskitalo}
\affiliation{Computational Cosmology Center, Lawrence Berkeley National Laboratory, Berkeley, CA 94720, USA}
\affiliation{Department of Physics, University of California, Berkeley, CA 94720, USA}
\affiliation{Space Sciences Laboratory, University of California, Berkeley, CA 94720, USA}

\author{S. Kikuchi}
\affiliation{Yokohama National University, Yokohama, Kanagawa 240-8501, Japan}

\author{T. Kisner}
\affiliation{Computational Cosmology Center, Lawrence Berkeley National Laboratory, Berkeley, CA 94720, USA}

\author{N. Krachmalnicoff}
\affiliation{International School for Advanced Studies (SISSA), Via Bonomea 265, 34136, Trieste, Italy}

\author{A. Kusaka}
\affiliation{Physics Division, Lawrence Berkeley National Laboratory, Berkeley, CA 94720, USA}
\affiliation{Department of Physics, The University of Tokyo, Tokyo 113-0033, Japan}
\affiliation{Kavli Institute for the Physics and Mathematics of the Universe (Kavli IPMU, WPI), Berkeley Satellite, the University of California, Berkeley 94720, USA}
\affiliation{Research Center for the Early Universe, School of Science, The University of Tokyo, Tokyo 113-0033, Japan}

\author{A. T. Lee}
\affiliation{Department of Physics, University of California, Berkeley, CA 94720, USA}
\affiliation{Physics Division, Lawrence Berkeley National Laboratory, Berkeley, CA 94720, USA}

\author{D. Leon}
\affiliation{Department of Physics, University of California, San Diego, CA 92093-0424, USA}

\author[0000-0001-5536-9241]{E. Linder}
\affiliation{Space Sciences Laboratory, University of California, Berkeley, CA 94720, USA}
\affiliation{Physics Division, Lawrence Berkeley National Laboratory, Berkeley, CA 94720, USA}

\author{L. N. Lowry}
\affiliation{Department of Physics, University of California, San Diego, CA 92093-0424, USA}

\author{A. Mangu}
\affiliation{Department of Physics, University of California, Berkeley, CA 94720, USA}

\author[0000-0003-0041-6447]{F. Matsuda}
\affiliation{Kavli Institute for the Physics and Mathematics of the Universe (Kavli IPMU, WPI), UTIAS, The University of Tokyo, Kashiwa, Chiba 277-8583, Japan}

\author[0000-0003-2176-8089]{Y. Minami}
\affiliation{High Energy Accelerator Research Organization (KEK), Tsukuba, Ibaraki 305-0801, Japan}

\author{M. Navaroli}
\affiliation{Department of Physics, University of California, San Diego, CA 92093-0424, USA}

\author[0000-0003-0738-3369]{H. Nishino}
\affiliation{Research Center for the Early Universe, School of Science, The University of Tokyo, Tokyo 113-0033, Japan}

\author{A. T. P. Pham}
\affiliation{School of Physics, University of Melbourne, Parkville, VIC 3010, Australia}

\author[0000-0001-9807-3758]{D. Poletti}
\affiliation{International School for Advanced Studies (SISSA), Via Bonomea 265, 34136, Trieste, Italy}
\affiliation{Institute for Fundamental Physics of the Universe (IFPU), Via Beirut 2, 34151, Grignano (TS), Italy}
\affiliation{The National Institute for Nuclear Physics, INFN, Sezione di Trieste Via Valerio 2, I-34127, Trieste, Italy}

\author[0000-0002-0689-4290]{G. Puglisi}
\affiliation{Kavli Institute for Particle Astrophysics and Cosmology,SLAC National Accelerator Laboratory,2575 Sand Hill Rd, Menlo Park , CA 94025}

\author[0000-0003-2226-9169]{C. L. Reichardt}
\affiliation{School of Physics, University of Melbourne, Parkville, VIC 3010, Australia}

\author{Y. Segawa}
\affiliation{SOKENDAI (The Graduate University for Advanced Studies), Shonan Village, Hayama, Kanagawa 240-0193, Japan}
\affiliation{High Energy Accelerator Research Organization (KEK), Tsukuba, Ibaraki 305-0801, Japan}

\author[0000-0001-7480-4341]{M. Silva-Feaver}
\affiliation{Department of Physics, University of California, San Diego, CA 92093-0424, USA}

\author[0000-0001-6830-1537]{P. Siritanasak}
\affiliation{Department of Physics, University of California, San Diego, CA 92093-0424, USA}

\author{N. Stebor}
\affiliation{Department of Physics, University of California, San Diego, CA 92093-0424, USA}

\author[0000-0002-9777-3813]{R. Stompor}
\affiliation{AstroParticule et Cosmologie (APC), Univ Paris Diderot, CNRS/IN2P3, CEA/Irfu, Obs de Paris, Sorbonne Paris Cit\'e, France}

\author[0000-0001-8101-468X]{A. Suzuki}
\affiliation{Physics Division, Lawrence Berkeley National Laboratory, Berkeley, CA 94720, USA}

\author{O. Tajima}
\affiliation{Department of Physics, Kyoto University, Kyoto 606-8502, Japan}

\author[0000-0001-9461-7519]{S. Takakura}
\affiliation{Kavli Institute for the Physics and Mathematics of the Universe (Kavli IPMU, WPI), UTIAS, The University of Tokyo, Kashiwa, Chiba 277-8583, Japan}

\author{S. Takatori}
\affiliation{SOKENDAI (The Graduate University for Advanced Studies), Shonan Village, Hayama, Kanagawa 240-0193, Japan}
\affiliation{High Energy Accelerator Research Organization (KEK), Tsukuba, Ibaraki 305-0801, Japan}

\author{D. Tanabe}
\affiliation{SOKENDAI (The Graduate University for Advanced Studies), Shonan Village, Hayama, Kanagawa 240-0193, Japan}
\affiliation{High Energy Accelerator Research Organization (KEK), Tsukuba, Ibaraki 305-0801, Japan}

\author{G. P. Teply}
\affiliation{Department of Physics, University of California, San Diego, CA 92093-0424, USA}

\author{C. Tsai}
\affiliation{Department of Physics, University of California, San Diego, CA 92093-0424, USA}

\author[0000-0002-3942-1609]{C. Verges}
\affiliation{AstroParticule et Cosmologie (APC), Univ Paris Diderot, CNRS/IN2P3, CEA/Irfu, Obs de Paris, Sorbonne Paris Cit\'e, France}

\author[0000-0001-5109-9379]{B. Westbrook}
\affiliation{Department of Physics, University of California, Berkeley, CA 94720, USA}
\affiliation{Radio Astronomy Laboratory, University of California, Berkeley, CA 94720, USA}

\author[0000-0002-5878-4237]{Y. Zhou}
\affiliation{Department of Physics, University of California, Berkeley, CA 94720, USA}

\collaboration{1000}{(The \pb\ Collaboration)}

\begin{abstract}

We present a measurement of the $B$-mode polarization power spectrum of the cosmic 
microwave background~(CMB) using taken from July~2014 to December~2016 with the \pb\ experiment. 
The CMB power spectra are measured using observations at 150~GHz 
with an instantaneous array sensitivity of \arraynet\ on a 670 square degree patch of sky centered 
at \patchloc. A continuously rotating half-wave plate is used to modulate polarization and to suppress 
low-frequency noise. We achieve \mapdepth\ effective polarization map noise with a knee in sensitivity 
of \ellknee, where the inflationary gravitational wave signal is expected to peak. The measured $B$-mode 
power spectrum is consistent with a \lcdm\ lensing and single dust component foreground model 
over a range of multipoles $50 \leq \ell \leq 600$. The data disfavor zero $C_\ell^{BB}$ at $2.2\sigma$ using 
this $\ell$ range of \pb\ data alone. We cross-correlate our data with \planck\ high frequency maps and 
find the low-$\ell$ $B$-mode power in the combined dataset to be consistent with thermal dust emission. We place an 
upper limit on the tensor-to-scalar ratio \rupperlimit\ at 95\% confidence level after marginalizing over 
foregrounds. 

\end{abstract}

\keywords{cosmic microwave background radiation, cosmic inflation, large-scale structure of the universe}

\section{Introduction}

The cosmic microwave background (CMB) was last scattered at redshift $z\sim1100$ when the primordial plasma recombined to form neutral hydrogen transparent to radiation. The anisotropy of the CMB has become one of the leading probes in precision cosmology.
The temperature anisotropies in the CMB have been constrained to high precision over a large range of angular scales by many
experiments including \wmap\ and \planck\ \citep{2013ApJS..208...20B, 2018arXiv180706205P} at large angular scales and SPT and \act\ at small angular scales \citep{Story:2012wx,Das:2013zf}. The polarization anisotropy of the CMB provides additional information and is the focus of most current CMB experiments.
The linear polarization is traditionally decomposed into even and odd parity modes referred to as $E$ and $B$-modes.
$E$-mode polarization is sourced by both scalar and tensor perturbations in the early universe. 
In contrast, $B$-modes are not created by primordial scalar perturbations and are only expected to be generated by tensor perturbations or gravitational lensing of $E$-mode polarization
by intervening large scale structure. The $B$-mode signal is subdominant to the $E$-mode power spectrum on all angular scales in the standard cosmological model and its measurement presents a significant experimental challenge.

Evidence for the lensing $B$-mode power spectrum from CMB polarization alone was first reported by the \pb\ collaboration in \citet{pb2014a} herein PB14. Subsequent direct measurements of $B$-mode polarization have been reported by \pb\ \citep{pb2017a} herein PB17, \bicep2\ \citep{bkI}, SPTpol \citep{Keisler2015}, \keckarray\ \citep{bkV}, and \actpol\ \citep{2016arXiv161002360L}. 

We report a measurement of the CMB $B$-mode power spectrum for $50 \leq \ell \leq 600$ as measured by the \pb\ experiment. The target of this analysis is the degree scale range of the $B$-mode signal ($\ell \approx 100$) where the primordial gravitational wave perturbation signal is expected to peak. The lensing $B$-mode signal peaks at ten times smaller angular scales ($\ell \approx 1000$) and is the subject of a forthcoming analysis using the same experimental data. We also measure the cross correlation of our data with the \planck\ data set to estimate the level of Galactic foreground contamination. Similar cross correlation analyses have been presented by the \bicep2\ and \keckarray\ collaboration \citep{bkp,2018PhRvL.121v1301B} herein BK15 and the \abs\ collaboration \citep{2018JCAP...09..005K}.

This paper is structured as follows. In Section~\ref{instrument}, we describe the \pb\ instrument and the data set used in this analysis, paying particular attention to the addition of a continuously rotating half wave plate (HWP) to the optical path to modulate the sky polarization. In Section~\ref{calibration}, we describe the pre-processing and low-level calibrations applied to the data. In Section~\ref{analysis}, we outline the procedure for converting raw detector time ordered data (TOD) into maps and power spectra including high level calibrations. In Section~\ref{systematics}, we describe internal consistency checks performed on the data as well as simulations of known systematics. In Section~\ref{results}, we present the measured $B$-mode power spectrum, cross correlation with \planckeight\ data to estimate foreground contamination in our maps, and our constraints on cosmological parameters. We conclude in Section~\ref{conclusions}.

\section{Instrument and Data Set}\label{instrument}

\subsection{The \pb\ Instrument}

\pb\ is a CMB experiment installed on the 2.5\,m aperture Huan Tran Telescope in January 2012. 
The telescope is located at the James Ax Observatory at an elevation of 5{,}190~m in the Atacama Desert in Chile. 
The \pb\  receiver consists of seven wafers containing a total of 1274 transition edge sensor~(TES) bolometers cooled to 0.3~K observing the sky through lenslet-coupled double-slot dipole antennas.
More details on the receiver and telescope can be found in \citet{Arnold_SPIE2012} and \citet{Kermish_SPIE2012}.

In February 2014, we installed a continuously rotating HWP at the focus of the primary mirror to mitigate low-frequency noise. The HWP consists of an anti-reflection coated birefringent single crystal sapphire plate rotating at 2 revolutions per second. More details of the installation of the HWP in the telescope optical path and its performance in a series of test observations can be found in \citet{Takakura:2017ddx}, herein T17.

\begin{figure}
\begin{center}
\includegraphics[scale=0.44]{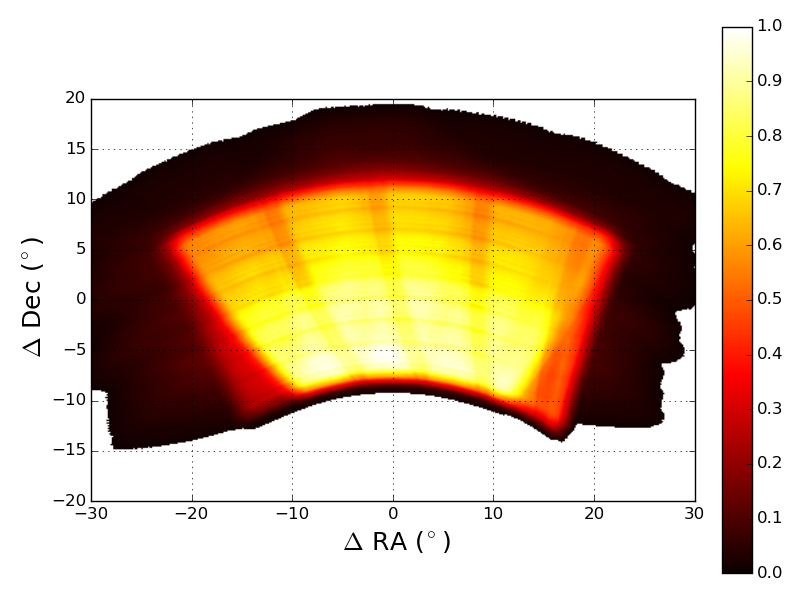}
\end{center}
\caption{Normalized map depth illustrating the scan pattern centered at \patchloc. The average effective map depth is \mapdepth\ in polarization. The vertical stripes are an artifact of breaks in the low elevation 
scans to retune the detectors. The horizontal stripes are an artifact of the elevation offsets 
used in the transit scan. The patch overlaps with the area observed by South Pole experiments.}
\label{4fdepth}
\end{figure}

\begin{table}
\begin{center}
\caption{Breakdown of observing time by season}  \label{seasontable}
\begin{tabular}{ c  c  c }
\hline
Season number & Beginning date & End date \\
\hline 
HWP comissioning 	& 4 May 2014 		& 24 July 2014 \\
Third season 		& 25 July 2014 		& 1 March 2015 \\
Fourth season		& 2 March 2015 		& 31 December 2015 \\
Fifth season 		& 1 January 2016 	& 30 December 2016 \\
\hline
\end{tabular} \par
\end{center}
\vspace{8pt}
\noindent Breakdown of observing seasons for all of the data used in this analysis. The seasons
are separated by periods of instrument maintenance. The first two seasons of data are described 
in PB14 and PB17 and are not used in this analysis.
\end{table}

\subsection{Scan Strategy and Observations}

The \pb\ observations used in this analysis target a 670 effective square degree patch of sky centered on \patchloc. This patch has significant overlap with the area mapped by South Pole experiments including \bicep, the \keckarray, and \sptpol. The patch coverage is shown in Figure~\ref{4fdepth}. Regular science observations with the HWP began on 25~July~2014 and continued until 30~December~2016. The data set is broken up into three seasons described in Table~\ref{seasontable}.

The scan strategy consists of three sets of scans repeated every sidereal day. We scan for a 4 hour 45 minute block as the patch rises above the horizon, a 3 hour 53 minute block at high elevation as the patch transits, and a 4 hour 45 minute block as the patch sets. The rising and setting scan occur at a boresight elevation of $30^\circ$ and $35 \fdg 2$ respectively. The elevation of the high elevation scan is stepped through ten offsets from $45\fdg 5$ to $65\fdg 5$ to provide even coverage of the patch. Each observation block is broken into hour-long scans referred to as \enquote{constant elevation scans} (CESs) after which the detectors are retuned. More details on the calibration procedures are given in Section~\ref{calibration}. The telescope is repointed between four hour long blocks but not between hour long CESs. 

During scans, the telescope is scanned at a constant velocity of $0 \fdg 4~\mathrm{s}^{-1}$ with a throw of $20^\circ$ and $35^\circ$ on the sky for the low elevation and high elevation scans, respectively. This results in approximately 70 subscans (defined as one left going or right going motion of the telescope) in each hour long CES.

As a result of the ten high-elevation scan offsets, a complete map of the field is produced once every ten days. Several of these ten day data subsets have been combined in post-processing due to low yield or incomplete coverage after data selection. This results in 38 approximately even splits of the data set. These splits form the basis for the cross spectra used in the power spectrum estimation. 

Data are discarded from two blocks of time due to mechanical problems with the weatherproof enclosure of the HWP and the eruption of a nearby volcano on 30 October 2015. Additionally, the thermal source used for relative gain calibration was replaced multiple times due to mechanical problems. The gain calibration is described in Section~\ref{gains}.

When the science patch is not available, the cryogenic refrigeration system is recycled, and we perform calibration measurements. These measurements are described in further detail in Section~\ref{calibration}. For most of the data set, the fridge was cycled every 24 sidereal hours; however starting in October of 2015, the fridge cycle was done every 48 sidereal hours to provide time to scan the Northern science patch referred to as RA12 in PB14 and PB17. Those data are not included in this analysis.


\section{Pre-processing and Calibration} \label{calibration}

This section describes the steps taken to generate the inputs to our data analysis pipelines. We record dropped data packets that are later flagged in the data selection. 

\subsection{HWP angle reconstruction} \label{preprocess}

We reconstruct the HWP angle and interpolate this to match the bolometer readout time stamps. We see a jitter in the reconstructed HWP angle on the order of $10^{-7}~\mathrm{rad}^2\,\mathrm{s}^{-1}$ due to a combination of physical angle jitter and reconstruction error which is expected to be subdominant at approximately $10^{-9}~\mathrm{rad}^2\,\mathrm{s}^{-1}$. We simulate the impact of this jitter in Section~\ref{systematicspipeline} and find the impact on our results to be small. 

\subsection{Pointing}

The telescope pointing reconstruction is performed in a very similar manner to PB14 and PB17. The telescope is used for dedicated raster observations of bright point sources selected from PCCS and ATCA catalogs \citep{2014A&A...571A..28P,at20g} prior to each science observation. The source selection has been modified from PB17 to better match the azimuth and elevation ranges used during the science scans. The observed position computed based on the telescope's azimuth and elevation encoder values are compared to the expected positions. The resulting azimuth and elevation offsets $\Delta\mathrm{Az}(\mathrm{Az}, \mathrm{El}), \Delta\mathrm{El}(\mathrm{Az}, \mathrm{El})$ are fit to an eight parameter model including terms for uneven heating of the telescope due to insolation \citep{FredPhD}. The fiducial model uses the solar radiation as in input parameter in contrast with PB17 which used ambient temperature. We find an root-mean-square (RMS) pointing model residual of $50^{\prime\prime}$.

We also construct pointing solutions that include the Crab Nebula (\taua) and Jupiter scans performed for polarization angle and beam calibrations in the same way, however these data are not used in the fiducial boresight pointing solution as they are observed in a significantly different range of azimuth and elevation from the science observations. We perform this calibration with several different subsets of pointing observations and parameter combinations. In Section~\ref{systematicspipeline} we show that the difference between these pointing solutions is negligible for the $\ell$ range considered in this paper.

Following results reported in PB14 and PB17, the detector beam offsets are derived from array raster scans over Jupiter. We find that the reconstructed offsets are consistent with previous results at the level of several tens of arcseconds. We explicitly cut detectors whose mean fitted beam offset differs from PB17 by more than one arcminute. This cut has a negligible impact on the overall data selection efficiency.

\subsection{Beams}

\begin{figure}
\begin{center}
\includegraphics[scale=0.44]{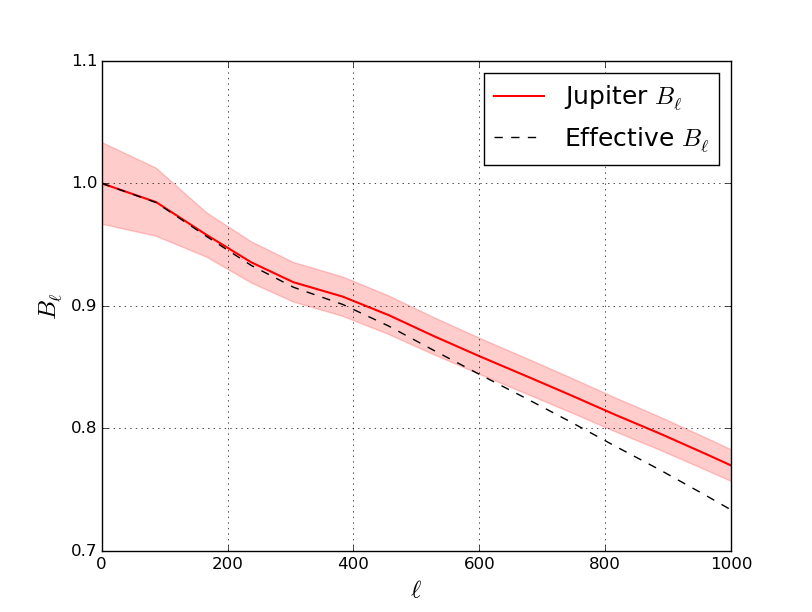}
\end{center}
\caption{Reconstructed beam window function from Jupiter observations. 
Statistical error in the pointing model adds additional supression of power 
at high-$\ell$. This is indicated by the dashed effective beam curve. The shaded area 
represents the statistical error in the Jupiter beam window function.}
\label{beamplot}
\end{figure}

Following PB14 and PB17, the instrumental beam and the associated window function $B_\ell$ are measured using dedicated raster observations of Jupiter. We take Jupiter observations with the HWP rotating nominally at 2 Hz at a scan speed of $0\fdg 2~\mathrm{s}^{-1}$ on the sky. The HWP synchronous structure is subtracted by masking off a $25^\prime$ disk in the map domain centered on Jupiter and fitting a first order polynomial to the time dependent amplitude of each HWP harmonic. After the HWP synchronous structure is subtracted the TOD are projected into 0\farcm 5 pixels on the sky. The beam window function is taken to be the average of the azimuthally averaged two dimensional Fourier transform after dividing out the Jupiter disk for each observation and correcting for the pixel window function analytically. The reconstructed beam window function is shown in Figure~\ref{beamplot}.
 
We find a beam window function that is similar to, but marginally wider than, PB17 with a median Gaussian $3\farcm 6$ full width at half maximum (FWHM). This difference is possibly due to imperfect focusing of the telescope since the addition of the sapphire HWP lengthens the optical path between the primary and the secondary mirrors or due to diffraction off the HWP structure itself. We see no weather dependence or variation between seasons in the fitted beam width.

Boresight and detector statistical pointing error adds an additional suppression of high-$\ell$ power in the maps. The pointing model predicts residual pointing errors of $50^{\prime \prime}$ RMS.  In Section~\ref{abscal} we fit an $\ell$-dependence to the  overall gain amplitude corresponding to the widening of the effective beam due to pointing error and find no statistical preference for an additional widening of the effective beam due to pointing error. As a result we convolve the beam function with a Gaussian corresponding to the best fit pointing model error from the $E$-mode spectrum and include the effective beam width uncertainty in our multiplicative error estimate.

An additional source of systematic uncertainty in the beam comes from detector crosstalk. We measure the detector beam window function using temperature data but compute polarization power spectra. As described in \citet{Crowley:2018eib}, crosstalk acts differently in temperature and polarization in the presence of a continuous HWP resulting in an effective beam mismatch. This is described in detail in Section~\ref{systematicspipeline}. We find this effect to be subdominant to the statistical uncertainty for all spectra.

We estimate the polarization response to Jupiter and decompose this into scalar, dipole and quadrupole terms. We subtract the scalar term as temperature to polarization leakage. This is described in Section~\ref{mapmaking}. In Section~\ref{systematicspipeline} we show the contamination due to higher order terms is negligible.

\subsection{Detector gains and time constants}
	
Following PB14 and PB17, we use a three step gain calibration to turn our TOD into physical temperature units with an additional correction for polarization efficiency. We measure a time-dependent gain calibration between detectors using a chopped thermal source before and after each set of four CESs. The gain is linearly interpolated between these measurements. We then calibrate these measurements to temperature on the sky using the same Jupiter scans conducted to measure the beam window function. Finally, we construct a CMB map and scale the overall amplitude of the polarization $E$-mode fluctuations to match the \planckeight\ data. Performing the absolute gain calibration with the  $E$-mode spectrum jointly constrains the overall gain and polarization efficiency. 

At several points during the observations the thermal source used in the relative calibration was replaced for mechanical reasons. The conversion from the thermal source amplitude to temperature on the sky is performed separately for each source. PB14 describes a correction in this analysis due to the position of a cold HWP in the receiver. The cold HWP was stepped almost daily during the first half of the first season of observations, however it was never moved during the three seasons comprising this data set. We simulate the systematic error introduced by uncertainty on the measured gain acting on the CMB signal and find the contamination in all polarization spectra to be negligible. This is described in Section \ref{systematicspipeline}.

The detector time constants are measured by sweeping the frequency of the chopped thermal calibration source from 4 Hz to 44 Hz. The source amplitude versus frequency is fit to a single pole transfer function. These time constants are interpolated linearly between the calibrations and deconvolved in the subsequent TOD processing. \label{gains}

\subsection{Polarization angle}
	
The detector polarization angles and efficiencies are derived from dedicated raster scans of \taua. The raster scan data is taken at $0\fdg 2~\mathrm{s}^{-1}$ scan speed on the sky with the HWP rotating nominally. We deconvolve the detector time constants measured from a chopped thermal source calibration immediately before and after the \taua\ raster scan. This is necessary because a phase lag between the detector TOD and the reconstructed HWP angle appears as a polarization angle error. After correcting for detector time constants, we find no significant correlation between the measured polarization angle and the precipitable water vapor (PWV) measured at the nearby Atacama Pathfinder Experiment\footnote{\url{http://www.apex-telescope.org/}} (\apex) site. Without time constant deconvolution such a correlation is produced by the dependence of the time constants on the atmospheric loading and therefore the local PWV.  The demodulated polarization TOD described in Section~\ref{demod} is fit to a beam convolved polarization map of \taua\ from a measurement taken with the IRAM $30~\mathrm{m}$ telescope \citep{2010A&A...514A..70A}. We find statistically consistent detector polarization angles in focal plane coordinates compared to PB17 and no significant variation between seasons. We find the difference in measured polarization angle between detectors in a pair to be $90\fdg 5 \pm 1\fdg 2$. This is consistent with the design value of $90 ^\circ$. In Section~\ref{systematicspipeline} we simulate the expected systematic contamination due to detector polarization angle uncertainty at this level and find the effect to be negligible.

The measured polarized flux from these scans is used to estimate the polarization efficiency of the telescope and receiver. The polarization efficiency is degraded in three ways in addition to the non-ideality of the cold HWP described in PB14 and PB17. 

These three effects are the non-ideality of the rotating HWP, the breaking of the Mizuguchi-Dragone (MD) condition \citep{1976isap.conf....2M, dragone} and non-zero bolometer time constants. The modulation efficiency of the warm HWP is estimated from coherent source lab measurements and design detector bandpasses \citep{Arnold_SPIE2012}. The polarization efficiency term due to the MD breaking is estimated from a GRASP\footnote{\url{https://www.ticra.com/software/grasp/}} physical optics simulation \citep{mdbreakingspie}. The detector time constant acts as a time domain low pass filter on the TOD which has a response less than unity at the polarization modulation frequency. We find the measured values from the \taua\ calibration to be consistent with the predictions but with a larger statistical error as expected. The non-ideality of the rotating HWP, cold HWP, and MD breaking polarization efficiency terms are intrinsic to the detectors and telescope geometry and are corrected by rescaling the TOD and noise weights with the predicted values. The average contributions to the overall polarization efficiency from the (rotating HWP, cold HWP, and MD breaking) are $\varepsilon \approx (0.98,~ 0.98,~0.93$), respectively. The polarization efficiency due to the detector time constant depends on the numerical value of the time constant which in turn depends on the detector bias point and optical loading. As a result this polarization efficiency term is corrected by deconvolving the measured time constant from the chopped thermal source before and after each set of four CESs. The average polarization efficiency contribution from the detector time constants is $\varepsilon \approx 0.98$.

We self-calibrate the overall polarization angle by fitting $C_\ell^{EB} = 0$ following \citet{Keating2013}. The overall polarization efficiency is degenerate with the absolute gain of the polarization maps and is set by matching the \pb~$E$-mode spectrum to the \planck~143 GHz $E$-mode data on our patch. This is described in more detail in Section \ref{abscal}. \label{polangle}
			
\section{Data Analysis} \label{analysis}
	
\input{data_analysis}

\subsection{TOD Processing}\label{demod}

This analysis uses a similar pre-processing and demodulation procedure as T17 and \citet{Takakura:2018wky}, herein T19. A brief overview is provided here for completeness. For simplicity terms relating to detector non-linearity and instrumental polarization are neglected here. The raw bolometer TOD in the telescope frame can be modeled following

\begin{equation}
\begin{split}
d_m(t) = I(t) + \varepsilon \mathrm{Re}\{[Q(t) + {i} U(t)]\, e^{-i(4\chi +2\theta_\mathrm{det}}) \} \\
+ A(\chi, t) + \mathcal{N}_m, \\
\end{split}
\end{equation}

\noindent where $\chi \approx \omega t$ is the HWP angle, $e^{-{i4\chi}} \equiv m(\chi)$ describes the polarization modulation of the HWP, $\varepsilon$ is a polarization efficiency, $\theta_\mathrm{det}$ is the detector angle with respect to instrument coordinates, $\mathcal{N}_m$ is the detector white noise, and $A(\chi, t)$ is a slowly varying HWP-synchronous structure in the TOD.

Prior to demodulation we subtract the HWP-synchronous structure $A(\chi, t)$ using an iterative 
method similar to \cite{2007ApJ...665...42J}. The HWP angle is reconstructed from the encoder. The HWP synchronous structure is estimated 
following \cite{2014RScI...85c9901K} and is decomposed into harmonics. Gaps in the TOD are filled 
with the HWP synchronous structure. At each harmonic $n \in \{1, 2, \ldots 7\}$, the TOD are 
bandpass filtered with a bandwidth of 1 Hz and demodulated to form a time dependent amplitude for that harmonic of the 
HWP synchronous structure. A linear drift is fit to the amplitude of each harmonic. 
The sum over $n$ of these templates is subtracted from each TOD and the process is iterated again. 
After the HWP structure has been subtracted, the polarization signal is reconstructed by 
multiplying the data by twice the conjugate of the modulation function $2m^*(\chi)$ and low pass 
filtering to recover the polarization signal,

\begin{equation}
d_d(t) = \varepsilon [Q(t) + {i} U(t)] e^{-2{i}\theta_\mathrm{det}} + \mathcal{N}_d.
\end{equation}

The TOD sampled at 8 Hz are used as the input to all subsequent analysis pipeline steps. Due to the high sample rate of the raw TOD and the number of Fourier Transform operations, including the demodulation in our main simulations would be computationally prohibitive. The window function associated with the 8 Hz sample rate is negligible for our $\ell$ range. The detector time constant deconvolution is performed on the demodulated polarization TOD.

Note that the noise in the demodulated TOD $\mathcal{N}_d$ is complex and both the real and imaginary components have twice the variance of the modulated detector white noise $\mathcal{N}_{m}$ due to the factor of two necessary to recover $Q$ and $U$ correctly. There is no intrinsic difference in the polarization sensitivity compared to the pair differencing case. The noise $\mathcal{N}_d$ can be well described as white noise plus a single low frequency component common to all detectors. This is described in Section~\ref{low_freq_noise}. The demodulation algorithm assumes perfect separation of intensity and polarization signals in time domain frequency, or in other words the HWP frequency is much higher than the scan speed divided by the beam size. In Section~\ref{systematicspipeline} we quantify the impact of imperfect separation in the real data and find the effect to be negligible. These demodulated data are used as the input to the subsequent data characterization and mapmaking pipelines.

We record the low pass filtered HWP structure subtracted TOD (the $0f$ or intensity component) and the demodulation at twice the HWP frequency (the $2f$ component) for data characterization and selection.
	
\subsection{Data Characterization and Selection}\label{dataselection}

\begin{table}
\begin{center}
\caption{Data selection efficiency} \label{efftable}
\begin{tabular}{ c  c }
\hline
Stage of data selection & Efficiency \\
\hline 
Time spent observing patch & 36.8 \% \\
Focal plane yield		& 50.7 \% \\
Glitch cuts and off bolometers &  29.2 \% \\
Individual bolometer PSD cuts  & 76.6 \% \\
Common mode PSD and map cuts & 79.6 \% \\
\hline
\end{tabular} \par
\end{center}
\vspace{8pt}
\noindent Data selection efficiency for this analysis. A total of 2985 hour long CESs are used in the final
science analysis. The focal plane yield is normalized to the nominal number of optical bolometers, \nbolo.
\end{table}

The data selection framework used in this analysis consists of several rounds of increasingly selective criteria to characterize low-frequency noise and mitigate possible systematic contamination. The stages can be roughly described as cutting out glitches and effects well localized in time, cutting data based on individual detector noise properties over the course of an observation, cutting data based on array noise properties, and cutting data based on map domain noise.

In all stages, the fundamental unit of data considered is the detector subscan. Data where the telescope is accelerating (turnarounds) are rejected completely. A table of efficiencies is shown in Table~\ref{efftable}. A total of 2985 CES observations are used in the final mapmaking from the full observation period from 25 July 2014 until 30 December 2016.

In the first stage, a set of time-domain glitch criterium similar to what was used in PB14 is applied to the pre-demodulated timestreams to find and remove high frequency features. The TOD are convolved with a kernel designed to pick up sharp temporal spikes in the data while nulling the HWP synchronous structure at multiples of 2 Hz. Subscans where the maximum  deviation of the convolved TOD is greater than ten times the standard deviation are discarded. This operation is performed on the full sample rate data to improve sensitivity to fast glitches. The post-demodulation intensity, $2f$ and polarization TOD are convolved with an additional series of kernels sensitive to fast glitches in the TOD, and subscans are cut following the same criteria.

After these flagging steps, an analysis is performed to remove the low-frequency array common mode glitches in the polarization data shown in T19 to be correlated with polarized emission from clouds. We begin by performing a principal component analysis (PCA) on the detector intensity TOD. Working detectors are selected by removing channels whose intensity timestreams are not dominated by the largest eigenmode of the detector covariance matrix due to atmospheric fluctuations. The $Q+iU$ timestreams from each detector are rotated into the instrument frame by multiplication with $e^{i2\theta_\mathrm{det}}$. These TOD are coadded to form a full focal plane common mode signal which is rotated again into minimum and maximum variance eigenmodes of the $Q \times U$ covariance matrix. Subscans where the ratio of the standard deviations of the two modes is greater than ten are cut. This corresponds to the TOD cloud detection criteria used in T19.

In the second stage, each demodulated timestream is processed through the first portion of the mapmaking TOD filters described in Section \ref{mapmaking}. The TOD are filtered by a first-order subscan polynomial, ground-fixed structure is subtracted, and monopole temperature to polarization leakage is removed. Turnarounds and subscans flagged by the glitch cuts are filled in with white noise matched to the RMS of the surrounding samples. The power spectral density (PSD) of the TOD is then measured and fit to a model consisting of white noise and a $1/f^2$ low frequency noise term. Detectors with anomalously high white noise floors, high low-frequency noise, or poor fits to the model are discarded. For most individual bolometers the low frequency noise is undetectably small after the subscan polynomial filter. Fourier domain lines in the PSD are noted and notched out in subsequent data processing. This notch filter is described in Section~\ref{mapmaking}. 

In the third stage, the common mode timestream is re-computed using the inverse variance individual detector weights, the additional data cuts, and Fourier notch filters defined in the previous stage. We fit the common mode PSD to a white noise and $1/f^\alpha$ term for instrument frame $Q$ and $U$ separately. White noise is largely uncorrelated between bolometers whereas the low frequency component is coherent across detectors meaning that averaging detectors results in a higher knee frequency. The power law index $\alpha$ is allowed to float because the higher knee frequency provides a sufficient lever arm to resolve the low frequency exponent. Observations where the common mode knee frequency $f_\mathrm{knee}$ in telescope frame $Q$ ($U$) is greater than 150 mHz (100 mHz) or where the common mode noise floor is anomalously high are rejected. We find median knee frequencies of 45 mHz (24 mHz) in the common mode $Q$ ($U$) PSDs and a median polarization white noise floor of $46~\mathrm{\mu K} \sqrt{\mathrm{s}}$ for $Q$ and $U$ individually in CMB temperature units. This corresponds to an array noise equivalent temperature of \arraynet\ that is comparable to PB14 and PB17. The knee frequency distributions differ because common mode $Q$ and $U$ noise are produced by different physical mechanisms. $Q$ low frequency noise can be produced by temperature noise brought into the polarization TOD by temperature to polarization leakage subtraction or by spurious gain drifts acting on the HWP $4f$ structure. In contrast, $U$ is out of phase with both the $4f$ HWP structure and the temperature to polarization leakage, meaning that low frequency noise requires a phase drift of the HWP structure in the TOD produced by effects such as detector time constant drift. It is worth noting that we do not see a correlation between the array knee frequencies and the amplitude of the atmospheric fluctuations in the temperature data.

In the fourth stage, the detector weights computed in the second stage and the common mode PSD defined in the third stage are used to create individual observation maps following the standard TOD filtering outlined in Section \ref{mapmaking} with the modification that the telescope frame polarization is treated as a scalar field and is not rotated into right ascension and declination coordinates. The maps are downsampled to degree pixels and a map $\chi^2$ is computed by comparing the fluctuation in the data to the expectation from the detector noise weights. Maps with an anomalously high $\chi^2$ are rejected as a check for low frequency pathologies that are not readily visible in the common mode PSDs. In practice this cut strongly overlaps with the common mode PSD criteria.
	
\begin{figure}
\begin{center}
\includegraphics[scale=0.44]{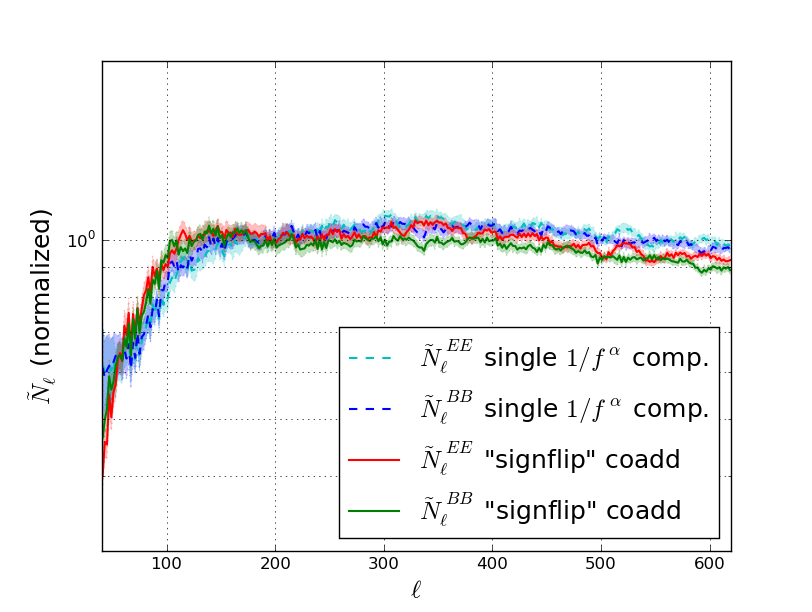}
\end{center}
\caption{Noise bias comparison for the full data set from the TOD noise model and the 
signflip coaddition pipeline. We see broadly consistent results between the two noise models. 
The fiducial power spectra use the \enquote{signflip} noise model. These spectra do not reflect the filter transfer 
function and beam window function correction described in Section~\ref{powerspectrum}. 
The \enquote{signflip} curves with these corrections are shown in Figure~\ref{nbfigure}. The shaded region 
represents the standard deviation of the simulated noise bias.} 
\label{nlpseudo}
\end{figure}

\subsection{Low Frequency Noise Model}\label{low_freq_noise}

We define two noise models for use in our simulation pipeline and show good overall agreement between the two. The noise in the data at low frequencies in the time domain can be described by a single common mode $1/f^{\,\alpha}$ component meaning higher order modes in the TOD covariance are negligible. We run the end-to-end analysis pipeline in two configurations; one using a TOD noise model and the other using random sign coaddition of individual CES maps to generate matched \enquote{signflip} noise realizations.

In the TOD noise model, we generate white noise on a per-detector basis assuming no correlations between bolometers and the noise weights derived in Section~\ref{dataselection}. The common mode low-frequency noise synthesized in telescope coordinates for $Q$ and $U$ separately and is matched to the amplitude and power law index $\alpha$ fit from the real data. This mode is then rotated into the polarization angle of each detector and added to the uncorrelated white noise.

In the \enquote{signflip} configuration, the noise realizations are generated by randomly assigning a +1 or -1 factor to each map during the coaddition of CESs. This creates noise realizations with the exact power spectrum and correlation structure of the real data assuming that the noise fluctuations between CESs are uncorrelated. This assumption can be broken by coherent drift in the amplitude of the ground synchronous structure as described in Section \ref{systematicspipeline}. We find this effect to be negligible.

We find good agreement between the \enquote{signflip} noise realizations and the TOD noise realizations using our cross spectrum estimator described in Section~\ref{powerspectrum}. The noise bias derived from both pipelines is shown in Figure~\ref{nlpseudo}. The full coadd and all null test splits described in Section~\ref{null_test} agree well except for one null spectrum. In the \enquote{top versus bottom bolometers} case that explicitly splits paired detectors, we observe excess variance and an anti-correlation in map space between the two halves that cannot be reproduced by the simple TOD noise model. Temperature noise aliasing into the polarization frequencies in the TOD can create a similar noise anti-correlation. This is naturally accounted for in the \enquote{signflip} noise realizations.

We use the random sign coaddition pipeline to generate the noise realizations used in our fiducial null tests and error bar estimation.
	
\begin{figure*}
\begin{center}
\includegraphics[scale=0.44]{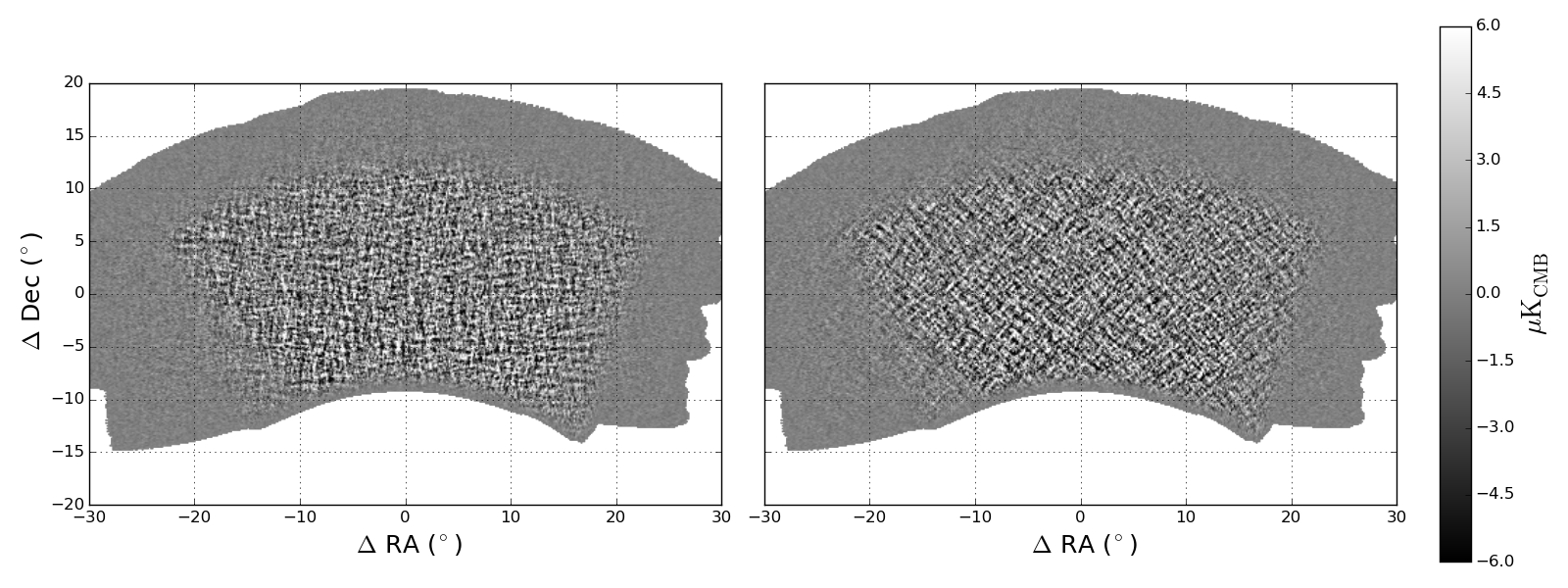}
\includegraphics[scale=0.44]{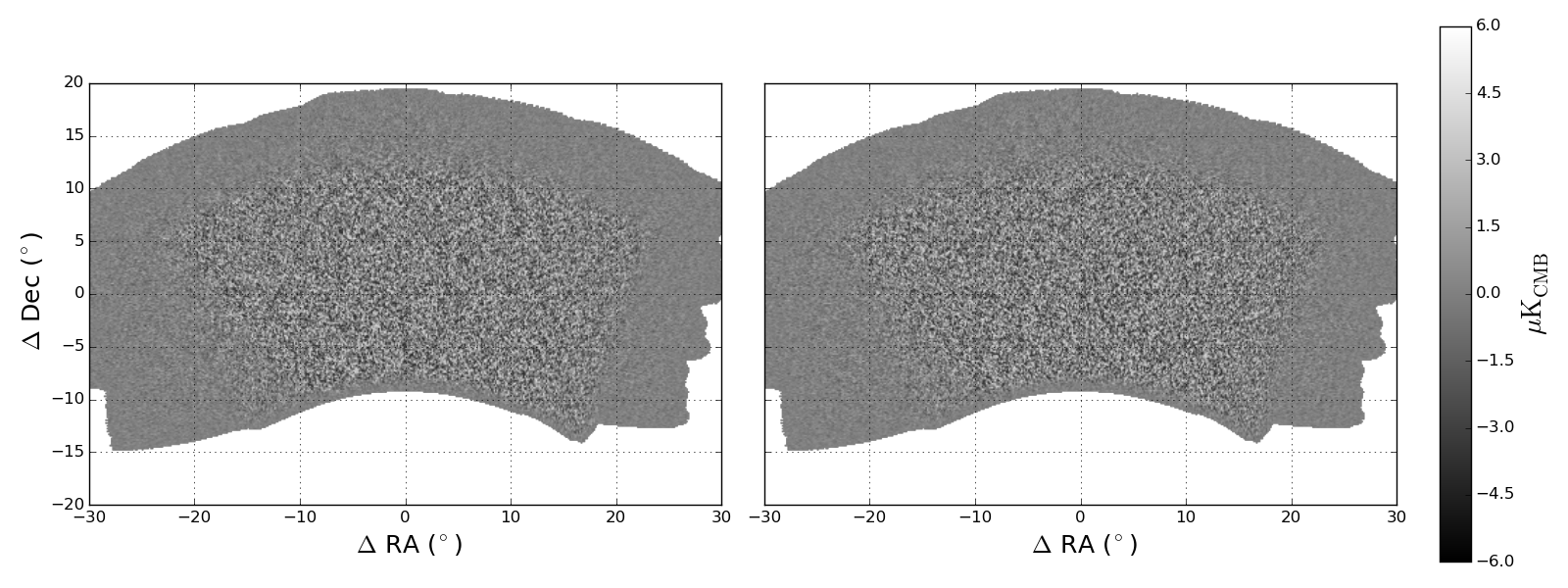}
\end{center}
\caption{\pb\ $Q$ and $U$ maps (top) and a sample noise realization (bottom) 
produced using the \enquote{signflip} coadd pipeline. The CMB $E$-mode signal is visible in the 
real maps as a checkerboard pattern in $Q$ and $U$. These noise realizations are used to 
estimate the band power covariance of the the final power spectrum
and the noise bias used in the foreground estimation pipeline.}
\label{maps}
\end{figure*}

\subsection{Mapmaking} \label{mapmaking}

In this analysis, we use a MASTER \enquote{filter and bin} mapmaker \citep{Hivon2002} where the TOD is filtered to suppress low frequency noise and projected onto the sky using inverse variance noise weights.

The mapmaking pipeline takes the demodulated data described in Section \ref{demod} and applies an additional stack of time domain filters. The ordering of the filtering is set to avoid bias in the fitted temperature leakage coefficients by spurious modes in the temperature and polarization TOD.

The first filtering operation works in the Fourier domain and low pass filters signal above 1.2 Hz. A set of notch filters are applied to Fourier domain glitches flagged in the data selection pipeline. A notch width of 10 mHz is used on every bolometer on the focal plane when a high significance glitch is seen in more than 50 detectors for a given CES. An identical set of notch filters are applied to the simulation data.

The second filtering operation removes a second order polynomial from each bolometer TOD for the whole CES. This mode is expected to be dominated by thermal fluctuations of the focal plane and cryogenics.

The third filter subtracts a ground-fixed template in $I$, $Q$, and $U$. The filter is constructed by averaging the telescope frame TOD in $14 \farcm 4$ azimuth bins and subtracting the resulting template from the TOD. This operation is performed for each bolometer and CES independently. Unlike PB14 and PB17, the same template is used for both left-going and right-going subscans. We expect that the
ground synchronous structure in our data is due to telescope sidelobes far from the main beam interacting with the surrounding terrain.  We conservatively assume no correlation between CESs and subtract a unique ground fixed template for each CES even though the ground fixed signal that we observe is generally stable between CESs. In Section \ref{systematicspipeline} we place an upper limit on the error introduced by possible time variability in the ground synchronous structure within each CES. We test varying the bin size and subtracting out a smoothed version of the template and find no significant difference in the final measured power spectrum.

Once these modes have been projected out of the data we perform a PCA similar to T17 to remove temperature to polarization leakage due to detector non-linearity and instrumental polarization from the off-axis telescope design. We see a weak frequency dependence in the temperature leakage coefficients below the telescope scan frequency and Fourier domain glitches in the TOD at high frequencies. As a result, to estimate the leakage coefficients we form a copy of the TOD and apply a low pass filter at 400 mHz
and subtract a first order polynomial from each subscan. We then compute a $3 \times 3$ covariance matrix between the $I$, $Q$, and $U$ TOD and average this between subscans. The
leakage coefficients are determined using the PCA described in T17 with this covariance matrix. We find the estimated leakage coefficients to be stable over the course of our observations. The temperature leakage is subtracted from the original polarization TOD without the subscan polynomial and low-pass filters. Since the leakage coefficient determination is heavily dominated by atmospheric fluctuations we do not expect this process to be significantly biased by cosmological signal. In Section~\ref{systematicspipeline} we simulate the error expected in the $B$-mode power spectrum due to multiplicative detector non-linearity and find the effect to be negligible. We do not include this leakage or the PCA filter in our main simulation pipeline. 

Following this a first order polynomial is subtracted from the $Q$ and $U$ TOD for each subscan in polarization to mitigate low frequency noise. No further processing is applied to the $I$ TOD as these data are not used in the subsequent analyses.

Finally, a filter is applied to the data to suppress the low frequency mode seen in all detectors. A low pass filtered version of the $Q$ and $U$ array common mode is subtracted from each bolometer TOD in the instrument frame. The spectral shape of the low pass filter is the inverse of the fit power spectral density of the stacked timestream derived in the data selection pipeline. The same filter is applied to simulated data. As described in Section~\ref{low_freq_noise} we do not see significant low frequency noise beyond a single focal plane common mode.

We project the TOD into $8^\prime$ pixels on the sky using a Lambert Cylindrical equal area projection. The resulting maps for the real data and a sample noise realization are shown in Figure~\ref{maps}. We achieve an effective map depth of \mapdepth\ after correcting for the beam and TOD filtering.

\vfill\null
			
\subsection{Power Spectrum Estimation} \label{powerspectrum}

The power spectrum estimation pipeline closely follows PB14 and \enquote{Pipeline A} in PB17 with several minor changes to improve numerical accuracy at low-$\ell$. A brief overview is provided for completeness.

The dataset is grouped into 38 bundles of approximately uniform weight and sky coverage. We form pseudo power spectra by taking cross spectra between these bundles to remove the noise bias,

\begin{equation}
\tilde{C}^{XY} = \frac{1}{\sum_{i \neq j} w_{i} w_{j}} \sum_{i \neq j} w_{i}w_{j} \mathbf{m}_{i}^{ X } \mathbf{m}_{j}^{Y*}
\end{equation}

\noindent where $\mathbf{m}$ and $w$ are the apodized Fourier transform and weight of each map bundle, respectively. The pseudo spectrum is averaged into bins of $\Delta \ell = 2$. Following PB14 we use a pure $B$-mode estimator based on \citet{2006PhRvD..74h3002S} and \citet{PureEstimator_Smith2006}. The apodization mask used in the pseudo spectrum is significantly more aggressively smoothed than PB17. We apply an $8^\circ$ cosine square edge taper and an $8^\circ$ Hamming window to the pixel weight map. This improves the numerical stability of the reconstructed power spectrum at $\ell \leq 100$. We do not mask point sources in the power spectrum estimation. As described in Section~\ref{pointsource} we do not see evidence for bright polarized point sources in higher resolution versions of our maps and we expect the contamination from unresolved polarized point sources to be negligible for this $\ell$ range.


The noise pseudo spectrum $\tilde{N}^{XY}$ is taken to be the pseudo spectrum of the sum of the apodized map bundles minus the cross pseudo spectrum $\tilde{C}^{XY}$. 

The pseudo spectrum is taken to be a linear function of the true underlying power spectrum on the sky

\begin{equation}
\tilde{C}_{\ell}=\sum_{\ell^{\prime}}\mathbf{K}_{\ell\ell^{\prime}}C_{\ell^{\prime}},
\end{equation}

\noindent where the transformation is given by

\begin{equation}
\mathbf{K}_{\ell \ell^{\prime}} = \mathbf{M}_{\ell \ell^{\prime}} F_{\ell^{\prime} } B_{\ell^{\prime}}^{2}.
\end{equation}

The mode mixing matrix $\mathbf{M}_{\ell \ell^{\prime}}$ describes the mixing of $\ell$ modes due to finite sky coverage and is estimated directly by computing the pseudo spectrum of narrowband noise realizations. The filter transfer function is found via an iterative procedure following PB14.

\begin{equation}
F_{\ell}^{n}=F_{\ell}^{n-1}+\frac{\tilde{C}_{\ell}-\sum_{\ell^{\prime}} \mathbf{M}_{\ell \ell^{\prime}} F_{\ell^{\prime}}^{n-1} C_{\ell^{\prime}} B_{\ell^{\prime}}^{2}}{C_{\ell} B_{\ell}^{2}},~ F_\ell^{0} = 1.
\end{equation}

In contrast to PB17 we cut the iterative series off at $n=3$ to avoid over-fitting fluctuations in the simulated pseudo spectra. This results in a negligible bias on the reconstructed power spectrum. The filter transfer function is calculated using simulated \lcdm\ $EE$-only and $BB$-only skies drawn from the \planckeight\ baseline TT, EE, TE + lowE + lensing \lcdm\ cosmology \citep{Aghanim:2018eyx} with $1^\prime$ pixels. We verify that the numerical value of the filter transfer function does not depend on the underlying cosmology used in the simulations.

We estimate the power spectrum in coarser bins of width $\Delta \ell = 50$. The binned estimate for the true power spectrum can be written using binning and interpolation operators $\mathbf{P}$ and $\mathbf{Q}$,

\begin{equation}
\hat{C}_{b}=\sum_{b^{\prime} \ell} \mathbf{K}_{b b^{\prime}}^{-1} \mathbf{P}_{b^{\prime} \ell} \tilde{C}_{\ell},
\label{psestimation}
\end{equation}

\begin{equation}
\mathbf{K}_{b b^{\prime}}=\sum_{\ell \ell^{\prime}} \mathbf{P}_{b \ell} \mathbf{M}_{\ell \ell^{\prime}} F_{\ell^{\prime}} B_{\ell^{\prime}}^{2} \mathbf{Q}_{\ell^{\prime} b^{\prime}}.
\end{equation}

The dependence of the binned spectrum on the underlying spectrum is given by the band power window functions $\mathbf{w}_{b \ell}$,

\begin{equation}
\hat{C}_{b}=\sum_{\ell} \mathbf{w}_{b \ell} C_{\ell},
\end{equation}

\begin{equation}
\mathbf{w}_{b \ell}=\sum_{b^{\prime} \ell^{\prime}} \mathbf{K}_{b b^{\prime}}^{-1} \mathbf{P}_{b^{\prime} \ell^{\prime}} \mathbf{K}_{\ell^{\prime} \ell}.
\end{equation}

\begin{figure}
\begin{center}
\includegraphics[scale=0.44]{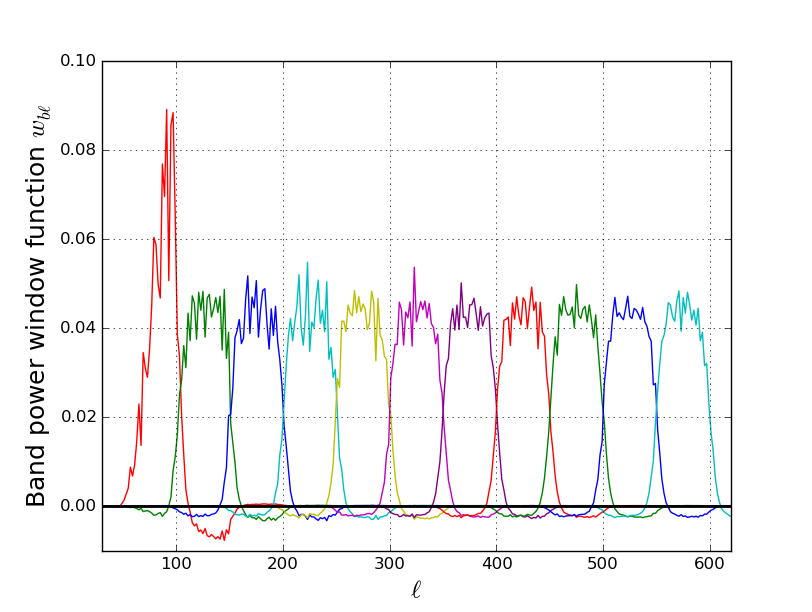}
\end{center}
\caption{$BB$ band power window functions. The shape of the lowest bandpower is due to 
the sensitivity degradation at low-$\ell$ due to timestream filtering.}
\label{bpwf}
\end{figure}

\noindent These band power window functions are shown in Figure~\ref{bpwf}. The shape of the lowest bandpower is due to sensitivity degradation at low-$\ell$. 

We have checked that the flat sky approximation does not introduce a significant bias in the measured power spectra.

The statistical uncertainty on the reconstructed power spectrum is taken to be the standard deviation of the reconstructed power spectrum of Monte Carlo (MC) simulations containing filtered \lcdm\ sky realizations and \enquote{signflip} noise realizations. We compute analytic uncertainties in the same way as PB14 and find those results to be consistent with our MC uncertainties.

For all spectra we follow the $D_\ell \equiv \ell (\ell + 1) C_\ell / (2\pi)$ convention unless otherwise noted.

\subsubsection{$E$-mode and $B$-mode mixing}

The timestream filtering mixes $E$-mode power into the $B$-mode spectrum. This is subtracted in the pseudo spectra following PB14,

\begin{equation}
\tilde{C}_{\ell}^{E \rightarrow B} = \frac{F_{\ell}^{E \rightarrow B}}{F_{\ell}^{E \rightarrow E }} \tilde{C}_ {\ell}^{E}. 
\label{ebleak}
\end{equation}

This reconstructs the correct central value of $C_\ell^{BB}$ however, it does not remove the excess variance in the $B$-mode spectrum. We find that the level of $B$-mode power introduced by timestream filtering is comparable to the expected lensing $B$-mode signal in our lowest band power. Since the fluctuations due to leaked $E$-modes are below the fluctuations of the noise bias, the contribution to the statistical uncertainties is small. We nonetheless account for this effect in our statistical uncertainty by subtracting the measured $E$-mode spectrum in each realization in our simulations. We do not perform any matrix based separation of $E$ and $B$-modes as described in \citet{2016ApJ...825...66B}.

\subsubsection{Quantifying low-$\ell$ statistical performance}

\begin{figure*}
\begin{center}	
\includegraphics[scale=0.44]{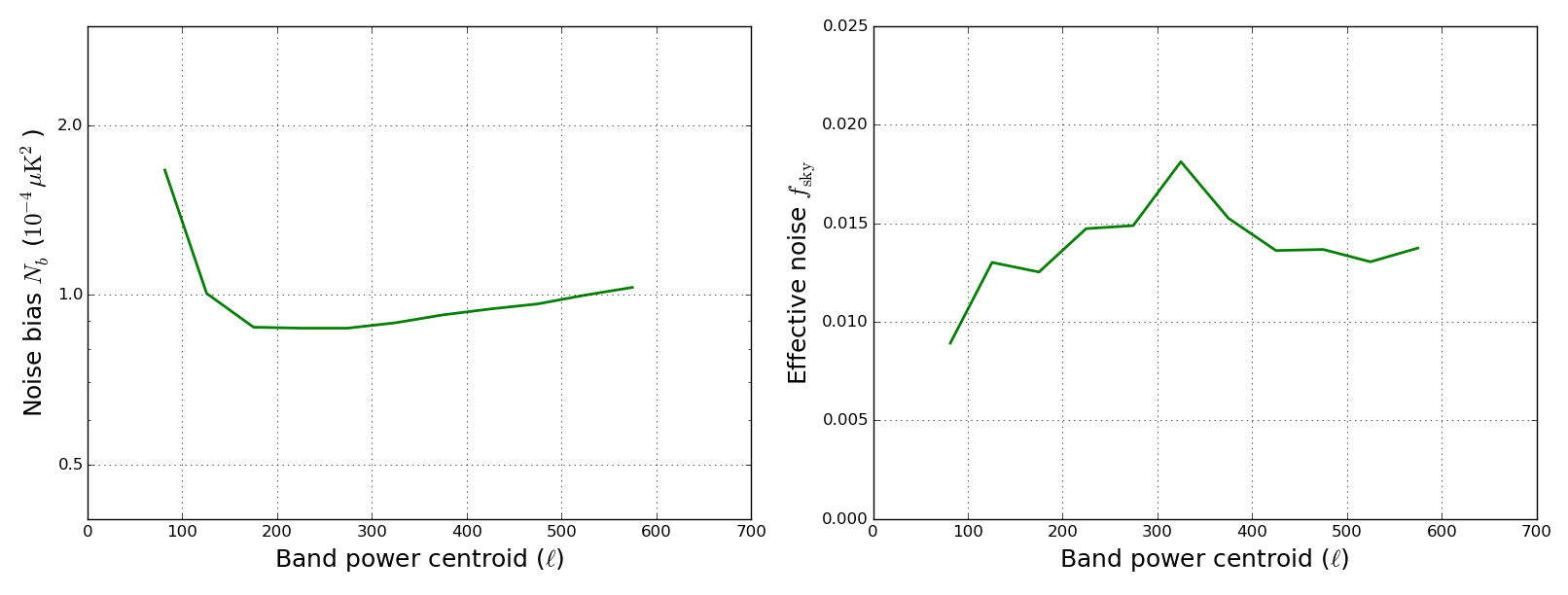}
\end{center}
\caption{Noise bias $N_b$ after correcting for TOD filtering and the beam window function (left),  
effective number of degrees of freedom $\nu_b$ per band power defined in Equation~\ref{nu_b_eq} 
written as an $\ell$-dependent 
$f_\mathrm{sky} = \nu_b/( 2 \ell \cdot \Delta \ell)$, where $\ell$ and $\Delta \ell$ refer to the nominal 
band power definitons (right). The degradation in $N_b$ at high $\ell$ 
is primarily due to the beam window function and the degradation at low $\ell$ is 
due to low frequency noise and timestream filtering. These curves are derived from the 
auto spectrum of the \enquote{signflip} noise realizations computed using the fiducial cross spectrum pipeline. 
The alternate auto spectrum estimate described in Appendix~\ref{alternateps} gives similar $N_b$ with a marginally larger effective $f_\mathrm{sky}$.
This plot can be directly compared to Figure 3 in BK15. Numerical values for 
these curves are given in Table~\ref{nbtable}.}
\label{nbfigure}
\end{figure*}

To accurately describe the low-$\ell$ statistical performance of \pb\ it is necessary to account for the effect of the TOD filtering and the beam window function in the measured noise bias. We do this following BK15 by referring the noise pseudo spectrum $\tilde{N}_\ell$ to a true power spectrum on the sky $N_b$ following Equation~\ref{psestimation}. We write the number of degrees of freedom per band power as an $\ell$-dependent effective sky area. These two quantities represent the noise contribution to the power spectrum statistical uncertainties. Figure~\ref{nbfigure} and Table~\ref{nbtable} show the results of this analysis. It should be noted that this is not exactly equivalent to an analytic estimate of the statistical errors because it does not account for terms arising from the CMB signal variance.

We additionally compute an $\ell$ knee that can be compared to the numbers presented in \citet{2019JCAP...02..056A}. We fit our estimated power spectrum uncertainties following the same formalism and find a knee of \ellknee. This $\ell$ knee can also be computed by incorporating the effective sky area degradation into the noise bias, $N_b / \sqrt{f_\mathrm{sky}}$ using the numbers in Table~\ref{nbtable} giving consistent results. 

						
\begin{table*} 
\begin{center}
\caption{Noise bias and effective degrees of freedom} \label{nbtable}
\begin{tabular}{ c  c  c  c }
\hline
Nominal band definition $\ell$ & Band power centroid ($\ell$) & 
Noise bias $N_b ~ (10^{-4} \mathrm{\mu K}^2)$ & Effective noise $f_\mathrm{sky}$ \\
\hline
$50 < \ell \leq 100$  & 81.5    & 1.665  & 0.0089  \\
$100 < \ell \leq 150$ & 125.6  & 1.006  & 0.0130  \\
$150 < \ell \leq 200$ & 175.1  & 0.875  & 0.0125  \\
$200 < \ell \leq 250$ & 224.8  & 0.872  & 0.0147  \\
$250 < \ell \leq 300$ & 274.4  & 0.872  & 0.0149  \\
$300 < \ell \leq 350$ & 324.8  & 0.891  & 0.0181  \\
$350 < \ell \leq 400$ & 374.9  & 0.921  & 0.0152  \\
$400 < \ell \leq 450$ & 425.0  & 0.943  & 0.0136  \\
$450 < \ell \leq 500$ & 474.8  & 0.963  & 0.0137  \\
$500 < \ell \leq 550$ & 524.7  & 0.998  & 0.0130  \\
$550 < \ell \leq 600$ & 574.7  & 1.030  & 0.0137  \\
\hline
\end{tabular} 
 \par
\end{center}
\vspace{8pt}
\noindent Data points for the curves shown in Figure~\ref{nbfigure}. 
These data use the \enquote{signflip} noise model and the fiducial internal 
cross power spectrum estimator. These numbers are written in $C_\ell$ units 
and do not include the factor of $\ell (\ell +1) / 2 \pi$ 
used in the rest of this work.
\end{table*}


\subsection{Map level calibration}

\begin{figure*}
\begin{center}	
\includegraphics[scale=0.44]{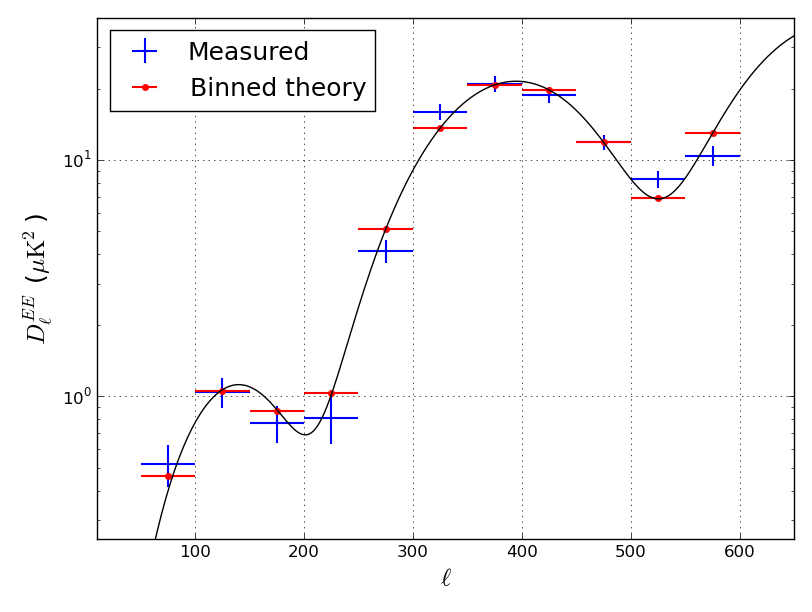}~\includegraphics[scale=0.44]{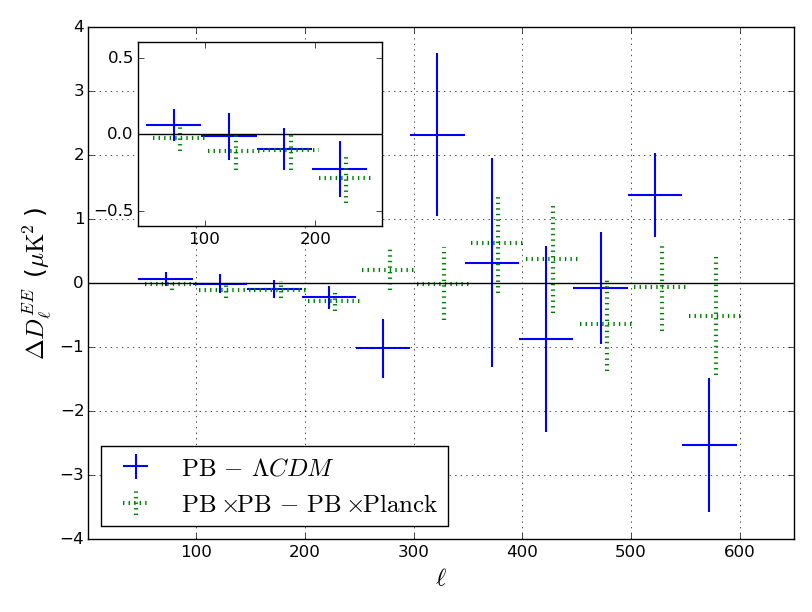}
\end{center}
\caption{The $E$-mode band powers after absolute gain calibration compared to the 
best fit \planckeight\ \lcdm\ cosmology (left), the residuals compared to the 
binned theory and the null quantity formed by subtracting the debiased cross spectrum 
with filtered \planckeight\ 143 GHz maps (right). The $E$-mode spectrum 
is used as an overall gain and effective beam width calibration.
The lowest four bandpowers are shown in the inset.}
\label{eespec}
\end{figure*}

We perform an overall gain calibration by cross correlating our data to the \planck\ satellite. We use the \planckeight\ PR3 143 GHz full mission maps\footnote{\url{https://pla.esac.esa.int}} and process them through our filtering and mapmaking pipeline. We compute debiased spectra for both the \pb\ internal cross spectra using the fiducial power spectrum estimate and the fully coadded \pb\ maps correlated with the scanned \planck\ maps using the full coadd power spectrum estimator described in Apendix~\ref{alternateps}. We fit an overall gain calibration factor based on the ratio of the $E$-mode spectra,

\begin{equation}
\hat{g}_b = \hat{C}_b^{EE, \mathrm{PB}} ~\big{/}~ \hat{C}_b^{EE, \mathrm{PB} \times \mathrm{Planck}} .
\end{equation}

The uncertainty on this ratio is computed by MC simulation, holding the underlying sky realization fixed. We use 96 realizations of the Planck FFP10 noise model \citep{Aghanim:2018fcm} to approximate the Planck map noise. We fit this calibration to an $\ell$-dependent gain model accounting for a smearing of the beam profile following

\begin{equation}
g(\ell) = g_0 \, \exp \left[-\frac{\ell (\ell + 1)}{2}  \, \sigma^2 \right],
\end{equation}

\begin{equation}
\hat{g}_{b}=\sum_{\ell} \mathbf{w}_{b \ell} g_{\ell}.
\end{equation}

We find an overall gain calibration factor of \gainvalanderr\ in amplitude and an effective beam smearing of $\sigma^2 = 1.14 \pm 5.58 ~\mathrm{arcmin}^2$. We convolve the beam with the best fit value for the pointing model error and treat the uncertainties in $g_0$ and $\sigma^2$ as a calibration error term in our final results. We compute an alternate absolute gain calibration fitting the \pb\ spectrum to the best fit \planck\ \lcdm\ theory spectrum and find agreement with the fiducial calibration at the percent level across our $\ell$ range.

After applying this absolute calibration we compare $E$-mode spectrum to Planck and form a null spectrum

\begin{equation}
\hat{C}_{b, \mathrm{null}} = \hat{C}_b^{EE, \mathrm{PB}} - \hat{C}_b^{EE, \mathrm{PB} \times \mathrm{Planck}}.
\end{equation}

The uncertainty on this null spectrum is computed by MC. We find this null spectrum to be consistent with zero with $\chi^2/\nu = 7.0/9$ corresponding to a probability-to-exceed (PTE) of 64\%. When we compare our measured $E$-mode spectrum to the best fit \lcdm~theory, we observe a marginally significant discrepancy in our highest two $\ell$ bins. This appears to be due to an anisotropic feature seen in the two-dimensional power spectrum at approximately the size scale and orientation of the detector wafers. These fluctuations have no significant counterpart in the null tests described in Section \ref{null_test}. These fluctuations do not depend on any of the operations in the TOD filtering and mapmaking pipeline that have characteristic scales on the sky or frequencies in the time domain. The overall gain and beam calibration does not significantly shift if these two bins are removed from the analysis. We find that there is no significant shift in the $B$-mode spectrum when these regions of the Fourier plane are masked in the pseudo spectrum estimation. We show our measured $E$-mode spectrum as well as the residuals compared to \planck\ and the \lcdm\ theory in Figure \ref{eespec}.

After applying the overall gain calibration we self calibrate the overall instrument polarization angle following \citet{Keating2013}. We apply an overall polarization angle correction $\Delta \psi$ such that the measured $C_\ell^{EB}$ power is minimized,

\begin{equation}
-2 \ln (\mathcal{L}) = \sum_{b}\left[\frac{\hat{C}_{b}^{E B}-(1/2) \sin (4 \Delta \psi) \sum_\ell \mathbf{w}_{b\ell} C_{\ell}^{E E}}{\Delta \hat{C}_{b}^{E B}}\right]^{2} .
\end{equation}

\begin{figure}
\begin{center}
\includegraphics[scale=0.44]{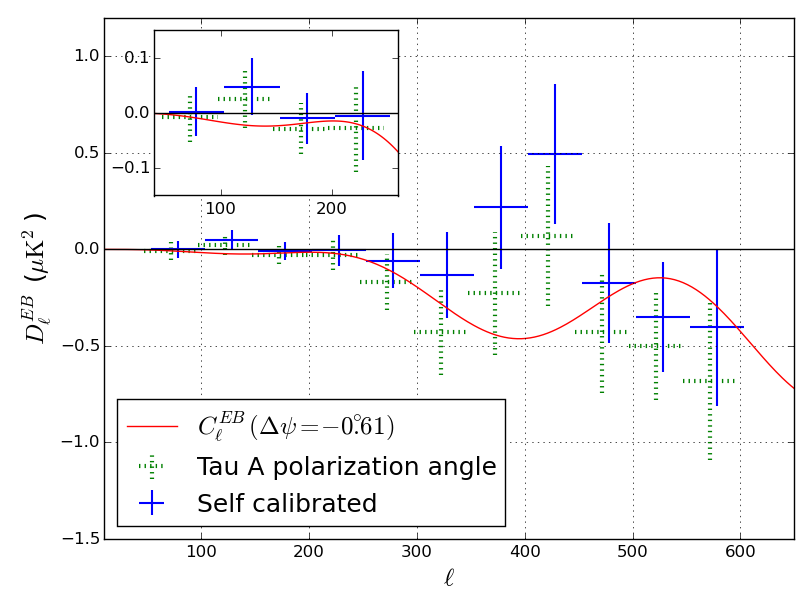}
\end{center}
\caption{$EB$ power spectrum based on the IRAM \taua\ measurement and after polarization angle self-calibration. We find the angle 
derived from self-calibration to be statistically consistent with PB17. We find our measured 
$EB$ spectrum to be consistent with zero after an overall polarization angle is subtracted with 
$\chi^2/\nu = 6.56/10$ corresponding a 77\% PTE. The lowest four bandpowers are shown in the inset.}
\label{eb_polangle}
\end{figure}

\vspace{5pt}
\noindent Here $\Delta \hat{C}_{b}^{E B}$ is the uncertainty in the reconstructed $EB$ spectrum from the MC simulations. We find an overall calibration $\Delta \psi$ = \rotvalanderr. After applying this calibration we find that $\hat{C}_b^{EB}$ is consistent with zero with a total $\chi^2$ PTE of 77\%. Our absolute angle calibration is compatible with PB17 which reported $\Delta \psi_\mathrm{~PB17} = -0 \fdg 79 \pm 0 \fdg 16$. This calibration is shown in Figure~\ref{eb_polangle}. It should be noted this calibration is independent of PB17.

The self-calibrated polarization angle correction is applied in the map domain. The absolute gain computed with the $EE$ spectrum differs slightly from the MC simulations used to establish the band power errors and covariances. To account for this in the final power spectrum uncertainties, we assume that the total variance in each band power is the sum of the signal variance and the noise variance as derived from signal-only and noise-only simulations, respectively. The noise variance is rescaled to the best fit absolute gain calibration.

We quantify a calibration uncertainty in the measured power spectrum using the uncertainty in the overall gain calibration $g_0$, pointing model error $\sigma^2$, polarization angle self calibration $\Delta \psi$, and statistical uncertainty on the beam window function $\Delta B_\ell$. The correlation between $g_0$ and $\sigma^2$ is modeled as a simple Gaussian covariance. The three effects are added in quadrature to form a total calibration error. This represents an upper limit as the overall gain calibration and beam window function are degenerate. The numerical values of these calibration errors in the estimated power spectrum are shown in Section~\ref{autospectrum}.

\label{abscal}

\section{Systematic Uncertainties} \label{systematics}

\input{systematics}

\subsection{Null Tests}

We perform a set of null tests to establish the internal consistency of the dataset and search for possible systematic contamination in the final power spectra. In general, it is not possible to construct difference maps between halves of the data with zero signal due to anisotropic scanning and filtering effects. This effect becomes particularly pronounced at low-$\ell$. As a result, we follow the formalism developed originally by the \quiet\ collaboration \citep{PhDT_Bischoff} and used in PB14 and PB17. We include the filtering and mode mixing explicitly in the construction of the null spectrum

\begin{equation}
\hat{C}_{\ell}^{\rm null} = \hat{C}_\ell^{A} + \hat{C}_\ell^{B} - 2\hat{C}_\ell^{AB},
\end{equation}

\noindent where $\hat{C}_\ell^{A}$ and $\hat{C}_\ell^{B}$ ($\hat{C}_\ell^{AB}$) are the debiased autospectrum for each half of the split (the cross spectrum between the splits). The spectra are computed using the same cross spectrum formalism and map bundles used in the main pipeline.

The filter transfer function and mode mixing matrix are computed in the same way as the fiducial power spectrum pipeline using 92 $EE$-only and 92 $BB$-only \planckeight~\lcdm~input maps for each test. Only map regions present in both halves of the split are used to form the pseudo spectrum and mode mixing matrix. The null spectra are computed using the same $\ell$ binning as the final power spectrum. $EE$, $EB$, and $BB$ null spectra are compared to 192 $EE+BB$ signal and noise simulations. We find that this number of simulations is adequate for percent-level statistical uncertainties on the null spectrum PTE values and filter transfer functions across our full $\ell$ range. 

For our fiducial null test statistics, we use noise realizations generated with the \enquote{signflip} pipeline. There is no significant difference from the PTE values computed with the TOD noise model with the exception of the \enquote{top versus bottom} null test that explicitly separates detector pairs. This is due to the presence of an additional anticorrelated noise term when detector pairs are separated that is not included in the TOD noise model described in Section 
\ref{low_freq_noise}.\label{null_test}

\begin{figure*}
\begin{center}
\includegraphics[scale=0.44]{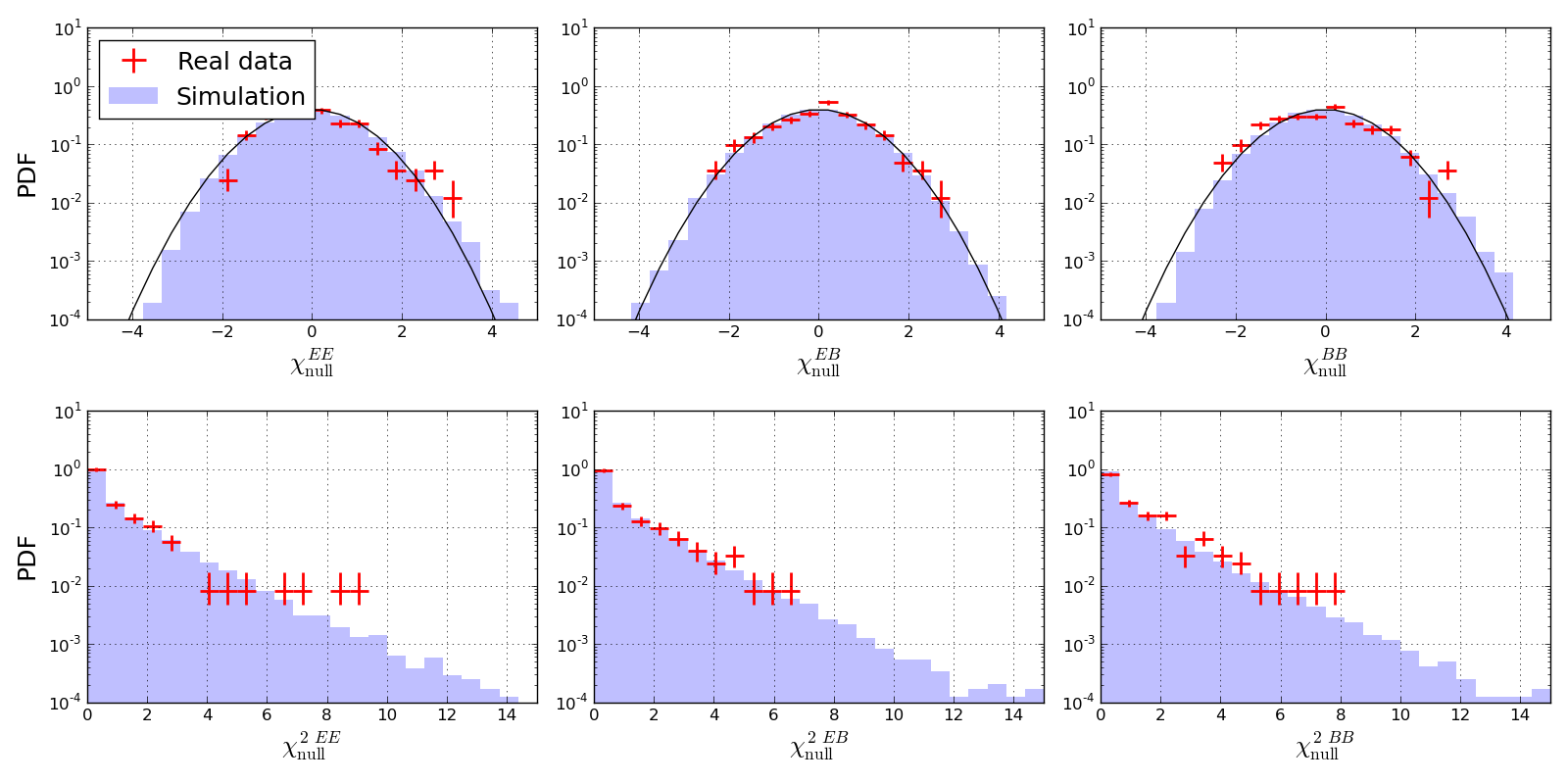}
\end{center}
\caption{One dimensional $\chi_\mathrm{null}= C_\mathrm{null} / \sigma_\mathrm{null, MC}$ 
distribution from the fiducial set of null test splits (top), the same data shown as 
$\chi^2_\mathrm{null}$ (bottom). No statistically significant outliers are seen in these data. 
Error bars on the real data histograms represent 68\% Poisson confidence intervals. 
The solid line in the upper panels shows a unit variance Gaussian as an approximation to the 
real distribution. Note that the summary statistics do not assume a Gaussian distribution for $\chi_\mathrm{null}$ 
since the data are only compared to the simulations.}
\label{chi_null}
\end{figure*}

\begin{table*}
\begin{center}
\caption{Null test PTE values}  \label{pte_by_test}
\begin{tabular}{ c c c c }
\hline
 & \textbf{Total $EE$ $\chi^2$ PTE} & \textbf{Total $EB$ $\chi^2$ PTE} & \textbf{Total $BB$ $\chi^2$ PTE} \\
\hline				
\textbf{Null test summed over $\ell$ bins} & & & \\
First half versus second half & 			86.5 \% & 43.2 \% & 79.7 \%\\
Rising versus middle and setting & 		85.4 \% & 70.8 \% & 6.8  \%\\
Middle versus rising and setting & 		63.0 \% & 63.0 \% & 39.1 \%\\
Setting versus rising and middle & 		50.5 \% & 53.1 \% & 19.3 \%\\
Left-going versus right-going subscans & 	36.5 \% & 26.0 \% & 6.2  \%\\
High gain versus low gain CESs & 	45.3 \% & 83.3 \% & 3.1  \%\\
High PWV versus low PWV &		 	50.5 \% & 33.9 \% & 28.6 \%\\
Common mode $Q$ knee frequency & 	57.8 \% & 36.5 \% & 26.6 \%\\
Common mode $U$ knee frequency & 	91.7 \% & 54.7 \% & 79.7 \%\\
Mean temperature leakage by bolometer &	78.6 \% & 54.2 \% & 56.8 \%\\
$2f$ amplitude by bolometer & 			5.7  \% & 32.8 \% & 83.9 \%\\
$4f$ amplitude by bolometer & 			35.4 \% & 18.2 \% & 33.3 \%\\
$Q$ versus $U$ pixels & 				72.4 \% & 64.6 \% & 28.1 \%\\
Sun above or below the horizon & 		76.6 \% & 91.1 \% & 32.2 \%\\
Moon above or below the horizon & 		76.0 \% & 77.6 \% & 57.3 \%\\
Top half versus bottom half & 			79.7 \% & 35.4 \% & 27.1 \%\\
Left half versus right half & 				53.6 \% & 33.9 \% & 76.0 \%\\
Top versus bottom bolometers &			36.5 \% & 55.7 \% & 35.4 \%\\
\hline
\textbf{$\ell$ bin summed over null tests} &			&	  &	   \\
$50 \leq \ell \leq 100$ &				31.8 \%  & 58.3 \%  & 44.8  \%\\
$100 < \ell \leq 150$ &			64.1 \%  & 14.1 \%  & 24.5  \%\\
$150 < \ell \leq 200$ &			61.5 \%  & 46.9 \%  & 96.9  \%\\
$200 < \ell \leq 250$ &			71.4 \%  & 74.0 \%  & 28.6  \%\\
$250 < \ell \leq 300$ &			83.9 \%  &  7.3 \%  & 26.6  \%\\
$300 < \ell \leq 350$ &			50.5 \%  & 92.7 \%  &  6.8  \%\\
$350 < \ell \leq 400$ &			64.1 \%  & 97.9 \%  & 92.2  \%\\	
$400 < \ell \leq 450$ &			44.3 \%  & 84.4 \%  &  5.2  \%\\
$450 < \ell \leq 500$ &			96.9 \%  & 63.5 \%  &  3.1  \%\\
$500 < \ell \leq 550$ &			68.8 \%  & 84.5 \%  & 49.0  \%\\
$550 < \ell \leq 600$ &			49.5 \%  & 16.1 \%  & 49.5  \%\\
\hline
\end{tabular} \par
\end{center}
\vspace{8pt}
\noindent PTE values for the total $\chi^2$ of each null spectrum summed 
over $\ell$ bins and each $\ell$ bin summed over null spectra. None of the 
null spectra indicate significant problems. The PTE values are computed 
directly from the 192 signal+noise simulations and are therefore quantized at the 0.5\% level.
\end{table*}

\subsubsection{Null test data splits}

We split the data along 18 largely uncorrelated axes designed to probe a wide range of possible sources of systematic contamination. Where possible the data set is split into halves with equal weight. In the following we briefly described these splits.

\begin{itemize}
\item \enquote{First half versus second half}: the dataset is split into two equal-weight halves chronologically to probe for time dependent miscalibration or changes in the instrument.

\item \enquote{Rising versus middle and setting}, \enquote{middle versus rising and setting}, \enquote{setting versus rising and middle}: the three different CES types are split in all possible combinations to detect elevation-dependent miscalibration or residual ground synchronous signal.

\item \enquote{Left-going versus right-going subscans}: the dataset is split in half according to the direction of motion of the telescope to test for microphonic or magnetic pickup in the data.

\item \enquote{High gain versus low gain observations}: the dataset is split into observations with above and below average mean detector gain coefficients to search for problems with the gain calibration.

\item \enquote{High PWV versus low PWV}: the dataset is split by PWV as measured by the nearby \apex\ radiometer to check for loading or weather dependent effects.

\item \enquote{Common mode $Q$ knee frequency}, \enquote{common mode $U$ knee frequency}: the dataset is split into observations with high and low knee frequencies in the telescope frame $Q$ and $U$ common mode signal to check for problems in the treatment of low frequency contamination. The $Q$ knee frequency split overlaps with the cloud detection criteria from T19. Both splits are largely uncorrelated with the PWV split.

\item \enquote{Mean temperature to polarization leakage by channel}: split the dataset into detectors that see small and large temperature leakage coefficients to test the subtraction and search for residual contamination.

\item \enquote{$2f$ amplitude by channel}, \enquote{$4f$ amplitude by channel}: split the data by HWP signal amplitude to check for problems removing the HWP structure or systematic contamination coupling into the data through these terms.

\item \enquote{Q versus U pixels}: each detector wafer is fabricated with two sets of polarization angles. We split the data into the two pixel types to check for problems in the device fabrication.

\item \enquote{Sun above or below the horizon}, \enquote{Moon above or below the horizon}: we split observations based on whether or not the sun or moon is up to check for residual sidelobe contamination.

\item \enquote{Top half versus bottom half}, \enquote{left half versus right half}: we split detectors by the boresight axis of the telescope to check for optical distortion and problems due to  far sidelobes.

\item \enquote{Top versus bottom bolometers}: with a continuous HWP each bolometer TOD independently measures $Q$ and $U$. We explicitly separate detector pairs to check for temperature aliasing or device mismatch.

\end{itemize}

We have also considered season-by-season data splits, however these are not included in the final suite because they are highly correlated with the first half versus second half split. Additionally, we have examined average $2f$ and $4f$ amplitudes as well as average temperature to polarization leakage by  observation but do not include these due to redundancy with the weather and gain splits. We also considered sun and moon boresight distance splits but do not include them due to overlap with the sun and moon being above the horizon. None of the excluded splits indicated significant problems in earlier iterations of the pipeline and removing redundant tests improves sensitivity to outliers in the splits used.

\begin{figure}
\begin{center}
\includegraphics[scale=0.44]{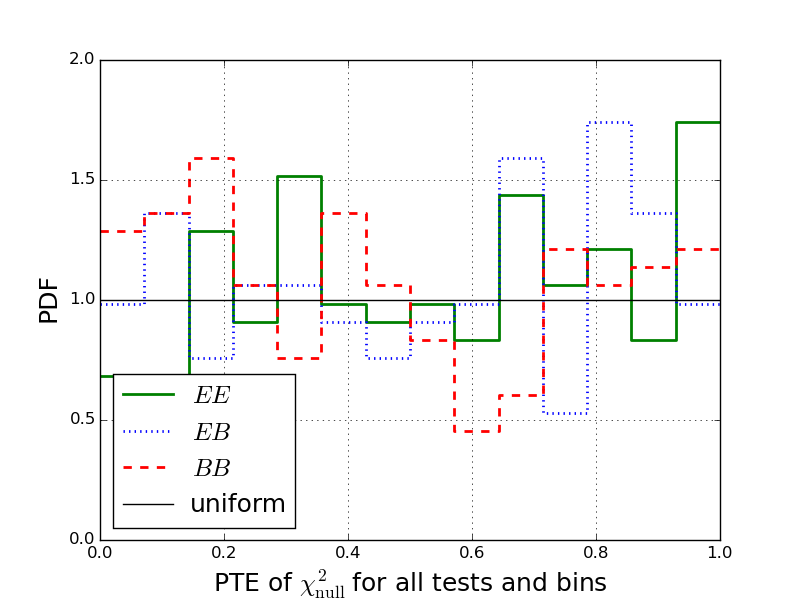}
\end{center}
\caption{Distribution of PTE values for each bin in each test. The distribution for all three spectra are consistent with uniform.}
\label{pte_ks}
\end{figure}

\begin{table} 
\begin{center}
\caption{Null test PTE values} \label{null_test_stats}
\begin{tabular}{ c c  c c }
\hline
Null statistic 	& 			$EE$ PTE & $EB$ PTE & $BB$ PTE \\
\hline
Average $\chi$ overall & 		73.4 \% & 80.7 \% & 9.9  \%\\
Most extreme $\chi^2$ by bin & 		96.9 \% & 43.8 \% & 31.7 \%\\
Most extreme $\chi^2$ by test & 	70.3 \% & 98.4 \% & 57.3 \%\\
Most extreme $\chi^2$ overall & 	48.4 \% & 84.9 \% & 66.1 \%\\
Total $\chi^2$ overall & 		90.6 \% & 78.7 \% & 12.5 \%\\
\hline
\textbf{Lowest statistic} & 		\textbf{85.4} \% & \textbf{86.4} \% & \textbf{33.9} \%\\
\hline
KS test on all bins & 			10.1 \% & 60.4 \% & 15.9 \%\\
KS test on all spectra & 		6.4  \% & 31.8 \% & 10.5 \%\\
KS test overall & 			35.9 \% & 27.5 \% & 14.6 \%\\
\hline
\end{tabular} \par
\end{center}
\vspace{8pt}
\noindent PTE values for each of the high level null test statistics.
\end{table}

\subsubsection{Null test statistics and analysis}

For each bin in each null spectrum we compute the statistic $\chi_\mathrm{null} \equiv \hat{C}_b^\mathrm{null} / \sigma (\hat{C}_b^\mathrm{null})$ where $\sigma (\hat{C}_b^\mathrm{null})$ is the standard deviation of the MC null spectra. We use both $\chi_\mathrm{null}$ and $\chi^2_\mathrm{null}$ because the former is more sensitive to systematic biases and the latter is more sensitive to outliers. Figure \ref{chi_null} shows the $\chi_\mathrm{null}$ and $\chi^2_\mathrm{null}$ distributions compared to the expectation from MC simulations. The PTE of the total $\chi^2_\mathrm{null}$ summed over $\ell$ bins for each test and summed over tests for each bin is shown in Table \ref{pte_by_test}.

To probe for systematic contamination and consistency of the noise model we compute five statistics on the $\chi_\mathrm{null}$ values. These statistics were determined before computing spectra with the real data.

\begin{enumerate}
\item \enquote{Average $\chi$ overall}: the mean value of $\chi_\mathrm{null}$ for all tests and bins
\item \enquote{Most extreme $\chi^2$ by bin}: the most extreme $\chi^2_\mathrm{null}$ when summing spectra over tests
\item \enquote{Most extreme $\chi^2$ by test}: the most extreme  $\chi^2_\mathrm{null}$ when summing spectra over bins
\item \enquote{Most extreme $\chi^2$ overall}: the most extreme $\chi^2_\mathrm{null}$ for all bins and tests
\item \enquote{Total $\chi^2$ overall}: the sum of $\chi^2_\mathrm{null}$ for all spectra
\end{enumerate}

For each statistic we compute a PTE by comparing the real data to same statistic computed with the MC realizations. Directly comparing the data to the simulations accounts for any correlations that exist between bins and tests in the computation of the PTE values. We define an additional statistic $P_\mathrm{low}$ that is the lowest of the five PTE values. We require that the PTE of $P_\mathrm{low}$ be greater than 5\%. The numerical value for these statistics can be seen in Table \ref{null_test_stats}. All spectra ($EE$, $EB$ and $BB$) pass these criteria.

Additionally, we require that the PTE of the $\chi^2_\mathrm{null}$ values by test, by bin, and overall be consistent with a uniform distribution using a Kolmogorov-Smirnov (KS) test to test for systematic mismatch between the real and simulated uncertainties. Figure \ref{pte_ks} shows the PTE distribution for all bins and tests. We find the PTE distributions to be consistent with uniform.

\subsection{Simulation of Systematic Errors} \label{systematicspipeline}

In addition to the null tests, we simulate several known sources of systematic error. We find upper bounds on the contamination introduced by these effects to be subdominant to the statistical error in all spectra.

The systematic error estimate uses a modified version of the simulation pipeline developed for the null tests and the power spectrum estimation. For most systematic effects, a signal-only \lcdm~sky is scanned to form simulated TOD. These TOD are then distorted following a model of the given effect and then filtered and projected onto the sky using the fiducial mapmaker. We compute pseudo spectra from these distorted maps and refer them to the underlying sky using the same mode coupling and filter transfer functions used for the fiducial power spectrum estimate. Since several systematics are suppressed by the overall gain and polarization angle calibration, we perform these overall calibrations in each simulation. We do not model the $\ell$-dependent gain model applied to the real data and only fit the overall gain $g_0$. In parallel, the same mapmaking and power spectrum estimation is performed without distorting the TOD to form a reference simulation. The systematic error is taken to be the absolute difference between the contaminated and reference power spectra.

\begin{figure*}
\begin{center}
\includegraphics[scale=0.36]{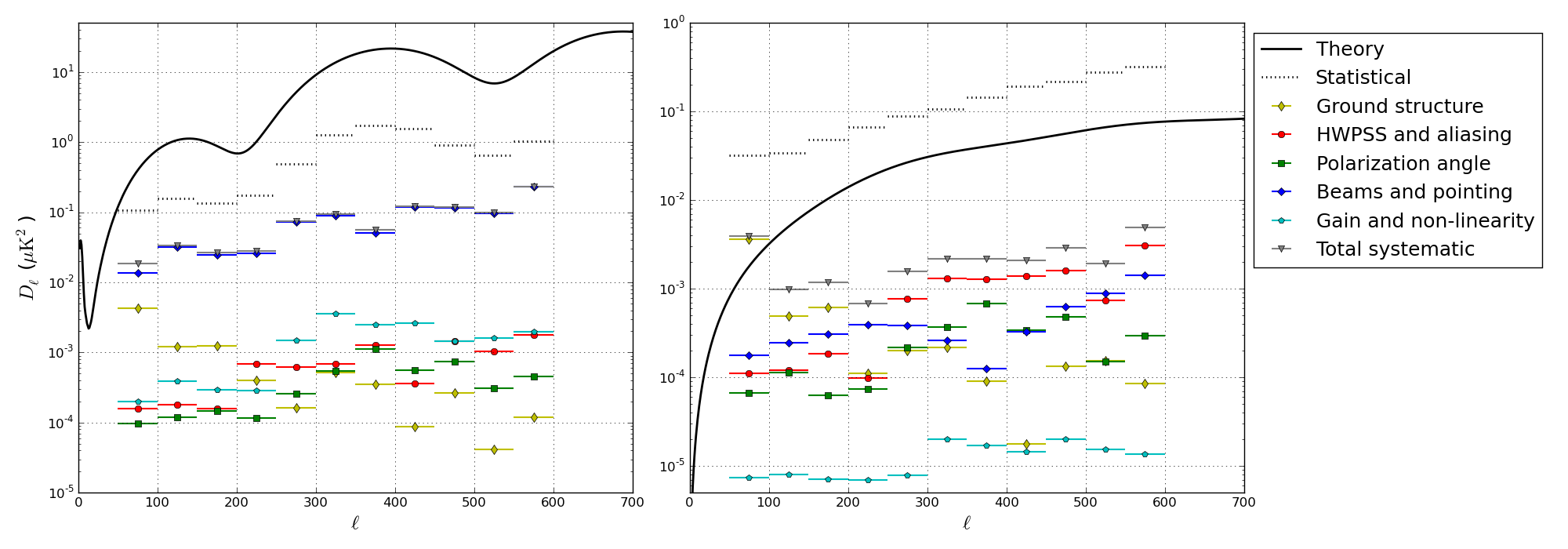}
\end{center}
\caption{Systematic contributions from all simulated effects grouped thematically. The ground structure curve
indicates the sum of contamination from the detector gain and linear drift models of ground contamination. The 
HWPSS and aliasing curve includes $0f$ and $2f$ signal aliasing as well as the HWP imperfections. The 
polarization angle curve shows the sum of individual and overall polarization angle uncertainties after self calibration. The 
beams and pointing curve includes pointing model uncertainty and effects related to detector crosstalk. The 
gain and non-linearity curve includes relative gain uncertainty and detector gain and time constant drift acting on the CMB signal. 
The dominant effect in $EE$ is the misestimation of the effective polarization beam due to detector crosstalk while the dominant 
systematic in $BB$ is the uncertainty in the ground structure subtraction. It should be noted that this
 is a conservative estimate driven by significant model uncertainties. Future experiments may be able to 
 suppress this effect significantly through careful study and control of ground pickup. 
 The total systematic error is formed assuming all systematics add linearly in power.}
\label{systematicsplot}
\end{figure*}

To form an overall systematic error estimate we linearly add the power spectrum contamination from each set of simulations. In practice we expect that each source of error will be largely uncorrelated with the others meaning this is a conservative upper limit. The total systematic error estimate as well as the contributions from groups of effects in $EE$ and $BB$ are shown in Figure~\ref{systematicsplot}. We find the expected systematic contamination in the $EB$ spectrum to be likewise subdominant to our statistical errors. 

\subsubsection{Gain miscalibration, time constant drift, detector non-linearity}

We simulate the error introduced by finite uncertainty on the relative gain calbration of each detector. We estimate the statistical uncertainty in each bolometer relative gain calibration to be 4.7\% due to the amplitude of the chopped thermal source and detector noise. 

The primary impact of detector non-linearity is the additive temperature to polarization leakage. However, there is a smaller multiplicative term from the gain variation acting on the CMB signal. We simulate this using a downsampled version of the normalized $4f$ amplitude (phase) as a tracer of the detector small signal gain (time constant) and inject the non-linear response (time dependent polarization angle error) into the simulation timestreams. Time constant drift and detector non-linearity modulating the ground synchronous structure is modeled separately.

\subsubsection{Polarization angle error}

We estimate the impact on the reconstructed power spectrum assuming that the calibrated detector polarization angle errors are Gaussian distributed around the true values with standard deviation $1\fdg 2$. This error estimate is taken from the scatter of the difference in polarization angles measured for the two detectors within a pair. This polarization angle uncertainty is comparable to the value used in PB17. The systematic effect is strongly suppressed by the absolute polarization angle calibration. We also estimate the effect of an overall polarization angle miscalibration by rotating the polarization angle of the input sky $0\fdg 5$ RMS based on the quoted systematic uncertainty in the \taua\ polarization angle from \citet{2010A&A...514A..70A}. The residual error from an overall polarization angle shift is not identically zero after self calibration because the map making operation is not invariant under polarization angle rotation due to different $Q$ and $U$ common mode filtering.

\subsubsection{Boresight pointing error}

We quantify the systematic impact of the imperfect knowledge of the boresight pointing by considering several candidate pointing solutions. We perform the input map scanning with the fiducial pointing model and project the TOD to the sky with an alternate pointing solution derived using different subsets of the point source observations or model parameters. We find the largest discrepancy from the fiducial pointing solution results from the inclusion of Jupiter data in the pointing model fit. We conservatively quote that residual  in our systematic error estimate. This is one of the largest systematic uncertainties in our $E$-mode spectrum, however this effect is significantly less than our statistical error over our $\ell$ range. This effect has a significantly smaller impact on the $B$-mode spectrum.

\subsubsection{Crosstalk}

We observe an electrical coupling between detectors read out on the same cables. We assume that this crosstalk is constant through the entire dataset and linear. We estimate the amplitude of this effect using the observations of Jupiter used for the gain and beam calibration. Crosstalk appears in these data as an apparent negative copy of the beam shape several tens of arcminutes away from the main beam. We estimate the amplitude of these images using a matched filter and construct a matrix representing the coupling of signal in detector $i$ to observed signal in detector $j$,

\begin{equation}
d_{j,\mathrm{observed}}=d_{j,\mathrm{real}}+\sum_{i \neq j} \mathbf{L}_{i j} d_{i,\mathrm{real}}.
\end{equation}

We find the matrix $\mathbf{L}$ to be sparse and do not see significant crosstalk for detectors read out on different cables. The median non-zero off-diagonal element of the matrix $\mathbf{L}$ is 1\% and the median row sum representing the ratio of cross talked power to power in the main beam is approximately 4\%. 

We estimate this systematic error as the sum of two effects. We first quantify error introduced in the beam calibration. This is due to the fact that crosstalk is strongly suppressed in polarization due to the different polarization angles of each detector \citep{Crowley:2018eib} in HWP experiments. As a result the temperature and polarization see different effective beam profiles. The second effect that we simulate is the direct distortion of the polarization signal by injecting crosstalk at the TOD level in signal-only simulations. We find the beam miscalibration effect to be the dominant systematic in our $E$-mode spectrum. It is possible to strongly suppress crosstalk by inverting the mixing matrix $\mathbf{L}$ in the data processing pipeline as done in \citep{2018ApJ...852...97H}, however we do not take this step as the expected contamination is already below our statistical error.

\subsubsection{Ground Pickup} \label{ground}

We observe a ground-fixed structure in the TOD that is on the order of 100 $\mathrm{\mu K}$ after subscan polynomial filtering. The majority of the effect is subtracted by binning each detector TOD in azimuth and subtracting that mode for each CES. This structure is likely optical and due to far sidelobes sensing the surrounding terrain. We simulate possible variation in this ground-fixed structure within each CES and estimate the impact of the residual on the final $B$-mode power spectrum.  We find this to be the dominant source of possible systematic contamination in our lowest $B$-mode spectrum $\ell$ bin.

One physically motivated model for this effect is the detector non-linearity modulating stable ground pickup. We simulate this using the low-pass filtered $4f$ amplitude as a gain tracer. As shown in T17 the dominant source of $4f$ amplitude variation is detector non-linearity modulating the detector small signal gain. This gain function modulates the ground fixed structure producing imperfect subtraction by the TOD filters.

In addition to the detector gain model, we simulate several linear drift models of instability in the ground template. We simulate the apparent ground structure amplitude drifting during the course of a CES and place an upper limit on this model using the TOD directly. For each bolometer and CES type (rising, middle, and setting) we fit $Q+iU$ for each subscan as a tenth order Legendre polynomial series. We then average these coefficients across all CESs to build a set of polarization templates for the ground synchronous structure. The TOD is then fit to this template for each subscan and averaged across detectors to fit a ground amplitude as a function of subscan number within an CES. The slope of this amplitude is then computed for each CES. We do not see any correlation of the ground amplitude or the slope of the amplitude with local solar or sidereal time and place an upper limit of 1\% temperature drift correlated between CESs. We simulate a ground synchronous signal amplitude increase of 1\% in each CES to place an approximate upper limit on the $BB$ power created by this model.

We also expect that the ground fixed signal amplitude drift within a CES will have a component uncorrelated with solar or sidereal time. We simulate this by assuming that all of the variance in the ground amplitude slope is due to physical temperature drift of the apparent ground fixed signal. This model appears primarily as low-$\ell$ noise in the map domain. We quantify the estimated residual after suppression by the cross spectrum estimator. It should be noted that this represents the maximum possible contamination for this model that is consistent with the data and receives some contribution from the TOD noise. The majority of the possible systematic contamination due to ground fixed signal comes from this mode. The physically motivated gain modulation predicts a significantly smaller level of contamination.

We also validate that the null test statistics are sensitive to these models of drift in the observed ground structure amplitude. Statistically significant contamination due to imperfect ground template subtraction results in null test failures confined to the lowest $\ell$ bin of the CES type (rising, middle, setting) and half focal plane (left versus right and top versus bottom) splits. While less sensitive than the template approach, the null tests are significantly more model independent in that the contamination need not follow the linear drift or $4f$ gain models. Since no such failure is observed we can be confident that this systematic error is indeed subdominant to our statistical errors.

\subsubsection{HWP Signal Aliasing} \label{aliasing}

An additional source of variance in the polarization data is produced by the imperfect separation of temperature and polarization in time domain frequency. While small, the (beam-convolved) temperature signal on the sky will have power that aliases into the polarization band centered at the HWP $4f$. Additionally, there is non-zero leakage of the temperature signal into the sidebands of the HWP $2f$ amplitude. We simulate these effects by scanning a temperature-only beam convolved sky and injecting the aliased signal at the 0.6\% level via the $2f$ and directly via the $0f$ into the polarization TOD. We find the contamination in both the $E$ and $B$-mode channels to be negligible. We do not simulate the impact of the TOD filtering on this systematic and simply bin the pseudo spectrum of the aliased maps.

\subsubsection{HWP Imperfections} \label{hwpspot}

We observe a small air bubble in the anti-reflection coating on our warm HWP. We find the HWP synchronous structure associated with this spot to be stable with time. We simulate the aliased power from the total non-$4f$ harmonics coupled with non-linearity, time constant drift, and gain error and find the excess variance added to the data to be negligible.

\subsubsection{Temperature to Polarization Leakage} \label{leakage_beams}

The temperature to polarization leakage produced by the off-axis telescope design has
both scalar and higher order terms. Following \citet{2016RScI...87i4503E} we break the leakage into scalar, dipole, and quadrupole terms. We measure the higher order modes using the scans of Jupiter performed for the beam calibration. We find the contamination from these higher order modes in the polarization power spectra to be negligible. More details on this analysis can be found in \citet{SatoruPhD}.

\subsubsection{Cross Polarization from MD breaking} \label{crosspol}

The systematic contamination due to cross polarization from the breaking of the MD condition by the HWP located at the focus of the primary is expected to be negligible. \citet{mdbreakingspie} simulated the impact of this cross polarization for the \pb-2 receiver assuming the observation strategy used in this analysis. Due to the focal plane dependence of the cross polarization and the smaller field of view of the \pb\ instrument the contamination will be smaller than the expectation for \pb-2.

\section{Power Spectra and Parameter Constraints} \label{results}

In this section we present our power spectrum results, an estimate 
of the foreground contamination in our spectra and an upper limit on the tensor-to-scalar ratio $r$.

\subsection{\pb\ $BB$ power spectrum}\label{autospectrum}

\begin{figure*}
\begin{center}
\includegraphics[scale=0.65]{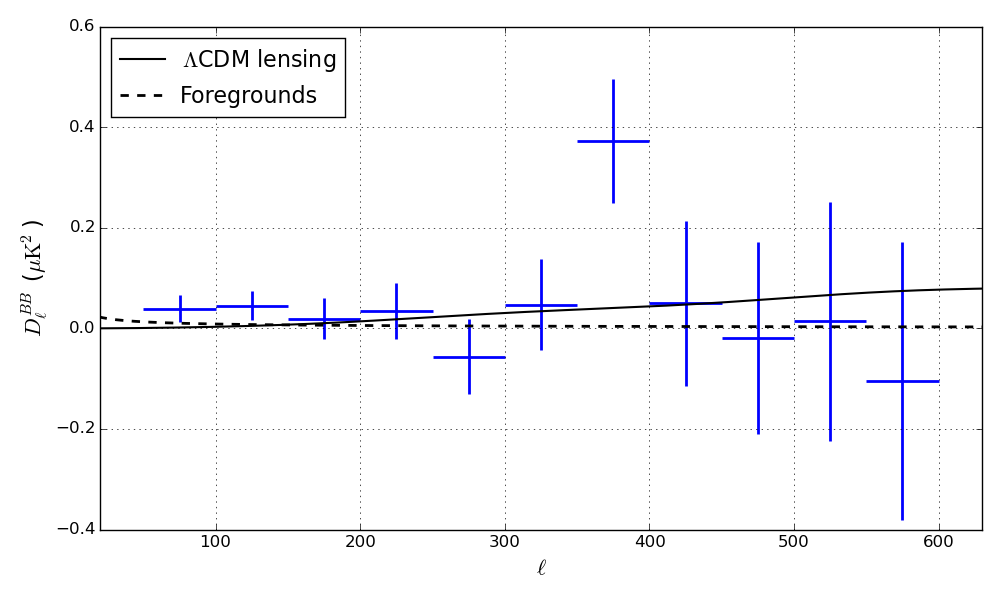}
\end{center}
\caption{Measured $B$-mode spectrum using the fiducial cross spectrum pipeline. 
The error bars shown reflect only the statistical uncertainties. The numerical bandpowers 
including systematic and calibration errors can be found in Table~\ref{spectrumtable}. The 
dashed foreground curve represents the best fit dust power at 150 GHz described in 
Section~\ref{results_components}.}
\label{bb_fiducial}
\end{figure*}

Our estimated $B$-mode power spectrum is shown in Figure~\ref{bb_fiducial}. We observe a modest excess above the \lcdm\ lensing expectation in our lowest two $\ell$ bins. 

The band power uncertainties from all terms (statistical, systematic, and calibration) are shown in Table~\ref{spectrumtable}. We find the statistical error to be dominant in all $\ell$ bins.

We compute an estimate of the overall amplitude of our observed $B$-mode signal relative to previous measurements. For the purposes of this calculation we assume that the underlying sky consists of a lensing CMB component corresponding to the \planckeight\ \lcdm\ lensing $B$-mode spectrum and a foreground component modeled by a power law $D_{\ell,\mathrm{dust}} = 9 \times 10^{-3} ~ (\ell / 80)^{-0.6}~\mathrm{\mu K}^2$ taken from the BK15 spectral decomposition at 150~GHz. We find a reduced $\chi^2$ of $11.6/11$ compared to this model indicating good agreement. Naively fitting for an overall $B$-mode amplitude rescaling this template we find $A_{BB} = 1.8 \pm 0.8$ disfavoring the null $BB$ hypothesis at $2.2 \sigma$. This estimate neglects the slightly non-Gaussian shape of the band power distributions, however that is accounted for in our cosmological parameter constraints shown in subsequent sections.

We additionally compute a naive uncertainty on $r$ from our \lcdm\ lensing only ($r=0$) simulations. We find $\sigma(r) = 0.34$ neglecting the non-Gaussian shape of the likelihood. 

\begin{table*}
\begin{center}
\caption{Band powers and uncertainties}  \label{spectrumtable}
\begin{tabular}{ c  c c c c }
\hline
~Nominal band definition~ & ~$D_\ell^{BB}$ $(\mathrm{\mu K}^2)$~ & ~$D_\ell^{BB}$ stat error $(\mathrm{\mu K}^2)$~ & ~$D_\ell^{BB}$ syst error $(\mathrm{\mu K}^2)$~ & ~$D_\ell^{BB}$ cal error $(\mu K^2)$~ \\
\hline
$50 < \ell \leq 100$ & 0.0390   & 0.0268 & 0.0040 & 0.0027 \\
$100 < \ell \leq 150$ & 0.0449   & 0.0288 & 0.0010 & 0.0023 \\
$150 < \ell \leq 200$ & 0.0194   & 0.0415 & 0.0012 & 0.0009 \\
$200 < \ell \leq 250$ & 0.0345   & 0.0559 & 0.0007 & 0.0014 \\
$250 < \ell \leq 300$ & -0.0566  & 0.0747 & 0.0015 & 0.0019 \\
$300 < \ell \leq 350$ & 0.0471   & 0.0910 & 0.0022 & 0.0019 \\
$350 < \ell \leq 400$ & 0.3731   & 0.1236 & 0.0022 & 0.0168 \\
$400 < \ell \leq 450$ & 0.0503   & 0.1641 & 0.0021 & 0.0031 \\
$450 < \ell \leq 500$ & -0.0189  & 0.1907 & 0.0029 & 0.0017 \\
$500 < \ell \leq 550$ & 0.0143   & 0.2377 & 0.0019 & 0.0015 \\
$550 < \ell \leq 600$ & -0.1037  & 0.2765 & 0.0049 & 0.0126 \\
\hline
\end{tabular}  \par
\end{center}
\vspace{8pt}
\noindent Measured bandpowers as well as the statistical, 
combined systematic, and calibration uncertainty error estimates. The calibration and systematic
error estimates are described in Section~\ref{abscal} and Section~\ref{systematicspipeline}, respectively.
\end{table*}
 
\begin{figure*}
\begin{center}	
\includegraphics[scale=0.44]{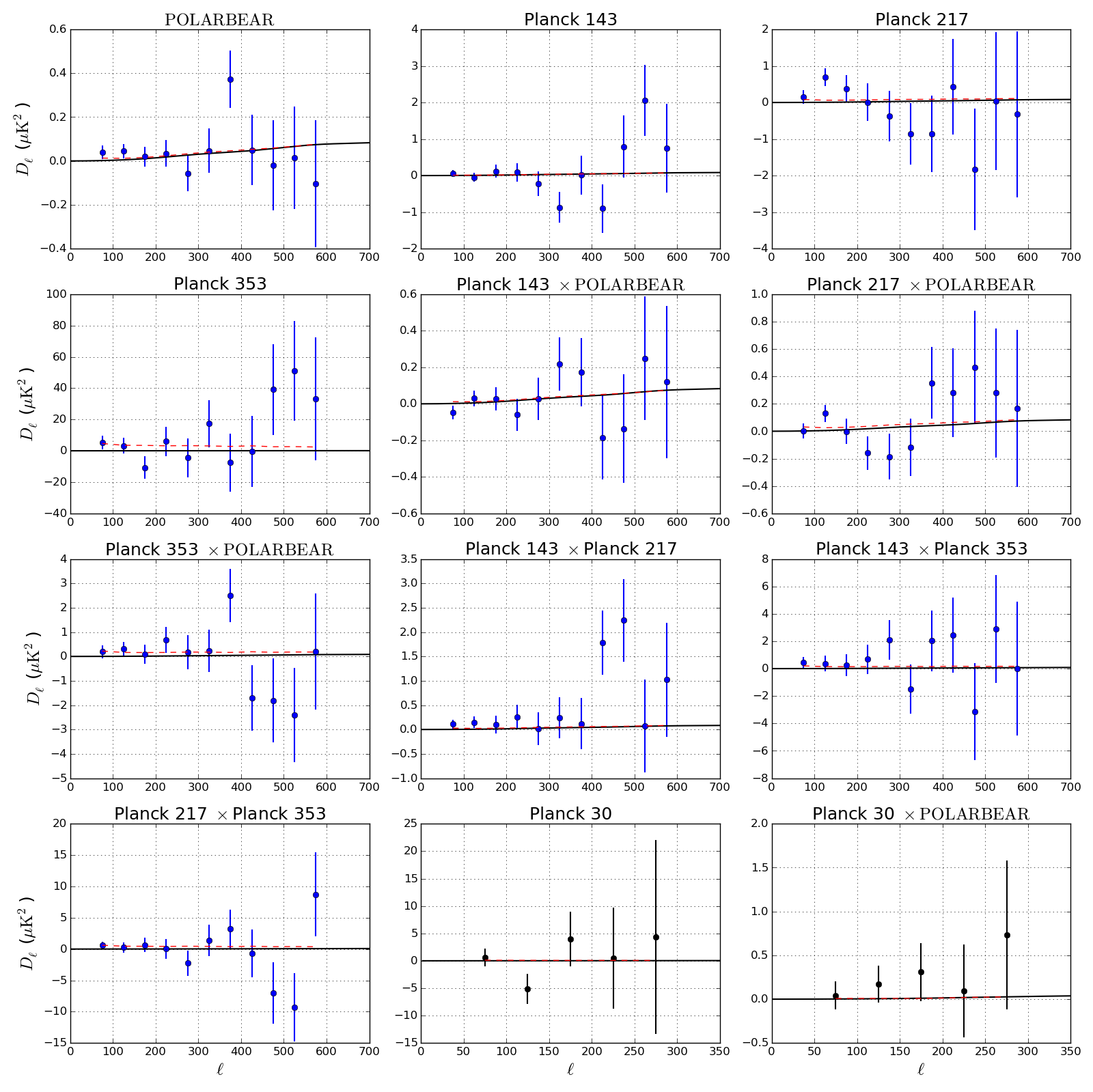}
\end{center}
\caption{All cross and auto spectra measured in comparison with \planck\ maps. Spectra including the 
\planck\ 30 GHz maps are indicated with black points and are not used the fiducial $r$ likelihood. Error 
bars show the fixed $\nu_b$ approximation. The red dashed curves indicate the best-fit CMB + foreground 
model. All spectra are shown in CMB temperature units. The black curve indicates the \lcdm\ lensing expectation.}
\label{all_crosses}
\end{figure*}

\subsection{Cross correlation with \planck~ HFI maps}

We compute cross spectra with three \planckeight~HFI maps to quantify the contribution of galactic dust to our $B$-mode auto spectrum. We form the ten unique auto and cross spectra between four maps,

\begin{itemize}
\item \pb\ 150 GHz map
\vspace{-3pt}
\item \planck~143 GHz frequency map 
\vspace{-3pt}
\item \planck~217 GHz frequency map
\vspace{-3pt}
\item \planck~353 GHz frequency map.
\end{itemize}

We process the \planck~PR3 full mission frequency maps through the \pb~observing pipeline to create \planck~maps as seen by \pb. Additionally, we process 96 noise realizations from the \planck~FFP10 noise simulations for each frequency to establish the \planck~noise bias. The \pb~noise bias is estimated using 192 \enquote{signflip} noise realizations. 

The spectra involving the \planck\ maps are formed using the alternate full auto spectrum pipeline described in Appendix~\ref{alternateps} where we simply compute the power spectrum of the fully coadded map rather than the cross spectrum of the ten day data subsets. We compute the measured debiased power spectrum $\hat{C}_b$ for display purposes by subtracting the mean auto spectrum of the FFP10 noise simulations from the auto spectrum of the real map. This is done to avoid relying on same-frequency cross spectra in the \planckeight~PR3 maps which are known to be contaminated by systematics at the lowest $\ell$ values \citep{Aghanim:2018fcm}.

The \pb\ auto spectrum and noise bias is computed using the fiducial cross spectrum pipeline. Using the auto spectrum of the fully coadded map in place of the internal cross spectrum formalism does not significantly shift the final parameter estimate, however we use the fiducial power spectrum pipeline for consistency with previous results and robustness to noise bias misestimation.

The six cross spectra between frequency bands are formed directly from the cross spectra of the fully coadded maps assuming that systematics and noise are uncorrelated between frequencies and experiments. 

We use a quasi-analytic approach to estimate uncertainty in the cross and auto spectra motivated by the computational cost of running the number of simulations necessary to directly estimate the full covariance matrix. We assume that every spectrum follows a reduced $\chi$-squared distribution with the same effective number of degrees of freedom per band power denoted $\nu_b$ determined by the patch geometry and TOD filtering. We estimate the number of degrees of freedom from the fractional uncertainty in the auto spectrum of the scanned noise realizations, 

\begin{equation}
\nu_b = 2 \bigg{(} \frac{\hat{C}_b}{\sigma ( \hat{C}_b ) }\bigg{)}^2 .
\label{nu_b_eq}	
\end{equation}

For the auto spectra we estimate the uncertainty of the measured power spectrum $\hat{C}_b$ using the best fit signal power spectrum $C_b$ and mean noise bias $N_b$,

\begin{equation}
\Delta \hat{C}_b = \sqrt{\frac{2}{\nu_b}}\, (C_b + N_b).
\label{errorbar_auto}
\end{equation}

Equivalently, the uncertainty in the measured cross spectrum can be written in the form,

\begin{equation}
\Delta \hat{C}_b^{AB} =  \sqrt{\frac{1}{\nu_b}\bigg{(}(C_{b,A} + N_{b,A})(C_{b,B} + N_{b,B}) + C_{b,AB}^{~2} \bigg{)}}.
\label{errorbar_cross}
\end{equation}

It is important to note that these uncertainties are plotted for visualization purposes and do not directly enter the likelihood model. We perform an end-to-end validation of the pipeline by simulating all cross spectra using 12 CMB realizations for each value of $r \in \{0.0, 0.1, 0.2\}$ and a single PySM dust realization \citep{2017MNRAS.469.2821T} scaled to match the measured dust emission reported in BK15. 

The fixed $\nu_b$ model is an approximation to the real data uncertainties because signal and noise will in general have different $\nu_b$ values. This is due to the two significant effects. The approximately isotropic \planck\ noise in our patch has systematically fewer degrees of freedom per band power than the anisotropic \pb\ noise. Additionally, the $E \rightarrow B$ leakage subtraction in Equation \ref{ebleak} suppresses the effective number of degrees of freedom at low-$\ell$. We combine the \pb\ and \planck\ derived $\nu_b$ by taking the geometric mean. In Section \ref{components} we discuss this choice and define a goodness of fit metric for our likelihood model that is sensitive to systematic misestimation of the auto and cross spectrum uncertainties between the simulations and the real data. \label{cross_spectrum}

\subsection{Cross correlation with \planck~ LFI maps}

We compute an upper limit on polarized $B$-mode Galactic synchrotron emission in our field at 150 GHz.

The power spectrum estimation formalism follows the formalism used to compute the cross spectra with high frequency \planck\ data. The auto spectrum of the \planck\ 30 GHz full mission map \citep{2018arXiv180706206P} as seen by \pb\ is computed and the noise bias is subtracted using the \planck\ FFP10 noise realizations. We validate our pipeline using a set of CMB simulations with $r=0.0$ added to a PySM map of Galactic synchrotron emission as a fiducial signal model for the 30 GHz channel. Our simulations assume that dust foregrounds are negligible at 30 GHz and that synchrotron foregrounds are negligible at 150 GHz. We use the same fixed $\nu_b$ approximation as the high frequency case using the signal cross spectrum $C_{b,AB}$ computed from the PySM models. Due to the \planck\ beam size at 30 GHz we only consider the first five band powers corresponding to $\ell \leq 300$. 

We model the synchrotron emission as a power law in $\ell$ and frequency. Specifically, we assume that the power spectrum takes the form

\begin{equation}
D_{\ell, \mathrm{sync}} = A_\mathrm{sync}\bigg{(}\frac{\ell}{\ell_0} \bigg{)}^{\alpha_\mathrm{sync}}
\bigg{(}\frac{\nu}{\nu_0}\bigg{)}^{2\beta_\mathrm{sync}},
\end{equation}

\noindent in brightness units where the 2 is due to the fact that $D_{\ell}$ is quadratic in signal amplitude. We assumed fixed values for the power law indices $\alpha_\mathrm{sync} = -1.18$ taken from the highest Galactic latitudes in a recent measurement by S-PASS in conjunction with \wmap\ and \planck\ data \citep{2018A&A...618A.166K}. We assume a value of $\beta_\mathrm{sync} = -3.2$ from the same analysis which is consistent with the prior used in BK15.  Following previous work we choose pivots $\ell_0 = 80$ and $\nu_0 = 23$ GHz. We construct a one parameter likelihood following \citet{2008PhRvD..77j3013H}, herein HL08, relating the amplitude $A_\mathrm{sync}$ to the auto spectrum of the \planck\ 30 GHz map assuming a fixed \lcdm\ CMB lensing component with $r=0$. We integrate over the Planck 30 GHz average bandpass from the LFI reduced instrument model  available from the \planck\ Legacy Archive and the \pb\ design bandpass.

We find the likelihood of $A_\mathrm{sync}$ peaks at zero with a 95\% upper limit of  $A_\mathrm{sync,23~GHz} \leq 12.5\,\mathrm{\mu K}^2$. Using the fixed prior on $\alpha_\mathrm{sync}$ and $\beta_\mathrm{sync}$,
this amplitude corresponds to a 95\% upper limit on the synchrotron contamination at our frequency band and $\ell=80$ of $2.7 \times 10^{-4}\,\mathrm{\mu K}^2$. We do not include synchrotron contamination in our fiducial $r$ likelihood model as it is deeply subdominant to dust foregrounds at 150 GHz.

We find the cross spectrum between \planck~30 GHz and \pb\ to be consistent with null. We do not directly use this spectrum in our estimate of synchrotron contamination as this adds no meaningful constraining power on the synchrotron spectral index.

All twelve spectra (ten with \planck~HFI and two with \planck~LFI) from the real data are 
shown in Figure~\ref{all_crosses}. The HFI spectra are used as the primary input to our 
foreground and cosmological parameter constraints described in Section \ref{components}.

\subsection{Contamination from polarized point sources} \label{pointsource}

Polarized point sources contribute to the level of $B$-mode power in our maps as a Poisson noise term. At frequencies $\lesssim 150$ GHz
the source number counts are expected to be dominated by blazars and flat spectrum quasars. \actpol\ has reported 26 detections of radio sources during the first two seasons on a comparable fraction of the sky at the same frequency \citep{2019MNRAS.486.5239D}. Fourteen of these sources showed linear polarization at 
more than $3 \sigma$ significance. The mean polarization fraction of these sources is $0.028 \pm 0.005$, in agreement with previous studies in \citet{2018ApJ...858...85P} and \citet{2017MNRAS.469.2401B}. 

We search our maps for statistically significant polarized point sources. We first apply a matched spatial filter to a high resolution version of our maps similar to \citet{Vieira2010} and \citet{Marriage2011}. We detect 19 point sources in intensity above $5\sigma$, however do not detect any sources in polarization at the same significance. This null result is consistent with forecasts at our frequency \citep{2011A&A...533A..57T}. We therefore consider the emission from polarized sources below our detection flux.

We estimate the level of $B$-mode power in our maps from undetected radio sources using the \textsc{PS4C} forecasting tools presented in \citet{2018ApJ...858...85P}. We convert our sensitivity into a $5\sigma$ equivalent detection flux of 60~mJy in polarization. The upper limit on the contamination at $\ell=80$ is estimated to be  $D_{\ell} = 1.6 \times 10^{-4} \mathrm{\mu K}^2$. This is approximately equivalent to $\Delta r \lesssim 0.002$. We therefore neglect the point source contamination in our parameter constraints.

\begin{figure*}
\begin{center}
\includegraphics[width=0.49\textwidth]{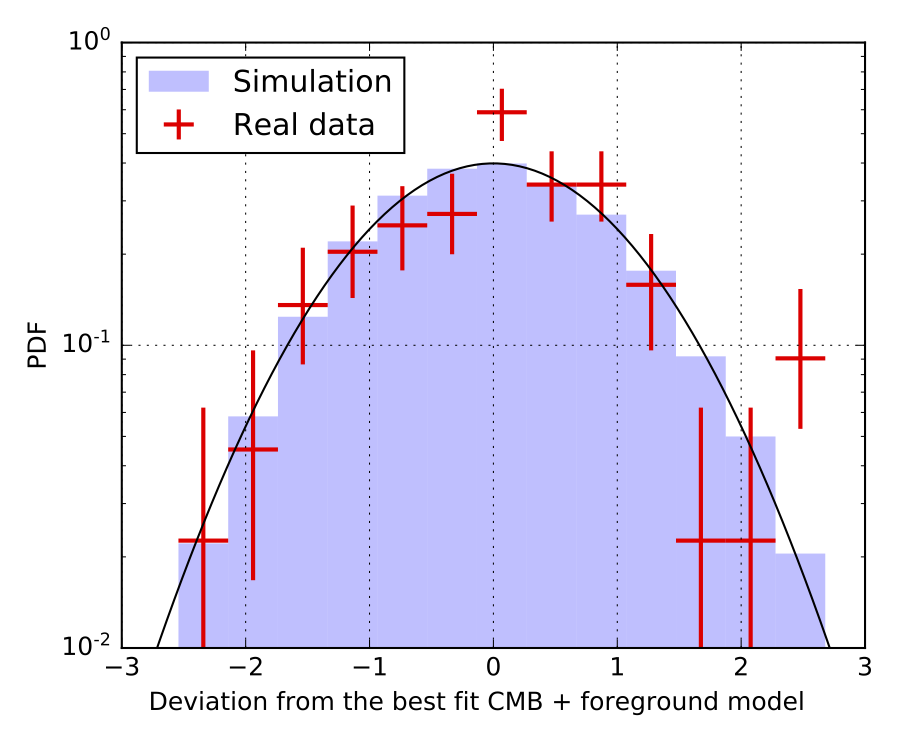}
\includegraphics[width=0.49\textwidth]{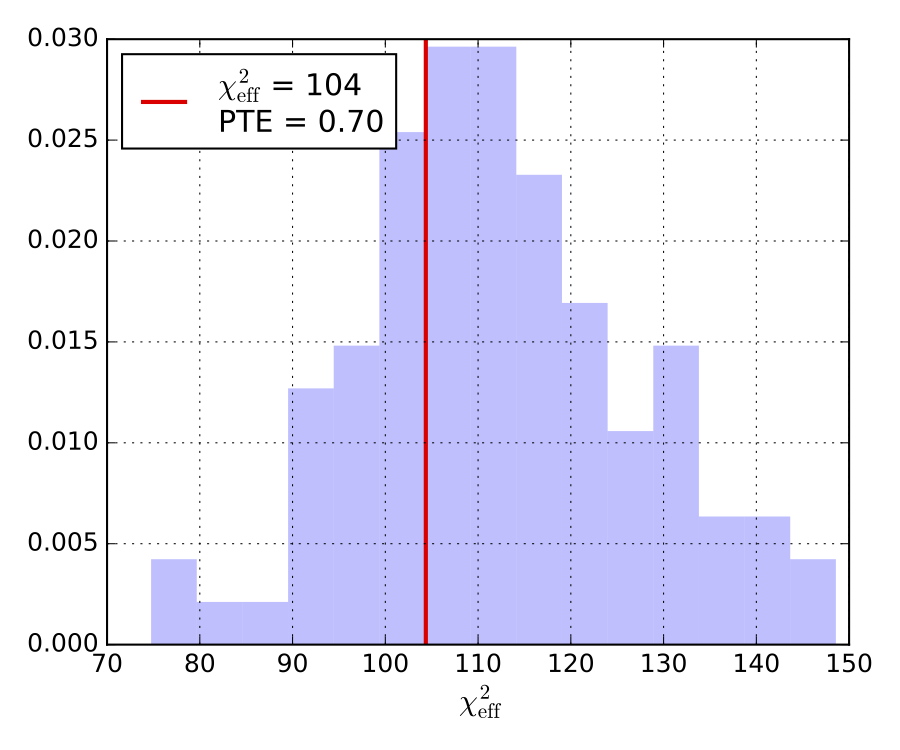}
\end{center}
\caption{Normalized difference between the measured cross spectra
and the best fit CMB+foreground model shown in units of standard deviation (left).  
Effective $\chi_\mathrm{eff}^2$ of the data fit to the model as defined in 
Equation \ref{effective_chisq_equation} compared to simulations (right). 
The distribution shows the 96 simulations and the vertical line indicates the real data.}
\label{error_crosses}
\end{figure*}

\begin{figure*}
\begin{center}	
\includegraphics[scale=0.75]{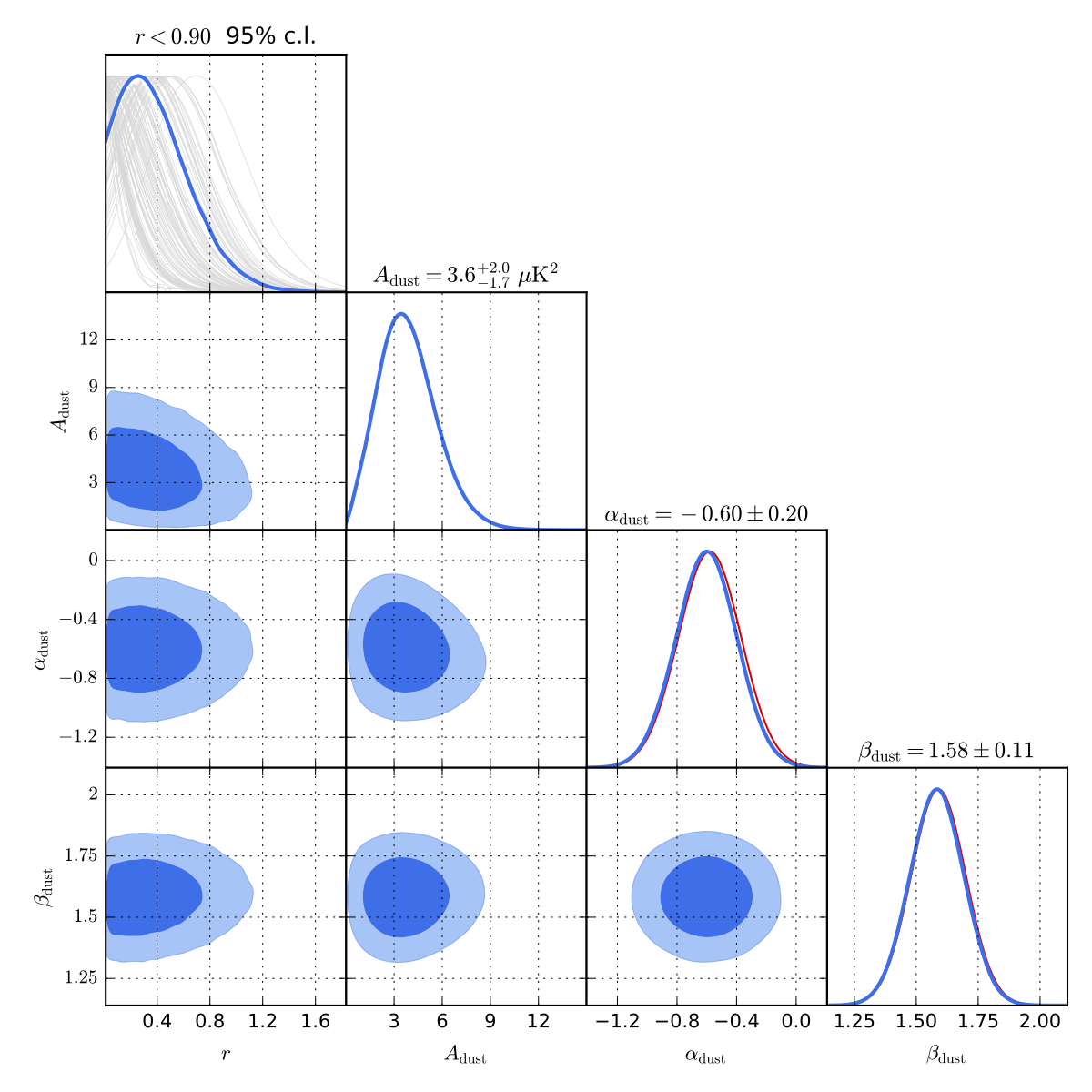}
\end{center}
\caption{Two- and one-dimensional marginal posteriors for the four free parameters
 $r$, $A_\mathrm{dust}$, $\alpha_\mathrm{dust}$, and $\beta_\mathrm{dust}$.
We compare the posteriors to simulations (light gray) for $r$ and with the 
priors (red lines) for the final two parameters. The contours indicate 68\% and 95\% of the probability weight.}
\label{triangle_plot}
\end{figure*}

\subsection{Parameter constraints} \label{components}

We fit the $B$-mode power spectrum from our data and cross correlation with \planck\ high frequency data to a CMB
and single dust component as done by \citet{bkp}, using a likelihood model
similar to HL08 and \citet{2008arXiv0803.1814C}. For each
bandpower, we write a $n_\mathrm{freq} \times n_\mathrm{freq}$ matrix of the 
measured cross spectra $\mathbf{\hat{C}}_b$, where $n_\mathrm{freq}=4$ is the 
number of frequency channels considered.
The diagonal elements of this matrix contain the noise bias.  For the Planck channels,
this is simply a consequence of computing the auto spectrum of the full mission
frequency maps. For \pb, this is obtained by adding the noise bias to the fiducial $B$-mode power
spectrum described in Section~\ref{powerspectrum}. This makes the estimation of $r$ more robust
to misestimations of the \pb\ noise bias than simply using the auto spectrum of 
the full coadd. We simulate artificially changing the \pb\ noise bias by 10\% and 
find a small shift in the uncertainty but no significant shift in the reconstructed $r$ value.
 
In the case that the underlying fields
are isotropic, Gaussian and measured on the full sky, $\mathbf{\hat{C}}_b$ follows a Wishart distribution with
$\sum_{\ell} \mathbf{P}_{b\ell}(2\ell + 1)$ degrees of freedom per band power where $\mathbf{P}_{b\ell}$
is the binning operator defined in Section~\ref{powerspectrum}. We assume 
that our $\mathbf{\hat{C}}_b$ follows the same distribution but with an effective number of degrees of
freedom $\nu_b$ estimated using Equation~\ref{nu_b_eq} for our partial sky area. This is an
approximation because the effective number of degrees of freedom 
for the \planck\ and the \pb\ maps differ somewhat due to the anisotropy of the \pb\ noise. 
The choice between these sets of $\nu_b$ values has little influence on the results; in both cases the simulations
show that there is no bias on the estimation of $r$ and that the width of the marginal posterior on 
$r$ is compatible with the dispersion of the best-fit values. We choose to use the 
geometric mean of the two $\nu_b$ estimates for the analysis that we
describe in the rest of this section.

Under these assumptions, the likelihood $\mathcal{L}$ of 
a true spectrum $\mathbf{C}_b$ given measured $\mathbf{\hat{C}}_b$ is given by

\begin{equation}
-2 \ln \mathcal{L} = \sum_{b} \nu_{b} \bigg{\{} \operatorname{Tr}
[\mathbf{\hat{C}}_{b} \mathbf{C}_{b}^{-1}]- \ln | \mathbf{\hat{C}}_{b}
\mathbf{C}_{b}^{-1} |  -n_\mathrm{freq} \bigg{\}},
\end{equation}

\noindent up to a constant offset. We model the underlying $\mathbf{C}_{b}$ as a sum of CMB, dust, 
and noise components,

\begin{equation}
\mathbf{C}_{b} = \mathbf{C}_{b}^{\mathrm{CMB}}+\mathbf{C}_{b}^{\mathrm{dust}}+\mathbf{N}_{b}.
\end{equation}

\noindent Every element of the CMB cross spectrum matrix $\mathbf{C}_{b}^{\mathrm{CMB}}$ is equal 
within a bandpower because all spectra are computed in CMB temperature units. 
The CMB receives contributions from lensing and tensor $B$-mode signals,

\begin{equation}
\mathbf{C}_{b}^{\mathrm{CMB}}=r \mathbf{C}_{b}^{\mathrm{tens}}+A_{\mathrm{lens}} \mathbf{C}_{b}^{\mathrm{lens}}.
\end{equation}

\noindent Here $r$ is the tensor-to-scalar ratio, $A_{\mathrm{lens}}$ is the normalized amplitude of the \lcdm\ lensing signal,
$\mathbf{C}_{b}^{\mathrm{tens}}$ is the binned tensor $B$-mode signal, and $\mathbf{C}_{b}^{\mathrm{lens}}$ is the binned lensing signal. 
The dust component is treated as  a power
law in $\ell$ and a modified black body in frequency following 
\citet{2016A&A...586A.133P, 2018arXiv180104945P} and subsequent work. We define a vector 
$f(\beta_\mathrm{dust}, T_\mathrm{dust})$ that represents the dust emission at for each frequency bandpass
converted into CMB temperature units, where $\beta_\mathrm{dust}$ and
$T_\mathrm{dust}$ are the spectral index and temperature of the modified black
body, respectively. The dust component of $\mathbf{C}_{b}$ for frequencies $i,j$ can be written as

\begin{equation}
\mathbf{C}_{b,ij}^{\mathrm{dust}} = A_{\mathrm{dust}}(f_i f_j)
\bigg{(}\frac{\ell}{\ell_0}\bigg{)}^{\,\alpha_\mathrm{dust}},
\end{equation}

\noindent where $A_{\mathrm{dust}}$ is the amplitude of the dust signal, and
$\alpha_\mathrm{dust}$ is the power law index in $\ell$. We assume a pivot value
of $\ell_0 = 80$ and normalize $f$ such that $A_{\mathrm{dust}}$ represents the dust 
emission at 353~GHz. It should be noted that dust foregrounds are not expected to
be Gaussian or to follow a power law in $\ell$. The physics of interstellar dust 
may result in a polarized frequency scaling significantly different from a single modified
black body, a more complex $\ell$ dependence, or spatial decorrelation 
between high frequencies and CMB channels. However, we do not have the sensitivity 
to meaningfully constrain more complex foreground models with this dataset and therefore only 
consider the fiducial case.

For each frequency channel we integrate over the instrument bandpass. The \pb\ channel uses the design bandpasses
from \citet{Arnold_SPIE2012} and the Planck channels use the HFI reduced instrument model.

The noise component $\mathbf{N}_{b}$ is entirely diagonal as the noise between 
frequency bands and experiments is expected to be uncorrelated. The value of the 
diagonal elements is taken to be the power spectrum of the \enquote{signflip} (FFP10) noise realizations 
for the \pb\ (\planck) frequency channels using the cross spectrum (auto spectrum) pipelines.

The model contains six parameters, $r$, $A_\mathrm{lens}$,
$\alpha_\mathrm{dust}$, $\beta_\mathrm{dust}$, $T_\mathrm{dust}$, and
$A_\mathrm{dust}$. We fix the values of $A_\mathrm{lens}=1$,
$T_\mathrm{dust}=19.6 \mathrm{K}$ and allow the other four to float. We apply Gaussian
priors on $\alpha_\mathrm{dust} = -0.58 \pm 0.21$ and $\beta_\mathrm{dust} =
1.59 \pm 0.11$ respectively based on the marginal posterior from BK15 and the
estimate of patch-to-patch variation found by \citet{2016A&A...586A.133P}. In our fiducial 
likelihood we impose the prior $r \geq 0$. 

We define a goodness of fit criterion following HL08

\begin{equation}
\chi_\mathrm{eff}^{2} = -2 \ln \mathcal{L},
\label{effective_chisq_equation}
\end{equation}

\noindent and analogously find that,
in the limit that $\nu_b \gg n_\mathrm{freq}$ and that the number of fit parameters is 
negligible compared to the total number of bins across all spectra,
$\chi_\mathrm{eff}^{2}$ has expectation value and
variance

\begin{equation}
\langle \chi_\mathrm{eff}^{2} \rangle \simeq n_\mathrm{bins} \frac{n_\mathrm{freq}
(n_\mathrm{freq} + 1)}{2}
\end{equation}

\begin{equation}
\operatorname{var} ( \chi_\mathrm{eff}^{2} ) \simeq 2 \langle \chi_\mathrm{eff}^{2}\rangle,
\end{equation}

\noindent where $n_\mathrm{bins}$ is the number of $\ell$ bins in each spectrum.\footnote{If we relax the approximation that $\nu_b \gg n_\mathrm{freq}$,
the values of $\langle \chi_\mathrm{eff}^{2}
\rangle$ and $\operatorname{var} ( \chi_\mathrm{eff}^{2} )$ differ from this limit 
by a few percent. Correcting for the number of fit parameters is
nontrivial because of the priors we impose. For these reasons we compare the
$\chi_\mathrm{eff}^{2}$ we obtain from the data with the distribution we obtain
from the simulations, and not an analytical expectation.}
This is consistent with a $\chi^2$ distribution with a number of degrees of freedom equal to the total 
number of bins across all unique spectra in $\mathbf{\hat{C}}$, in our case 110.
Figure \ref{error_crosses} shows that the simulated $\chi_\mathrm{eff}^{2}$ values
are consistent with expectations. The value of $\chi_\mathrm{eff}^{2}$ from the real data 
lies in the middle of the simulations with PTE = 70\%.

Our parameter constraints are shown in Figure~\ref{triangle_plot}. Our posteriors 
on $\alpha_\mathrm{dust}$ and $\beta_\mathrm{dust}$ are dominated by the
input prior meaning the data have little additional constraining power on these parameters. 
The prior on $\alpha_\mathrm{dust}$ is not critical for the estimation of $r$ 
since the dust power in the bins and spectra with the most constraining power on $r$ is
largely set by $A_\mathrm{dust}$ and $\beta_\mathrm{dust}$. We include this prior
because it results in a more efficient exploration of the parameter space by 
the Markov Chain Monte Carlo (MCMC). Using a prior on $\beta_\mathrm{dust}$ is more important
for our $r$ constraint since it is necessary for the MCMC chains to converge reliably. 
This prior improves our upper limit on $r$ by $\sim 30 \%$ in simulations. 
We find evidence for dust $B$-modes in our patch with a best fit value for $A_\mathrm{dust}$
of $3.6\,\mathrm{\mu K^2}$. This is marginally lower than the value reported in BK15. 
The expected difference between the two results is non-trivial to compute given the partial overlap
in data sets and different observation strategies. We exclude zero dust foregrounds 
with 99\% confidence. We find a 95\% confidence 
level upper limit on the primordial tensor-to-scalar ratio of \rupperlimit\ after 
marginalizing over foreground parameters.

We estimate the impact of our instrumental systematic and calibration errors on
the final $r$ constraint as follows. We add the upper bounds on systematic
contamination reported in Table \ref{spectrumtable} to a reference theoretical
power spectrum containing CMB ($r=0$), dust and noise. We analyze these
spectra following the real data and find $\Delta r_\mathrm{syst} = 0.02$.
We additionally generate random multiplicative calibration errors
according to the levels reported in Table \ref{spectrumtable} and run the component
separation. We find that the bias on the best-fit value
of $r$ is smaller than $\Delta r_\mathrm{cal} \leq 0.01$ and the effect on the estimated
uncertainty is less than 10\%. We therefore neglect multiplicative calibration effects
and only consider additive systematic errors. We do not attempt to quantify the impact of \planck\
systematics in our results.

Finally, for a sensitivity comparison to other experiments, we remove the prior on $r$ and provide a best-fit estimate as 
the maximum of the marginal posterior. The statistical uncertainty is estimated as the most narrow interval 
that contains 68\% of the integral of the distribution. We find $r = 0.26\ ^{+0.32}_{-0.30}\ ({\rm stat.})\ \pm 0.02\ ({\rm syst.})$
where the statistical error refers to the narrowest interval containing 68\% of the probability weight.
 \label{results_components}

\section{Conclusions}

We present a measurement of the CMB $B$-mode power spectrum from the multipole range $50 \leq \ell \leq 600$ using three seasons of \pb\ 
data taken with a continuously rotating HWP. We observed a $670$ effective square degree patch located near the southern celestial pole that 
significantly overlaps with observations by South Pole experiments including \bicep\ and \keckarray. Our data achieves an effective map depth of \mapdepth.

The use of a continuously rotating HWP for polarization modulation provides a powerful mitigation of low-frequency noise.
We demonstrate control of low-frequency noise in the data without significant sensitivity degradation above $\ell = 90$. 
We show that our data are consistent with a simple TOD model consisting of a single source of low frequency noise in the time domain.

We establish that the data are cleaned of systematic errors through a suite of jackknife null tests and direct simulation of known effects. 
We find that all expected sources of systematic contamination are below the statistical uncertainties.

We disfavor zero $C_\ell^{BB}$ at $2.2\sigma$ using \pb\ data alone. We observe a modest excess above the \lcdm\ lensing expectation in our lowest $\ell$ bins that is consistent with published foreground levels.

We further compute the cross spectrum of our data with the publicly available \planckeight\ high frequency maps and show that
the low-$\ell$ $B$-mode signal is consistent with Galactic dust emission. We find that our data are consistent with a single dust foreground model. We 
place an upper limit on the cosmological tensor-to-scalar ratio \rupperlimit\ at 95\% confidence level considering only statistical errors.

This paper builds on the results of the \abs\ experiment \citep{2018JCAP...09..005K} and demonstrates another degree-scale $B$-mode measurement including the deepest CMB maps yet produced using continuous polarization modulation. The \pb\ experiment has demonstrated measurements from the degree scales shown in this analysis to the arcminute scales shown in PB14 and PB17.  Analyses using the angular resolution of \pb\ to probe the CMB at smaller angular scales using the same data set as this analysis in preparation. 
Future experiments including the Simons Array \citep{2016JLTP..184..805S, MasayaSPIE} and Simons Observatory \citep{2019JCAP...02..056A} 
will build on these results with substantially improved statistical power. \label{conclusions}

\appendix

\section{Alternate auto spectrum pipeline} \label{alternateps}

We define an alternate power spectrum pipeline for use in sections 
comparing our data to \planck\ data for parameter constraints. In this pipeline we use the 
auto spectrum of the fully coadded map in place of the cross spectra between map bundles. 
The power spectrum estimation follows the fiducial pipeline exactly with the substitution

\begin{equation}
\tilde{C}^{XY} =  \mathbf{m}^{ X }\mathbf{m}^{Y*},
\end{equation}

\noindent where $\mathbf{m}$ is the Fourier transform of the apodized fully coadded map. 
We recompute the filter transfer function $F_\ell$ for this pipeline and find numerical values 
that differ at the level of a few percent from the fiducial filter transfer function. 

As a consistency check we can estimate the \pb\ auto spectrum using this formalism 
and the \enquote{signflip} noise realizations to estimate the noise bias. We find compatible 
results between this approach and the fiducial power spectrum pipeline by comparing 
the numerical difference between the two power spectra for the real data to the difference 
between the two spectra in our MC simulation set. This comparison gives $\chi^2 / \nu = 8.5/11$ 
indicating good agreement. We find a marginally larger effective number of degrees of freedom 
per band power and therefore marginally smaller statistical errors compared to the fiducial power 
spectrum estimate. This is likely due to imperfect mode overlap between the 38 map bundles due 
to variation in the data selection and common mode TOD filter. We use the fiducial internal cross 
spectrum for the \pb\ auto spectrum for all parameter constraints but use the alternate auto spectrum for \planck\ 
data and cross spectra between the experiments.

\acknowledgements

The \pb\ project is funded by the National Science Foundation under grants AST-0618398 
and AST-1212230.  The analysis presented here was also supported by Moore Foundation 
grant number 4633, the Simons Foundation grant number 034079, and the Templeton Foundation 
grant number 58724.
The James Ax Observatory operates in the Parque Astron\'omico Atacama in Northern Chile under 
the auspices of the Comisi\'on Nacional de Investigaci\'on Cient\'ifica y Tecnol\'ogica de Chile (CONICYT).
AK acknowledges the support by JSPS Leading Initiative for Excellent Young Researchers (LEADER) and by the JSPS KAKENHI Grant Numbers JP16K21744 and 18H05539.
CB, NK, and DP acknowledge support from the ASI-COSMOS Network (cosmosnet.it) and from the INDARK INFN Initiative (web.infn.it/CSN4/IS/Linea5/InDark).
GF acknowledges the support of the European Research Council under the European Union's Seventh Framework Programme (FP/2007-2013) / ERC Grant Agreement No. [616170] and of the UK STFC grant ST/P000525/1. 
HN acknowledges JSPS KAKENHI grant JP26800125.
JC is supported by the European Research Council under the European Union's Seventh Framework Programme (FP/2007-2013) / ERC Grant Agreement No. [616170].
MA acknowledges support from CONICYT UC Berkeley-Chile Seed Grant (CLAS fund) Number 77047, Fondecyt project 1130777 and 1171811, DFI postgraduate scholarship program and DFI Postgraduate Competitive Fund for Support in the Attendance to Scientific Events.
MD acknowledges funding from the Natural Sciences and Engineering Research Council of Canada and Canadian Institute for Advanced Research.
MH acknowledges the support from the JSPS KAKENHI Grant Numbers JP26220709 and JP15H05891.
NK acknowledges the support from JSPS Core-to-Core Program, A. Advanced Research Networks.
OT acknowledges the SPIRITS grant in the Kyoto University, and JSPS KAKENHI JP26105519.
ST was supported by Grant-in-Aid for JSPS Research Fellow JP14J01662 and JP18J02133.
YC acknowledges the support from the JSPS KAKENHI Grant Number 18K13558, 18H04347, 19H00674.
The APC group acknowledges the travel support from the Labex Univearths grant.
The Melbourne group acknowledges support from the University of Melbourne and an Australian Research Council's Future Fellowship (FT150100074).
This work was supported by the World Premier International Research Center Initiative (WPI), MEXT, Japan.
This research used resources of the Central Computing System, owned and operated by the Computing Research Center at KEK. 
Support from the Ax Center for Experimental Cosmology at UC San Diego is gratefully acknowledged.
Work at LBNL is supported in part by the U.S. Department of Energy, Office of Science, Office of High Energy Physics, under contract No. DE-AC02-05CH11231.
This research used resources of the National Energy Research Scientific Computing Center, which is supported by the Office of Science of the U.S. Department of Energy under Contract No. DE-AC02-05CH11231.
We acknowledge many useful conversations with Nathan Whitehorn. 
We acknowledge the use of the \texttt{emcee} package \citep{emcee}.
Some of the results in this paper have been derived using the HEALPix \citep{Gorski2005} package.

\bibliographystyle{aasjournal}

\bibliography{largepatch}

\newcommand{\noopsort}[1]{}
\begin{thebibliography}{}
\expandafter\ifx\csname natexlab\endcsname\relax\def\natexlab#1{#1}\fi
\providecommand{\url}[1]{\href{#1}{#1}}
\providecommand{\dodoi}[1]{doi:~\href{http://doi.org/#1}{\nolinkurl{#1}}}
\providecommand{\doeprint}[1]{\href{http://ascl.net/#1}{\nolinkurl{http://ascl.net/#1}}}
\providecommand{\doarXiv}[1]{\href{https://arxiv.org/abs/#1}{\nolinkurl{https://arxiv.org/abs/#1}}}

\bibitem[{{Arnold} {et~al.}(2012){Arnold}, {Ade}, {Anthony}, {Barron},
  {Boettger}, {Borrill}, {Chapman}, {Chinone}, {Dobbs}, {Errard}, {Fabbian},
  {Flanigan}, {Fuller}, {Ghribi}, {Grainger}, {Halverson}, {Hasegawa},
  {Hattori}, {Hazumi}, {Holzapfel}, {Howard}, {Hyland}, {Jaffe}, {Keating},
  {Kermish}, {Kisner}, {Le Jeune}, {Lee}, {Linder}, {Lungu}, {Matsuda},
  {Matsumura}, {Miller}, {Meng}, {Morii}, {Moyerman}, {Myers}, {Nishino},
  {Paar}, {Quealy}, {Reichardt}, {Richards}, {Ross}, {Shimizu}, {Shimmin},
  {Shimon}, {Sholl}, {Siritanasak}, {Spieler}, {Stebor}, {Steinbach},
  {Stompor}, {Suzuki}, {Tomaru}, {Tucker}, \& {Zahn}}]{Arnold_SPIE2012}
{Arnold}, K., {Ade}, P.~A.~R., {Anthony}, A.~E., {et~al.} 2012, in Society of
  Photo-Optical Instrumentation Engineers (SPIE) Conference Series, Vol. 8452,
  Society of Photo-Optical Instrumentation Engineers (SPIE) Conference Series,
  \dodoi{10.1117/12.927057}

\bibitem[{{Aumont} {et~al.}(2010){Aumont}, {Conversi}, {Thum}, {Wiesemeyer},
  {Falgarone}, {Mac{\'{\i}}as-P{\'e}rez}, {Piacentini}, {Pointecouteau},
  {Ponthieu}, {Puget}, {Rosset}, {Tauber}, \& {Tristram}}]{2010A&A...514A..70A}
{Aumont}, J., {Conversi}, L., {Thum}, C., {et~al.} 2010, \aap, 514, A70,
  \dodoi{10.1051/0004-6361/200913834}

\bibitem[{{Bennett} {et~al.}(2013){Bennett}, {Larson}, {Weiland}, {Jarosik},
  {Hinshaw}, {Odegard}, {Smith}, {Hill}, {Gold}, {Halpern}, {Komatsu}, {Nolta},
  {Page}, {Spergel}, {Wollack}, {Dunkley}, {Kogut}, {Limon}, {Meyer}, {Tucker},
  \& {Wright}}]{2013ApJS..208...20B}
{Bennett}, C.~L., {Larson}, D., {Weiland}, J.~L., {et~al.} 2013, \apjs, 208,
  20, \dodoi{10.1088/0067-0049/208/2/20}

\bibitem[{{BICEP2 Collaboration and Keck Array Collaboration: {Ade}, P.~A.~R.}
  {et~al.}(2016){BICEP2 Collaboration and Keck Array Collaboration: {Ade},
  P.~A.~R.}, {Ahmed}, {Aikin}, {Alexander}, {Barkats}, {Benton}, {Bischoff},
  {Bock}, {Bowens-Rubin}, {Brevik}, {Buder}, {Bullock}, {Buza}, {Connors},
  {Crill}, {Duband}, {Dvorkin}, {Filippini}, {Fliescher}, {Grayson}, {Halpern},
  {Harrison}, {Hildebrandt}, {Hilton}, {Hui}, {Irwin}, {Kang}, {Karkare},
  {Karpel}, {Kaufman}, {Keating}, {Kefeli}, {Kernasovskiy}, {Kovac}, {Kuo},
  {Leitch}, {Lueker}, {Megerian}, {Namikawa}, {Netterfield}, {Nguyen},
  {O'Brient}, {Ogburn}, {Orlando}, {Pryke}, {Richter}, {Schwarz}, {Sheehy},
  {Staniszewski}, {Steinbach}, {Sudiwala}, {Teply}, {Thompson}, {Tolan},
  {Tucker}, {Turner}, {Vieregg}, {Weber}, {Wiebe}, {Willmert}, {Wong}, {Wu}, \&
  {Yoon}}]{2016ApJ...825...66B}
{BICEP2 Collaboration and Keck Array Collaboration: {Ade}, P.~A.~R.}, {Ahmed},
  Z., {Aikin}, R.~W., {et~al.} 2016, \apj, 825, 66,
  \dodoi{10.3847/0004-637X/825/1/66}

\bibitem[{{BICEP2 Collaboration and Keck Array Collaboration: {Ade}, P.~A.~R.}
  {et~al.}(2018){BICEP2 Collaboration and Keck Array Collaboration: {Ade},
  P.~A.~R.}, {Ahmed}, {Aikin}, {Alexander}, {Barkats}, {Benton}, {Bischoff},
  {Bock}, {Bowens-Rubin}, {Brevik}, {Buder}, {Bullock}, {Buza}, {Connors},
  {Cornelison}, {Crill}, {Crumrine}, {Dierickx}, {Duband}, {Dvorkin},
  {Filippini}, {Fliescher}, {Grayson}, {Hall}, {Halpern}, {Harrison},
  {Hildebrandt}, {Hilton}, {Hui}, {Irwin}, {Kang}, {Karkare}, {Karpel},
  {Kaufman}, {Keating}, {Kefeli}, {Kernasovskiy}, {Kovac}, {Kuo}, {Larsen},
  {Lau}, {Leitch}, {Lueker}, {Megerian}, {Moncelsi}, {Namikawa}, {Netterfield},
  {Nguyen}, {O'Brient}, {Ogburn}, {Palladino}, {Pryke}, {Racine}, {Richter},
  {Schillaci}, {Schwarz}, {Sheehy}, {Soliman}, {St.~Germaine}, {Staniszewski},
  {Steinbach}, {Sudiwala}, {Teply}, {Thompson}, {Tolan}, {Tucker}, {Turner},
  {Umilt{\`a}}, {Vieregg}, {Wandui}, {Weber}, {Wiebe}, {Willmert}, {Wong},
  {Wu}, {Yang}, {Yoon}, \& {Zhang}}]{2018PhRvL.121v1301B}
---. 2018, Physical Review Letters, 121, 221301,
  \dodoi{10.1103/PhysRevLett.121.221301}

\bibitem[{{Bischoff}(2010)}]{PhDT_Bischoff}
{Bischoff}, C. 2010, PhD thesis, University of Chicago

\bibitem[{{Bonavera} {et~al.}(2017){Bonavera}, {Gonz{\'a}lez-Nuevo},
  {Arg{\"u}eso}, \& {Toffolatti}}]{2017MNRAS.469.2401B}
{Bonavera}, L., {Gonz{\'a}lez-Nuevo}, J., {Arg{\"u}eso}, F., \& {Toffolatti},
  L. 2017, \mnras, 469, 2401, \dodoi{10.1093/mnras/stx1020}

\bibitem[{{Cardoso} {et~al.}(2008){Cardoso}, {Martin}, {Delabrouille},
  {Betoule}, \& {Patanchon}}]{2008arXiv0803.1814C}
{Cardoso}, J.-F., {Martin}, M., {Delabrouille}, J., {Betoule}, M., \&
  {Patanchon}, G. 2008, arXiv e-prints, arXiv:0803.1814.
\newblock \doarXiv{0803.1814}

\bibitem[{Crowley {et~al.}(2018)}]{Crowley:2018eib}
Crowley, K.~T., {et~al.} 2018, Proc. SPIE Int. Soc. Opt. Eng., 10708, 107083Z,
  \dodoi{10.1117/12.2313414}

\bibitem[{Das {et~al.}(2014)}]{Das:2013zf}
Das, S., {et~al.} 2014, JCAP, 1404, 014, \dodoi{10.1088/1475-7516/2014/04/014}

\bibitem[{{Datta} {et~al.}(2019){Datta}, {Aiola}, {Choi}, {Devlin}, {Dunkley},
  {D{\"u}nner}, {Gallardo}, {Gralla}, {Halpern}, {Hasselfield}, {Hilton},
  {Hincks}, {Ho}, {Hubmayr}, {Huffenberger}, {Hughes}, {Kosowsky},
  {L{\'o}pez-Caraballo}, {Louis}, {Lungu}, {Marriage}, {Maurin}, {McMahon},
  {Moodley}, {Naess}, {Nati}, {Niemack}, {Page}, {Partridge}, {Prince},
  {Staggs}, {Switzer}, {Wollack}, \& {Farren}}]{2019MNRAS.486.5239D}
{Datta}, R., {Aiola}, S., {Choi}, S.~K., {et~al.} 2019, \mnras, 486, 5239,
  \dodoi{10.1093/mnras/sty2934}

\bibitem[{{Dragone}(1978)}]{dragone}
{Dragone}, C. 1978, The Bell System Technical Journal, 2663

\bibitem[{{Essinger-Hileman} {et~al.}(2016){Essinger-Hileman}, {Kusaka},
  {Appel}, {Choi}, {Crowley}, {Ho}, {Jarosik}, {Page}, {Parker}, \&
  {Raghunathan}}]{2016RScI...87i4503E}
{Essinger-Hileman}, T., {Kusaka}, A., {Appel}, J.~W., {et~al.} 2016, Review of
  Scientific Instruments, 87, 094503, \dodoi{10.1063/1.4962023}

\bibitem[{{Foreman-Mackey} {et~al.}(2013){Foreman-Mackey}, {Hogg}, {Lang}, \&
  {Goodman}}]{emcee}
{Foreman-Mackey}, D., {Hogg}, D.~W., {Lang}, D., \& {Goodman}, J. 2013, PASP,
  125, 306, \dodoi{10.1086/670067}

\bibitem[{{G{\'o}rski} {et~al.}(2005){G{\'o}rski}, {Hivon}, {Banday},
  {Wandelt}, {Hansen}, {Reinecke}, \& {Bartelmann}}]{Gorski2005}
{G{\'o}rski}, K.~M., {Hivon}, E., {Banday}, A.~J., {et~al.} 2005, \apj, 622,
  759, \dodoi{10.1086/427976}

\bibitem[{{Hamimeche} \& {Lewis}(2008)}]{2008PhRvD..77j3013H}
{Hamimeche}, S., \& {Lewis}, A. 2008, \prd, 77, 103013,
  \dodoi{10.1103/PhysRevD.77.103013}

\bibitem[{{Hasegawa} {et~al.}(2018){Hasegawa}, {The POLARBEAR Collaboration},
  Ade, Aguilar, Akiba, Ali, \& Arnold}]{MasayaSPIE}
{Hasegawa}, M., {The POLARBEAR Collaboration}, Ade, P., {et~al.} 2018, in
  \procspie, Vol. 10708, Society of Photo-Optical Instrumentation Engineers
  (SPIE) Conference Series, \dodoi{10.1117/12.2311576}

\bibitem[{{Henning} {et~al.}(2018){Henning}, {Sayre}, {Reichardt}, {Ade},
  {Anderson}, {Austermann}, {Beall}, {Bender}, {Benson}, {Bleem}, {Carlstrom},
  {Chang}, {Chiang}, {Cho}, {Citron}, {Corbett Moran}, {Crawford}, {Crites},
  {de Haan}, {Dobbs}, {Everett}, {Gallicchio}, {George}, {Gilbert},
  {Halverson}, {Harrington}, {Hilton}, {Holder}, {Holzapfel}, {Hoover}, {Hou},
  {Hrubes}, {Huang}, {Hubmayr}, {Irwin}, {Keisler}, {Knox}, {Lee}, {Leitch},
  {Li}, {Lowitz}, {Manzotti}, {McMahon}, {Meyer}, {Mocanu}, {Montgomery},
  {Nadolski}, {Natoli}, {Nibarger}, {Novosad}, {Padin}, {Pryke}, {Ruhl},
  {Saliwanchik}, {Schaffer}, {Sievers}, {Smecher}, {Stark}, {Story}, {Tucker},
  {Vanderlinde}, {Veach}, {Vieira}, {Wang}, {Whitehorn}, {Wu}, \&
  {Yefremenko}}]{2018ApJ...852...97H}
{Henning}, J.~W., {Sayre}, J.~T., {Reichardt}, C.~L., {et~al.} 2018, \apj, 852,
  97, \dodoi{10.3847/1538-4357/aa9ff4}

\bibitem[{{Hivon} {et~al.}(2002){Hivon}, {G{\'o}rski}, {Netterfield}, {Crill},
  {Prunet}, \& {Hansen}}]{Hivon2002}
{Hivon}, E., {G{\'o}rski}, K.~M., {Netterfield}, C.~B., {et~al.} 2002, \apj,
  567, 2, \dodoi{10.1086/338126}

\bibitem[{{Johnson} {et~al.}(2007){Johnson}, {Collins}, {Abroe}, {Ade}, {Bock},
  {Borrill}, {Boscaleri}, {de Bernardis}, {Hanany}, \&
  {Jaffe}}]{2007ApJ...665...42J}
{Johnson}, B.~R., {Collins}, J., {Abroe}, M.~E., {et~al.} 2007, \apj, 665, 42,
  \dodoi{10.1086/518105}

\bibitem[{{Keating} {et~al.}(2013){Keating}, {Shimon}, \&
  {Yadav}}]{Keating2013}
{Keating}, B.~G., {Shimon}, M., \& {Yadav}, A.~P.~S. 2013, \apjl, 762, L23,
  \dodoi{10.1088/2041-8205/762/2/L23}

\bibitem[{{Keisler} {et~al.}(2015){Keisler}, {Hoover}, {Harrington}, {Henning},
  {Ade}, {Aird}, {Austermann}, {Beall}, {Bender}, {Benson}, {Bleem},
  {Carlstrom}, {Chang}, {Chiang}, {Cho}, {Citron}, {Crawford}, {Crites}, {de
  Haan}, {Dobbs}, {Everett}, {Gallicchio}, {Gao}, {George}, {Gilbert},
  {Halverson}, {Hanson}, {Hilton}, {Holder}, {Holzapfel}, {Hou}, {Hrubes},
  {Huang}, {Hubmayr}, {Irwin}, {Knox}, {Lee}, {Leitch}, {Li}, {Luong-Van},
  {Marrone}, {McMahon}, {Mehl}, {Meyer}, {Mocanu}, {Natoli}, {Nibarger},
  {Novosad}, {Padin}, {Pryke}, {Reichardt}, {Ruhl}, {Saliwanchik}, {Sayre},
  {Schaffer}, {Shirokoff}, {Smecher}, {Stark}, {Story}, {Tucker},
  {Vanderlinde}, {Vieira}, {Wang}, {Whitehorn}, {Yefremenko}, \&
  {Zahn}}]{Keisler2015}
{Keisler}, R., {Hoover}, S., {Harrington}, N., {et~al.} 2015, \apj, 807, 151,
  \dodoi{10.1088/0004-637X/807/2/151}

\bibitem[{{Kermish} {et~al.}(2012){Kermish}, {Ade}, {Anthony}, {Arnold},
  {Barron}, {Boettger}, {Borrill}, {Chapman}, {Chinone}, {Dobbs}, {Errard},
  {Fabbian}, {Flanigan}, {Fuller}, {Ghribi}, {Grainger}, {Halverson},
  {Hasegawa}, {Hattori}, {Hazumi}, {Holzapfel}, {Howard}, {Hyland}, {Jaffe},
  {Keating}, {Kisner}, {Lee}, {Le Jeune}, {Linder}, {Lungu}, {Matsuda},
  {Matsumura}, {Meng}, {Miller}, {Morii}, {Moyerman}, {Myers}, {Nishino},
  {Paar}, {Quealy}, {Reichardt}, {Richards}, {Ross}, {Shimizu}, {Shimon},
  {Shimmin}, {Sholl}, {Siritanasak}, {Spieler}, {Stebor}, {Steinbach},
  {Stompor}, {Suzuki}, {Tomaru}, {Tucker}, \& {Zahn}}]{Kermish_SPIE2012}
{Kermish}, Z.~D., {Ade}, P., {Anthony}, A., {et~al.} 2012, in Society of
  Photo-Optical Instrumentation Engineers (SPIE) Conference Series, Vol. 8452,
  Society of Photo-Optical Instrumentation Engineers (SPIE) Conference Series,
  \dodoi{10.1117/12.926354}

\bibitem[{{Krachmalnicoff} {et~al.}(2018){Krachmalnicoff}, {Carretti},
  {Baccigalupi}, {Bernardi}, {Brown}, {Gaensler}, {Haverkorn}, {Kesteven},
  {Perrotta}, {Poppi}, \& {Staveley-Smith}}]{2018A&A...618A.166K}
{Krachmalnicoff}, N., {Carretti}, E., {Baccigalupi}, C., {et~al.} 2018, \aap,
  618, A166, \dodoi{10.1051/0004-6361/201832768}

\bibitem[{{Kusaka} {et~al.}(2014){Kusaka}, {Essinger-Hileman}, {Appel},
  {Gallardo}, {Irwin}, {Jarosik}, {Nolta}, {Page}, {Parker}, \&
  {Raghunathan}}]{2014RScI...85c9901K}
{Kusaka}, A., {Essinger-Hileman}, T., {Appel}, J.~W., {et~al.} 2014, Review of
  Scientific Instruments, 85, 039901, \dodoi{10.1063/1.4867655}

\bibitem[{{Kusaka} {et~al.}(2018){Kusaka}, {Appel}, {Essinger-Hileman},
  {Beall}, {Campusano}, {Cho}, {Choi}, {Crowley}, {Fowler}, {Gallardo},
  {Hasselfield}, {Hilton}, {Ho}, {Irwin}, {Jarosik}, {Niemack}, {Nixon},
  {Nolta}, {Page}, {Palma}, {Parker}, {Raghunathan}, {Reintsema}, {Sievers},
  {Simon}, {Staggs}, {Visnjic}, \& {Yoon}}]{2018JCAP...09..005K}
{Kusaka}, A., {Appel}, J., {Essinger-Hileman}, T., {et~al.} 2018, Journal of
  Cosmology and Astro-Particle Physics, 2018, 005,
  \dodoi{10.1088/1475-7516/2018/09/005}

\bibitem[{{Louis} {et~al.}(2017){Louis}, {Grace}, {Hasselfield}, {Lungu},
  {Maurin}, {Addison}, {Ade}, {Aiola}, {Allison}, {Amiri}, {Angile},
  {Battaglia}, {Beall}, {de Bernardis}, {Bond}, {Britton}, {Calabrese}, {Cho},
  {Choi}, {Coughlin}, {Crichton}, {Crowley}, {Datta}, {Devlin}, {Dicker},
  {Dunkley}, {D{\"u}nner}, {Ferraro}, {Fox}, {Gallardo}, {Gralla}, {Halpern},
  {Henderson}, {Hill}, {Hilton}, {Hilton}, {Hincks}, {Hlozek}, {Ho}, {Huang},
  {Hubmayr}, {Huffenberger}, {Hughes}, {Infante}, {Irwin}, {Muya Kasanda},
  {Klein}, {Koopman}, {Kosowsky}, {Li}, {Madhavacheril}, {Marriage}, {McMahon},
  {Menanteau}, {Moodley}, {Munson}, {Naess}, {Nati}, {Newburgh}, {Nibarger},
  {Niemack}, {Nolta}, {Nu{\~n}ez}, {Page}, {Pappas}, {Partridge}, {Rojas},
  {Schaan}, {Schmitt}, {Sehgal}, {Sherwin}, {Sievers}, {Simon}, {Spergel},
  {Staggs}, {Switzer}, {Thornton}, {Trac}, {Treu}, {Tucker}, {Van Engelen},
  {Ward}, \& {Wollack}}]{2016arXiv161002360L}
{Louis}, T., {Grace}, E., {Hasselfield}, M., {et~al.} 2017, JCAP, 2017, 031,
  \dodoi{10.1088/1475-7516/2017/06/031}

\bibitem[{Marriage {et~al.}(2011)Marriage, Juin, \& Lin}]{Marriage2011}
Marriage, T., Juin, J., \& Lin, Y. 2011, The Astrophysical, 731, 1,
  \dodoi{10.1088/0004-637X/731/2/100}

\bibitem[{{Matsuda}(2017)}]{FredPhD}
{Matsuda}, F.~T. 2017, PhD thesis, University of California at San Diego

\bibitem[{Matsuda {et~al.}(2018)Matsuda, Takakura, Arnold, Boettger, Chinone,
  Hazumi, Keating, Kusaka, \& Lee}]{mdbreakingspie}
Matsuda, F.~T., Takakura, S., Arnold, K., {et~al.} 2018, in \procspie, Vol.
  10708, Society of Photo-Optical Instrumentation Engineers (SPIE) Conference
  Series, \dodoi{10.1117/12.2313177}

\bibitem[{{Mizugutch} {et~al.}(1976){Mizugutch}, {Akagawa}, \&
  {Yokoi}}]{1976isap.conf....2M}
{Mizugutch}, Y., {Akagawa}, M., \& {Yokoi}, H. 1976, ISA Proceedings, 2

\bibitem[{{Murphy} {et~al.}(2010){Murphy}, {Sadler}, {Ekers}, {Massardi},
  {Hancock}, {Mahony}, {Ricci}, {Burke-Spolaor}, {Calabretta}, {Chhetri}, {de
  Zotti}, {Edwards}, {Ekers}, {Jackson}, {Kesteven}, {Lindley}, {Newton-McGee},
  {Phillips}, {Roberts}, {Sault}, {Staveley-Smith}, {Subrahmanyan}, {Walker},
  \& {Wilson}}]{at20g}
{Murphy}, T., {Sadler}, E.~M., {Ekers}, R.~D., {et~al.} 2010, \mnras, 402,
  2403, \dodoi{10.1111/j.1365-2966.2009.15961.x}

\bibitem[{{Planck Collaboration} {et~al.}(2014){Planck Collaboration}, {Ade},
  {Aghanim}, {Arg{\"u}eso}, {Armitage-Caplan}, {Arnaud}, {Ashdown},
  {Atrio-Barandela}, {Aumont}, \& {Baccigalupi}}]{2014A&A...571A..28P}
{Planck Collaboration}, {Ade}, P.~A.~R., {Aghanim}, N., {et~al.} 2014, \aap,
  571, A28, \dodoi{10.1051/0004-6361/201321524}

\bibitem[{{Planck Collaboration} {et~al.}(2018{\natexlab{a}}){Planck
  Collaboration}, {Akrami}, {Arroja}, {Ashdown}, {Aumont}, {Baccigalupi},
  {Ballardini}, {Banday}, {Barreiro}, {Bartolo}, {Basak}, {Battye}, {Benabed},
  {Bernard}, {Bersanelli}, {Bielewicz}, {Bock}, {Bond}, {Borrill}, {Bouchet},
  {Boulanger}, {Bucher}, {Burigana}, {Butler}, {Calabrese}, {Cardoso},
  {Carron}, {Casaponsa}, {Challinor}, {Chiang}, {Colombo}, {Combet},
  {Contreras}, {Crill}, {Cuttaia}, {de Bernardis}, {de Zotti}, {Delabrouille},
  {Delouis}, {D{\'e}sert}, {Di Valentino}, {Dickinson}, {Diego}, {Donzelli},
  {Dor{\'e}}, {Douspis}, {Ducout}, {Dupac}, {Efstathiou}, {Elsner},
  {En{\ss}lin}, {Eriksen}, {Falgarone}, {Fantaye}, {Fergusson},
  {Fernandez-Cobos}, {Finelli}, {Forastieri}, {Frailis}, {Franceschi},
  {Frolov}, {Galeotta}, {Galli}, {Ganga}, {G{\'e}nova-Santos}, {Gerbino},
  {Ghosh}, {Gonz{\'a}lez-Nuevo}, {G{\'o}rski}, {Gratton}, {Gruppuso},
  {Gudmundsson}, {Hamann}, {Hand ley}, {Hansen}, {Helou}, {Herranz}, {Hivon},
  {Huang}, {Jaffe}, {Jones}, {Karakci}, {Keih{\"a}nen}, {Keskitalo}, {Kiiveri},
  {Kim}, {Kisner}, {Knox}, {Krachmalnicoff}, {Kunz}, {Kurki-Suonio}, {Lagache},
  {Lamarre}, {Langer}, {Lasenby}, {Lattanzi}, {Lawrence}, {Le Jeune}, {Leahy},
  {Lesgourgues}, {Levrier}, {Lewis}, {Liguori}, {Lilje}, {Lilley}, {Lindholm},
  {L{\'o}pez-Caniego}, {Lubin}, {Ma}, {Mac{\'\i}as-P{\'e}rez}, {Maggio},
  {Maino}, {Mand olesi}, {Mangilli}, {Marcos-Caballero}, {Maris}, {Martin},
  {Mart{\'\i}nez-Gonz{\'a}lez}, {Matarrese}, {Mauri}, {McEwen}, {Meerburg},
  {Meinhold}, {Melchiorri}, {Mennella}, {Migliaccio}, {Millea}, {Mitra},
  {Miville-Desch{\^e}nes}, {Molinari}, {Moneti}, {Montier}, {Morgante}, {Moss},
  {Mottet}, {M{\"u}nchmeyer}, {Natoli}, {N{\o}rgaard-Nielsen}, {Oxborrow},
  {Pagano}, {Paoletti}, {Partridge}, {Patanchon}, {Pearson}, {Peel}, {Peiris},
  {Perrotta}, {Pettorino}, {Piacentini}, {Polastri}, {Polenta}, {Puget},
  {Rachen}, {Reinecke}, {Remazeilles}, {Renzi}, {Rocha}, {Rosset}, {Roudier},
  {Rubi{\~n}o-Mart{\'\i}n}, {Ruiz-Granados}, {Salvati}, {Sandri}, {Savelainen},
  {Scott}, {Shellard}, {Shiraishi}, {Sirignano}, {Sirri}, {Spencer}, {Sunyaev},
  {Suur-Uski}, {Tauber}, {Tavagnacco}, {Tenti}, {Terenzi}, {Toffolatti},
  {Tomasi}, {Trombetti}, {Valiviita}, {Van Tent}, {Vibert}, {Vielva}, {Villa},
  {Vittorio}, {Wandelt}, {Wehus}, {White}, {White}, {Zacchei}, \&
  {Zonca}}]{2018arXiv180706205P}
{Planck Collaboration}, {Akrami}, Y., {Arroja}, F., {et~al.}
  2018{\natexlab{a}}, arXiv e-prints, arXiv:1807.06205.
\newblock \doarXiv{1807.06205}

\bibitem[{{Planck Collaboration} {et~al.}(2018{\natexlab{b}}){Planck
  Collaboration}, {Aghanim}, {Akrami}, {Ashdown}, {Aumont}, {Baccigalupi},
  {Ballardini}, {Banday}, {Barreiro}, {Bartolo}, {Basak}, {Battye}, {Benabed},
  {Bernard}, {Bersanelli}, {Bielewicz}, {Bock}, {Bond}, {Borrill}, {Bouchet},
  {Boulanger}, {Bucher}, {Burigana}, {Butler}, {Calabrese}, {Cardoso},
  {Carron}, {Challinor}, {Chiang}, {Chluba}, {Colombo}, {Combet}, {Contreras},
  {Crill}, {Cuttaia}, {de Bernardis}, {de Zotti}, {Delabrouille}, {Delouis},
  {Di Valentino}, {Diego}, {Dor{\'e}}, {Douspis}, {Ducout}, {Dupac}, {Dusini},
  {Efstathiou}, {Elsner}, {En{\ss}lin}, {Eriksen}, {Fantaye}, {Farhang},
  {Fergusson}, {Fernandez-Cobos}, {Finelli}, {Forastieri}, {Frailis},
  {Franceschi}, {Frolov}, {Galeotta}, {Galli}, {Ganga}, {G{\'e}nova-Santos},
  {Gerbino}, {Ghosh}, {Gonz{\'a}lez-Nuevo}, {G{\'o}rski}, {Gratton},
  {Gruppuso}, {Gudmundsson}, {Hamann}, {Hand ley}, {Herranz}, {Hivon}, {Huang},
  {Jaffe}, {Jones}, {Karakci}, {Keih{\"a}nen}, {Keskitalo}, {Kiiveri}, {Kim},
  {Kisner}, {Knox}, {Krachmalnicoff}, {Kunz}, {Kurki-Suonio}, {Lagache},
  {Lamarre}, {Lasenby}, {Lattanzi}, {Lawrence}, {Le Jeune}, {Lemos},
  {Lesgourgues}, {Levrier}, {Lewis}, {Liguori}, {Lilje}, {Lilley}, {Lindholm},
  {L{\'o}pez-Caniego}, {Lubin}, {Ma}, {Mac{\'\i}as-P{\'e}rez}, {Maggio},
  {Maino}, {Mandolesi}, {Mangilli}, {Marcos-Caballero}, {Maris}, {Martin},
  {Martinelli}, {Mart{\'\i}nez-Gonz{\'a}lez}, {Matarrese}, {Mauri}, {McEwen},
  {Meinhold}, {Melchiorri}, {Mennella}, {Migliaccio}, {Millea}, {Mitra},
  {Miville-Desch{\^e}nes}, {Molinari}, {Montier}, {Morgante}, {Moss}, {Natoli},
  {N{\o}rgaard-Nielsen}, {Pagano}, {Paoletti}, {Partridge}, {Patanchon},
  {Peiris}, {Perrotta}, {Pettorino}, {Piacentini}, {Polastri}, {Polenta},
  {Puget}, {Rachen}, {Reinecke}, {Remazeilles}, {Renzi}, {Rocha}, {Rosset},
  {Roudier}, {Rubi{\~n}o-Mart{\'\i}n}, {Ruiz-Granados}, {Salvati}, {Sandri},
  {Savelainen}, {Scott}, {Shellard}, {Sirignano}, {Sirri}, {Spencer},
  {Sunyaev}, {Suur-Uski}, {Tauber}, {Tavagnacco}, {Tenti}, {Toffolatti},
  {Tomasi}, {Trombetti}, {Valenziano}, {Valiviita}, {Van Tent}, {Vibert},
  {Vielva}, {Villa}, {Vittorio}, {Wand elt}, {Wehus}, {White}, {White},
  {Zacchei}, \& {Zonca}}]{Aghanim:2018eyx}
{Planck Collaboration}, {Aghanim}, N., {Akrami}, Y., {et~al.}
  2018{\natexlab{b}}, arXiv e-prints, arXiv:1807.06209.
\newblock \doarXiv{1807.06209}

\bibitem[{{Planck Collaboration} {et~al.}(2018{\natexlab{c}}){Planck
  Collaboration}, {Aghanim}, {Akrami}, {Ashdown}, {Aumont}, {Baccigalupi},
  {Ballardini}, {Banday}, {Barreiro}, {Bartolo}, {Basak}, {Benabed}, {Bernard},
  {Bersanelli}, {Bielewicz}, {Bond}, {Borrill}, {Bouchet}, {Boulanger},
  {Bucher}, {Burigana}, {Calabrese}, {Cardoso}, {Carron}, {Challinor},
  {Chiang}, {Colombo}, {Combet}, {Couchot}, {Crill}, {Cuttaia}, {de Bernardis},
  {de Rosa}, {de Zotti}, {Delabrouille}, {Delouis}, {Di Valentino}, {Diego},
  {Dor{\'e}}, {Douspis}, {Ducout}, {Dupac}, {Efstathiou}, {Elsner},
  {En{\ss}lin}, {Eriksen}, {Falgarone}, {Fantaye}, {Finelli}, {Frailis},
  {Fraisse}, {Franceschi}, {Frolov}, {Galeotta}, {Galli}, {Ganga},
  {G{\'e}nova-Santos}, {Gerbino}, {Ghosh}, {Gonz{\'a}lez-Nuevo}, {G{\'o}rski},
  {Gratton}, {Gruppuso}, {Gudmundsson}, {Hand ley}, {Hansen},
  {Henrot-Versill{\'e}}, {Herranz}, {Hivon}, {Huang}, {Jaffe}, {Jones},
  {Karakci}, {Keih{\"a}nen}, {Keskitalo}, {Kiiveri}, {Kim}, {Kisner},
  {Krachmalnicoff}, {Kunz}, {Kurki-Suonio}, {Lagache}, {Lamarre}, {Lasenby},
  {Lattanzi}, {Lawrence}, {Levrier}, {Liguori}, {Lilje}, {Lindholm},
  {L{\'o}pez-Caniego}, {Ma}, {Mac{\'\i}as-P{\'e}rez}, {Maggio}, {Maino}, {Mand
  olesi}, {Mangilli}, {Martin}, {Mart{\'\i}nez-Gonz{\'a}lez}, {Matarrese},
  {Mauri}, {McEwen}, {Melchiorri}, {Mennella}, {Migliaccio},
  {Miville-Desch{\^e}nes}, {Molinari}, {Moneti}, {Montier}, {Morgante}, {Moss},
  {Mottet}, {Natoli}, {Pagano}, {Paoletti}, {Partridge}, {Patanchon},
  {Patrizii}, {Perdereau}, {Perrotta}, {Pettorino}, {Piacentini}, {Puget},
  {Rachen}, {Reinecke}, {Remazeilles}, {Renzi}, {Rocha}, {Roudier}, {Salvati},
  {Sand ri}, {Savelainen}, {Scott}, {Sirignano}, {Sirri}, {Spencer}, {Sunyaev},
  {Suur-Uski}, {Tauber}, {Tavagnacco}, {Tenti}, {Toffolatti}, {Tomasi},
  {Tristram}, {Trombetti}, {Valiviita}, {Vansyngel}, {Van Tent}, {Vibert},
  {Vielva}, {Villa}, {Vittorio}, {Wandelt}, {Wehus}, \&
  {Zonca}}]{Aghanim:2018fcm}
---. 2018{\natexlab{c}}, arXiv e-prints, arXiv:1807.06207.
\newblock \doarXiv{1807.06207}

\bibitem[{{Planck Collaboration} {et~al.}(2018{\natexlab{d}}){Planck
  Collaboration}, {Akrami}, {Arg{\"u}eso}, {Ashdown}, {Aumont}, {Baccigalupi},
  {Ballardini}, {Banday}, {Barreiro}, {Bartolo}, {Basak}, {Benabed}, {Bernard},
  {Bersanelli}, {Bielewicz}, {Bonavera}, {Bond}, {Borrill}, {Bouchet},
  {Boulanger}, {Bucher}, {Burigana}, {Butler}, {Calabrese}, {Cardoso},
  {Colombo}, {Crill}, {Cuttaia}, {de Bernardis}, {de Rosa}, {de Zotti},
  {Delabrouille}, {Di Valentino}, {Dickinson}, {Diego}, {Donzelli}, {Ducout},
  {Dupac}, {Efstathiou}, {Elsner}, {En{\ss}lin}, {Eriksen}, {Fantaye},
  {Finelli}, {Frailis}, {Franceschi}, {Frolov}, {Galeotta}, {Galli}, {Ganga},
  {G{\'e}nova-Santos}, {Gerbino}, {Ghosh}, {Gonz{\'a}lez-Nuevo}, {G{\'o}rski},
  {Gratton}, {Gruppuso}, {Gudmundsson}, {Hand ley}, {Hansen}, {Herranz},
  {Hivon}, {Huang}, {Jaffe}, {Jones}, {Karakci}, {Keih{\"a}nen}, {Keskitalo},
  {Kiiveri}, {Kim}, {Kisner}, {Krachmalnicoff}, {Kunz}, {Kurki-Suonio},
  {Lamarre}, {Lasenby}, {Lattanzi}, {Lawrence}, {Leahy}, {Levrier}, {Liguori},
  {Lilje}, {Lindholm}, {L{\'o}pez-Caniego}, {Ma}, {Mac{\'\i}as-P{\'e}rez},
  {Maggio}, {Maino}, {Mand olesi}, {Mangilli}, {Maris}, {Martin},
  {Mart{\'\i}nez-Gonz{\'a}lez}, {Matarrese}, {Mauri}, {McEwen}, {Meinhold},
  {Melchiorri}, {Mennella}, {Migliaccio}, {Molinari}, {Montier}, {Morgante},
  {Moss}, {Natoli}, {Pagano}, {Paoletti}, {Partridge}, {Patanchon}, {Patrizii},
  {Peel}, {Perrotta}, {Pettorino}, {Piacentini}, {Polenta}, {Puget}, {Rachen},
  {Racine}, {Reinecke}, {Remazeilles}, {Renzi}, {Rocha}, {Roudier},
  {Rubi{\~n}o-Mart{\'\i}n}, {Salvati}, {Sandri}, {Savelainen}, {Scott},
  {Seljebotn}, {Sirignano}, {Sirri}, {Spencer}, {Suur-Uski}, {Tauber},
  {Tavagnacco}, {Tenti}, {Terenzi}, {Toffolatti}, {Tomasi}, {Trombetti},
  {Valiviita}, {Vansyngel}, {Van Tent}, {Vielva}, {Villa}, {Vittorio},
  {Wandelt}, {Watson}, {Wehus}, {Zacchei}, \& {Zonca}}]{2018arXiv180706206P}
{Planck Collaboration}, {Akrami}, Y., {Arg{\"u}eso}, F., {et~al.}
  2018{\natexlab{d}}, arXiv e-prints, arXiv:1807.06206.
\newblock \doarXiv{1807.06206}

\bibitem[{{Planck Collaboration} {et~al.}(2018{\natexlab{e}}){Planck
  Collaboration}, {Akrami}, {Ashdown}, {Aumont}, {Baccigalupi}, {Ballardini},
  {Band ay}, {Barreiro}, {Bartolo}, {Basak}, {Benabed}, {Bernard},
  {Bersanelli}, {Bielewicz}, {Bond}, {Borrill}, {Bouchet}, {Boulanger},
  {Bracco}, {Bucher}, {Burigana}, {Calabrese}, {Cardoso}, {Carron}, {Chiang},
  {Combet}, {Crill}, {de Bernardis}, {de Zotti}, {Delabrouille}, {Delouis}, {Di
  Valentino}, {Dickinson}, {Diego}, {Ducout}, {Dupac}, {Efstathiou}, {Elsner},
  {En{\ss}lin}, {Falgarone}, {Fantaye}, {Ferri{\`e}re}, {Finelli},
  {Forastieri}, {Frailis}, {Fraisse}, {Franceschi}, {Frolov}, {Galeotta},
  {Galli}, {Ganga}, {G{\'e}nova-Santos}, {Ghosh}, {Gonz{\'a}lez-Nuevo},
  {G{\'o}rski}, {Gruppuso}, {Gudmundsson}, {Guillet}, {Handley}, {Hansen},
  {Herranz}, {Huang}, {Jaffe}, {Jones}, {Keih{\"a}nen}, {Keskitalo}, {Kiiveri},
  {Kim}, {Krachmalnicoff}, {Kunz}, {Kurki-Suonio}, {Lamarre}, {Lasenby}, {Le
  Jeune}, {Levrier}, {Liguori}, {Lilje}, {Lindholm}, {L{\'o}pez-Caniego},
  {Lubin}, {Ma}, {Mac{\'\i}as-P{\'e}rez}, {Maggio}, {Maino}, {Mandolesi},
  {Mangilli}, {Martin}, {Mart{\'\i}nez-Gonz{\'a}lez}, {Matarrese}, {McEwen},
  {Meinhold}, {Melchiorri}, {Migliaccio}, {Miville-Desch{\^e}nes}, {Molinari},
  {Moneti}, {Montier}, {Morgante}, {Natoli}, {Pagano}, {Paoletti}, {Pettorino},
  {Piacentini}, {Polenta}, {Puget}, {Rachen}, {Reinecke}, {Remazeilles},
  {Renzi}, {Rocha}, {Rosset}, {Roudier}, {Rubi{\~n}o-Mart{\'\i}n},
  {Ruiz-Granados}, {Salvati}, {Sandri}, {Savelainen}, {Scott}, {Soler},
  {Spencer}, {Tauber}, {Tavagnacco}, {Toffolatti}, {Tomasi}, {Trombetti},
  {Valiviita}, {Vansyngel}, {Van Tent}, {Vielva}, {Villa}, {Vittorio}, {Wehus},
  {Zacchei}, \& {Zonca}}]{2018arXiv180104945P}
{Planck Collaboration}, {Akrami}, Y., {Ashdown}, M., {et~al.}
  2018{\natexlab{e}}, arXiv e-prints, arXiv:1801.04945.
\newblock \doarXiv{1801.04945}

\bibitem[{{Planck Collaboration Int. XXX}(2016)}]{2016A&A...586A.133P}
{Planck Collaboration Int. XXX}. 2016, A\&A, 586, A133,
  \dodoi{10.1051/0004-6361/201425034}

\bibitem[{{Puglisi} {et~al.}(2018){Puglisi}, {Galluzzi}, {Bonavera},
  {Gonzalez-Nuevo}, {Lapi}, {Massardi}, {Perrotta}, {Baccigalupi}, {Celotti},
  \& {Danese}}]{2018ApJ...858...85P}
{Puglisi}, G., {Galluzzi}, V., {Bonavera}, L., {et~al.} 2018, \apj, 858, 85,
  \dodoi{10.3847/1538-4357/aab3c7}

\bibitem[{{Smith}(2006)}]{2006PhRvD..74h3002S}
{Smith}, K.~M. 2006, \prd, 74, 083002, \dodoi{10.1103/PhysRevD.74.083002}

\bibitem[{{Smith} \& {Zaldarriaga}(2007)}]{PureEstimator_Smith2006}
{Smith}, K.~M., \& {Zaldarriaga}, M. 2007, \prd, 76, 043001,
  \dodoi{10.1103/PhysRevD.76.043001}

\bibitem[{Story {et~al.}(2013)}]{Story:2012wx}
Story, K.~T., {et~al.} 2013, Astrophys. J., 779, 86,
  \dodoi{10.1088/0004-637X/779/1/86}

\bibitem[{{Suzuki} {et~al.}(2016){Suzuki}, {Ade}, {Akiba}, {Aleman}, {Arnold},
  {Baccigalupi}, {Barch}, {Barron}, {Bender}, {Boettger}, {Borrill}, {Chapman},
  {Chinone}, {Cukierman}, {Dobbs}, {Ducout}, {Dunner}, {Elleflot}, {Errard},
  {Fabbian}, {Feeney}, {Feng}, {Fujino}, {Fuller}, {Gilbert}, {Goeckner-Wald},
  {Groh}, {Haan}, {Hall}, {Halverson}, {Hamada}, {Hasegawa}, {Hattori},
  {Hazumi}, {Hill}, {Holzapfel}, {Hori}, {Howe}, {Inoue}, {Irie}, {Jaehnig},
  {Jaffe}, {Jeong}, {Katayama}, {Kaufman}, {Kazemzadeh}, {Keating}, {Kermish},
  {Keskitalo}, {Kisner}, {Kusaka}, {Jeune}, {Lee}, {Leon}, {Linder}, {Lowry},
  {Matsuda}, {Matsumura}, {Miller}, {Mizukami}, {Montgomery}, {Navaroli},
  {Nishino}, {Peloton}, {Poletti}, {Puglisi}, {Rebeiz}, {Raum}, {Reichardt},
  {Richards}, {Ross}, {Rotermund}, {Segawa}, {Sherwin}, {Shirley},
  {Siritanasak}, {Stebor}, {Stompor}, {Suzuki}, {Tajima}, {Takada}, {Takakura},
  {Takatori}, {Tikhomirov}, {Tomaru}, {Westbrook}, {Whitehorn}, {Yamashita},
  {Zahn}, \& {Zahn}}]{2016JLTP..184..805S}
{Suzuki}, A., {Ade}, P., {Akiba}, Y., {et~al.} 2016, Journal of Low Temperature
  Physics, 184, 805, \dodoi{10.1007/s10909-015-1425-4}

\bibitem[{{Takakura}(2017)}]{SatoruPhD}
{Takakura}, S. 2017, PhD thesis, Osaka University

\bibitem[{{Takakura} {et~al.}(2017){Takakura}, {Aguilar}, {Akiba}, {Arnold},
  {Baccigalupi}, {Barron}, {Beckman}, {Boettger}, {Borrill}, {Chapman},
  {Chinone}, {Cukierman}, {Ducout}, {Elleflot}, {Errard}, {Fabbian}, {Fujino},
  {Galitzki}, {Goeckner-Wald}, {Halverson}, {Hasegawa}, {Hattori}, {Hazumi},
  {Hill}, {Howe}, {Inoue}, {Jaffe}, {Jeong}, {Kaneko}, {Katayama}, {Keating},
  {Keskitalo}, {Kisner}, {Krachmalnicoff}, {Kusaka}, {Lee}, {Leon}, {Lowry},
  {Matsuda}, {Matsumura}, {Navaroli}, {Nishino}, {Paar}, {Peloton}, {Poletti},
  {Puglisi}, {Reichardt}, {Ross}, {Siritanasak}, {Suzuki}, {Tajima},
  {Takatori}, \& {Teply}}]{Takakura:2017ddx}
{Takakura}, S., {Aguilar}, M., {Akiba}, Y., {et~al.} 2017, Journal of Cosmology
  and Astroparticle Physics, 2017, 008.
\newblock \url{http://stacks.iop.org/1475-7516/2017/i=05/a=008}

\bibitem[{{Takakura} {et~al.}(2019){Takakura}, {Aguilar-Fa{\'u}ndez}, {Akiba},
  {Arnold}, {Baccigalupi}, {Barron}, {Beck}, {Bianchini}, {Boettger}, \&
  {Borrill}}]{Takakura:2018wky}
{Takakura}, S., {Aguilar-Fa{\'u}ndez}, M.~A.~O., {Akiba}, Y., {et~al.} 2019,
  \apj, 870, 102, \dodoi{10.3847/1538-4357/aaf381}

\bibitem[{{The BICEP2 and Keck Array Collaborations: P.~A.~R.~Ade}
  {et~al.}(2015){The BICEP2 and Keck Array Collaborations: P.~A.~R.~Ade},
  {Ahmed}, {Aikin}, {Alexander}, {Barkats}, {Benton}, {Bischoff}, {Bock},
  {Brevik}, {Buder}, {Bullock}, {Buza}, {Connors}, {Crill}, {Dowell},
  {Dvorkin}, {Duband}, {Filippini}, {Fliescher}, {Golwala}, {Halpern},
  {Harrison}, {Hasselfield}, {Hildebrandt}, {Hilton}, {Hristov}, {Hui},
  {Irwin}, {Karkare}, {Kaufman}, {Keating}, {Kefeli}, {Kernasovskiy}, {Kovac},
  {Kuo}, {Leitch}, {Lueker}, {Mason}, {Megerian}, {Netterfield}, {Nguyen},
  {O'Brient}, {Ogburn}, {Orlando}, {Pryke}, {Reintsema}, {Richter}, {Schwarz},
  {Sheehy}, {Staniszewski}, {Sudiwala}, {Teply}, {Thompson}, {Tolan}, {Turner},
  {Vieregg}, {Weber}, {Willmert}, {Wong}, \& {Yoon}}]{bkV}
{The BICEP2 and Keck Array Collaborations: P.~A.~R.~Ade}, {Ahmed}, Z., {Aikin},
  R.~W., {et~al.} 2015, \apj, 811, 126, \dodoi{10.1088/0004-637X/811/2/126}

\bibitem[{{The BICEP2 Collaboration: P.~A.~R.~Ade} {et~al.}(2014){The BICEP2
  Collaboration: P.~A.~R.~Ade}, {Aikin}, {Barkats}, {Benton}, {Bischoff},
  {Bock}, {Brevik}, {Buder}, {Bullock}, {Dowell}, {Duband}, {Filippini},
  {Fliescher}, {Golwala}, {Halpern}, {Hasselfield}, {Hildebrandt}, {Hilton},
  {Hristov}, {Irwin}, {Karkare}, {Kaufman}, {Keating}, {Kernasovskiy}, {Kovac},
  {Kuo}, {Leitch}, {Lueker}, {Mason}, {Netterfield}, {Nguyen}, {O'Brient},
  {Ogburn}, {Orlando}, {Pryke}, {Reintsema}, {Richter}, {Schwarz}, {Sheehy},
  {Staniszewski}, {Sudiwala}, {Teply}, {Tolan}, {Turner}, {Vieregg}, {Wong}, \&
  {Yoon}}]{bkI}
{The BICEP2 Collaboration: P.~A.~R.~Ade}, {Aikin}, R.~W., {Barkats}, D.,
  {et~al.} 2014, \prl, 112, 241101, \dodoi{10.1103/PhysRevLett.112.241101}

\bibitem[{{The BICEP2/Keck and Planck Collaborations: P.~A.~R.~Ade}
  {et~al.}(2015){The BICEP2/Keck and Planck Collaborations: P.~A.~R.~Ade},
  {Aghanim}, {Ahmed}, {Aikin}, {Alexander}, {Arnaud}, {Aumont}, {Baccigalupi},
  {Banday}, \& et~al.}]{bkp}
{The BICEP2/Keck and Planck Collaborations: P.~A.~R.~Ade}, {Aghanim}, N.,
  {Ahmed}, Z., {et~al.} 2015, \prl, 114, 101301,
  \dodoi{10.1103/PhysRevLett.114.101301}

\bibitem[{{The Simons Observatory collaboration: {Ade}, Peter}
  {et~al.}(2019){The Simons Observatory collaboration: {Ade}, Peter}, ,
  {Aguirre}, {Ahmed}, {Aiola}, {Ali}, {Alonso}, {Alvarez}, {Arnold}, {Ashton},
  {Austermann}, {Awan}, {Baccigalupi}, {Baildon}, {Barron}, {Battaglia},
  {Battye}, {Baxter}, {Bazarko}, {Beall}, {Bean}, {Beck}, {Beckman},
  {Beringue}, {Bianchini}, {Boada}, {Boettger}, {Bond}, {Borrill}, {Brown},
  {Bruno}, {Bryan}, {Calabrese}, {Calafut}, {Calisse}, {Carron}, {Challinor},
  {Chesmore}, {Chinone}, {Chluba}, {Cho}, {Choi}, {Coppi}, {Cothard},
  {Coughlin}, {Crichton}, {Crowley}, {Crowley}, {Cukierman}, {D'Ewart},
  {D{\"u}nner}, {de Haan}, {Devlin}, {Dicker}, {Didier}, {Dobbs}, {Dober},
  {Duell}, {Duff}, {Duivenvoorden}, {Dunkley}, {Dusatko}, {Errard}, {Fabbian},
  {Feeney}, {Ferraro}, {Flux{\`a}}, {Freese}, {Frisch}, {Frolov}, {Fuller},
  {Fuzia}, {Galitzki}, {Gallardo}, {Tomas Galvez Ghersi}, {Gao}, {Gawiser},
  {Gerbino}, {Gluscevic}, {Goeckner-Wald}, {Golec}, {Gordon}, {Gralla},
  {Green}, {Grigorian}, {Groh}, {Groppi}, {Guan}, {Gudmundsson}, {Han},
  {Hargrave}, {Hasegawa}, {Hasselfield}, {Hattori}, {Haynes}, {Hazumi}, {He},
  {Healy}, {Henderson}, {Hervias-Caimapo}, {Hill}, {Hill}, {Hilton}, {Hilton},
  {Hincks}, {Hinshaw}, {Hlo{\v{z}}ek}, {Ho}, {Ho}, {Howe}, {Huang}, {Hubmayr},
  {Huffenberger}, {Hughes}, {Ijjas}, {Ikape}, {Irwin}, {Jaffe}, {Jain},
  {Jeong}, {Kaneko}, {Karpel}, {Katayama}, {Keating}, {Kernasovskiy},
  {Keskitalo}, {Kisner}, {Kiuchi}, {Klein}, {Knowles}, {Koopman}, {Kosowsky},
  {Krachmalnicoff}, {Kuenstner}, {Kuo}, {Kusaka}, {Lashner}, {Lee}, {Lee},
  {Leon}, {Leung}, {Lewis}, {Li}, {Li}, {Limon}, {Linder}, {Lopez-Caraballo},
  {Louis}, {Lowry}, {Lungu}, {Madhavacheril}, {Mak}, {Maldonado}, {Mani},
  {Mates}, {Matsuda}, {Maurin}, {Mauskopf}, {May}, {McCallum}, {McKenney},
  {McMahon}, {Meerburg}, {Meyers}, {Miller}, {Mirmelstein}, {Moodley},
  {Munchmeyer}, {Munson}, {Naess}, {Nati}, {Navaroli}, {Newburgh}, {Nguyen},
  {Niemack}, {Nishino}, {Orlowski-Scherer}, {Page}, {Partridge}, {Peloton},
  {Perrotta}, {Piccirillo}, {Pisano}, {Poletti}, {Puddu}, {Puglisi}, {Raum},
  {Reichardt}, {Remazeilles}, {Rephaeli}, {Riechers}, {Rojas}, {Roy}, {Sadeh},
  {Sakurai}, {Salatino}, {Sathyanarayana Rao}, {Schaan}, {Schmittfull},
  {Sehgal}, {Seibert}, {Seljak}, {Sherwin}, {Shimon}, {Sierra}, {Sievers},
  {Sikhosana}, {Silva-Feaver}, {Simon}, {Sinclair}, {Siritanasak}, {Smith},
  {Smith}, {Spergel}, {Staggs}, {Stein}, {Stevens}, {Stompor}, {Suzuki},
  {Tajima}, {Takakura}, {Teply}, {Thomas}, {Thorne}, {Thornton}, {Trac},
  {Tsai}, {Tucker}, {Ullom}, {Vagnozzi}, {van Engelen}, {Van Lanen}, {Van
  Winkle}, {Vavagiakis}, {Verg{\`e}s}, {Vissers}, {Wagoner}, {Walker}, {Ward},
  {Westbrook}, {Whitehorn}, {Williams}, {Williams}, {Wollack}, {Xu}, {Yu},
  {Yu}, {Zago}, {Zhang}, \& {Zhu}}]{2019JCAP...02..056A}
{The Simons Observatory collaboration: {Ade}, Peter}, , {Aguirre}, J., {et~al.}
  2019, Journal of Cosmology and Astro-Particle Physics, 2019, 056,
  \dodoi{10.1088/1475-7516/2019/02/056}

\bibitem[{{The \textsc{Polarbear} Collaboration: P.~A.~R.~Ade}
  {et~al.}(2014){The \textsc{Polarbear} Collaboration: P.~A.~R.~Ade}, {Akiba},
  {Anthony}, {Arnold}, {Atlas}, {Barron}, {Boettger}, {Borrill}, {Borys},
  {Chapman}, {Chinone}, {Dobbs}, {Elleflot}, {Errard}, {Fabbian}, {Feng},
  {Flanigan}, {Gilbert}, {Grainger}, {Halverson}, {Hasegawa}, {Hattori},
  {Hazumi}, {Holzapfel}, {Hori}, {Howard}, {Hyland}, {Inoue}, {Jaehnig},
  {Jaffe}, {Keating}, {Kermish}, {Keskitalo}, {Kisner}, {Le Jeune}, {Lee},
  {Leitch}, {Linder}, {Lungu}, {Matsuda}, {Matsumura}, {Meng}, {Miller},
  {Morii}, {Moyerman}, {Myers}, {Navaroli}, {Nishino}, {Paar}, {Peloton},
  {Poletti}, {Quealy}, {Rebeiz}, {Reichardt}, {Richards}, {Ross}, {Rotermund},
  {Schanning}, {Schenck}, {Sherwin}, {Shimizu}, {Shimmin}, {Shimon},
  {Siritanasak}, {Smecher}, {Spieler}, {Stebor}, {Steinbach}, {Stompor},
  {Suzuki}, {Takakura}, {Tikhomirov}, {Tomaru}, {Wilson}, {Yadav}, \&
  {Zahn}}]{pb2014a}
{The \textsc{Polarbear} Collaboration: P.~A.~R.~Ade}, {Akiba}, Y., {Anthony},
  A.~E., {et~al.} 2014, \prl, 112, 131302,
  \dodoi{10.1103/PhysRevLett.112.131302}

\bibitem[{{The \textsc{Polarbear} Collaboration: P.~A.~R.~Ade}
  {et~al.}(2017)}]{pb2017a}
{The \textsc{Polarbear} Collaboration: P.~A.~R.~Ade}, {et~al.} 2017, Astrophys.
  J., 848, 121, \dodoi{10.3847/1538-4357/aa8e9f}

\bibitem[{{Thorne} {et~al.}(2017){Thorne}, {Dunkley}, {Alonso}, \&
  {N{\ae}ss}}]{2017MNRAS.469.2821T}
{Thorne}, B., {Dunkley}, J., {Alonso}, D., \& {N{\ae}ss}, S. 2017, \mnras, 469,
  2821, \dodoi{10.1093/mnras/stx949}

\bibitem[{{Tucci} {et~al.}(2011){Tucci}, {Toffolatti}, {de Zotti}, \&
  {Mart{\'\i}nez-Gonz{\'a}lez}}]{2011A&A...533A..57T}
{Tucci}, M., {Toffolatti}, L., {de Zotti}, G., \& {Mart{\'\i}nez-Gonz{\'a}lez},
  E. 2011, \aap, 533, A57, \dodoi{10.1051/0004-6361/201116972}

\bibitem[{Vieira {et~al.}(2010)Vieira, Crawford, Switzer, Ade, Aird, Ashby,
  Benson, Bleem, Brodwin, Carlstrom, Chang, Cho, Crites, de~Haan, Dobbs,
  Everett, George, Gladders, Hall, Halverson, High, Holder, Holzapfel, Hrubes,
  Joy, Keisler, Knox, Lee, Leitch, Lueker, Marrone, McIntyre, McMahon, Mehl,
  Meyer, Mohr, Montroy, Padin, Plagge, Pryke, Reichardt, Ruhl, Schaffer, Shaw,
  Shirokoff, Spieler, Stalder, Staniszewski, Stark, Vanderlinde, Walsh,
  Williamson, Yang, Zahn, \& Zenteno}]{Vieira2010}
Vieira, J.~D., Crawford, T.~M., Switzer, E.~R., {et~al.} 2010, The
  Astrophysical Journal, 719, 763, \dodoi{10.1088/0004-637X/719/1/763}

\end{thebibliography}

\end{document}